\documentclass[english,iop,apj]{emulateapj}
\usepackage[T1]{fontenc}
\usepackage{verbatim}
\usepackage{graphicx}
\usepackage{subfigure}
\usepackage{tabularx}
\usepackage{amsmath}
\usepackage{natbib}
\usepackage{float}
\usepackage[section]{placeins}

\makeatletter
\slugcomment{}
\shorttitle{}
\shortauthors{}

\usepackage{txfonts}
\usepackage{cleveref}
\usepackage{wrapfig,booktabs}

\def\cB{\p{c\_B}}
\def\lmax{\p{l\_max}}
\def\lumax{\p{l\_umax}}

\def\accrate{\p{acc\_rate}}
\def\acceff{\p{acc\_eff}}
\def\acclim{\p{acc\_lim}}
\def\racc{\p{r\_acc}}
\def\rhofrac{\p{rho\_fr}}

\def\apjl{ApJL }
\def\aj{AJ }
\def\apj{ApJ }
\def\pasp{PASP }

\def\apjs{ApJS }

\def\aap{A\&A }
\def\nat{Nature }
\def\na{New Astronomy }

\def\apss{\ref@jnl{Ap\&SS}}
\def\jgr{\ref@jnl{J.~Geophys.~Res.}}

\def\mnras{MnRAS}
\def\qjras{\ref@jnl{QJRAS}}

\newcommand{\p}[1]{{\tt{#1}}}
\newcommand{\unit}[1]{\ensuremath{\, \mathrm{#1}}}
\newcommand{\ramses}{{\sc ramses}}          
\newcommand{\stagger}{{\sc stagger}}        

\newcommand{\MAacc}{M$_{\odot}\,$AU$^{-2}\,$yr$^{-1}$}
\newcommand{\Macc}{M$_{\odot}$yr$^{-1}$}

\newcommand{\Fig}[1]{Fig.~\ref{fig:#1}}    
\newcommand{\Figure}[1]{Figure \ref{fig:#1}}

\def\rhu{\unit{g}\unit{cm}^{-3}}
\def\kms{\unit{km}\unit{s}^{-1}}

\makeatother

\usepackage[english]{babel}

\begin{document}

\title{Zoom-Simulations of Protoplanetary Disks starting from GMC scales}

\author{Michael Kuffmeier$^\dagger$}

\affil{Centre for Star and Planet Formation, Niels Bohr Institute and Natural History Museum of Denmark, University of Copenhagen,
{\O}ster Voldgade 5-7, DK-1350 Copenhagen K, Denmark}

\email{$^\dagger$kueffmeier@nbi.ku.dk}

\author{Troels Haugb{\o}lle}

\affil{Centre for Star and Planet Formation, Niels Bohr Institute and Natural History Museum of Denmark, University of Copenhagen,
{\O}ster Voldgade 5-7, DK-1350 Copenhagen K, Denmark}

\author{{\AA}ke Nordlund}

\affil{Centre for Star and Planet Formation, Niels Bohr Institute and Natural History Museum of Denmark, University of Copenhagen,
{\O}ster Voldgade 5-7, DK-1350 Copenhagen K, Denmark}

\date{\today}

\begin{abstract}
We investigate the formation of protoplanetary disks around nine solar mass stars
formed in the context of a (40 pc)$^3$ Giant Molecular Cloud model,
using \ramses \ adaptive-mesh refinement simulations extending over
a scale range of about 4 million, from an outer scale of 40 pc down to cell sizes of 2 AU.
Our most important result is that the accretion process is heterogeneous in multiple ways;
in time, in space, and among protostars of otherwise similar mass.
Accretion is heterogeneous in time, in the sense that accretion rates vary during the evolution,
with generally decreasing profiles,
whose slopes vary over a wide range,
and where accretion can increase again if a protostar enters a region
with increased density and low speed.
Accretion is heterogeneous in space, because of the mass distribution,
with mass approaching the accreting star-disk system in filaments and sheets.
Finally, accretion is heterogeneous among stars,
since the detailed conditions and dynamics in the neighborhood of each star can vary widely.
We also investigate the sensitivity of disk formation to physical conditions,
and test their robustness by varying numerical parameters.
We find that disk formation is robust even when choosing the least favorable sink particle parameters,
and that turbulence cascading from larger scales is a decisive factor in disk formation.
We also investigate the transport of angular momentum,
finding that the net inward mechanical transport is compensated for mainly
by an outward directed magnetic transport,
with a contribution from gravitational torques usually subordinate to the magnetic transport.
\end{abstract}


\maketitle

\section{Introduction}

Since the first observations of exoplanets \citep{1992Natur.355..145W,1995Natur.378..355M} interest
in modeling planet formation has grown enormously.
All planets in the solar system and the majority of
detected exoplanets move in nearly co-planar orbits,
consistent with planets forming in circumstellar disks,
and these disks are therefore commonly referred to as protoplanetary disks.

Observations of disks around young stellar objects strongly indicate that
protoplanetary disks are connected to the early formation phase of the host star,
where vigorous accretion from a thick envelope is controlling the dynamics of
the disk-star system.
Hence, protoplanetary disk should ideally be modeled in the context of star formation.
Existing models typically start from more or less idealized initial
conditions, often in the form of a spherical stellar core with a Bonnor-Ebert like profile,
and follow the collapse during star- and protoplanetary
disk formation \citep{2004MNRAS.348L...1M,2006ApJ...645.1227M,2007ApJ...670.1198M,%
2011MNRAS.413.2767M,2012A&A...543A.128J,2013A&A...554A..17J,2010ApJ...714L..58T,%
2013ApJ...763....6T,2011ApJ...738..180L,2011MNRAS.417.1054S,2012MNRAS.422..347S,2013ApJ...767L..11K,2013ApJ...766...97M,2014MNRAS.437...77B,2016ApJ...822...11P}.
Although these cloud collapse models are useful
to study specific cases of star- and protoplanetary disk formation,
they neglect the
influence of the stellar environment. In fact, stars preferentially form
due to the collapse of pre-stellar cores in filamentary structures of
Giant Molecular Clouds (GMCs) \citep{1993prpl.conf..125B}, and as illustrated
by the recent simulations of \citet{2014ApJ...797...32P}, the accretion
rates are in general heavily influenced by external factors, resulting in formation
times that can vary with an order of magnitude or more for similar mass stars.
Therefore, modeling star formation under idealized assumptions, with isolated
boundary and initial conditions, is an oversimplification.

In this paper we present a fundamentally different approach to modeling
star and protoplanetary disk formation, accounting for the influence of the environment
{\em ab initio}. In order to avoid the dependency of strongly idealized
initial and boundary conditions, we start from GMC scales and follow
the process of individual star formation and protoplanetary disk formation
down to scales of 2 AU, properly anchoring the formation of individual stars
in the larger GMC context. We model the GMC in
a cubic box of size (40 pc)$^3$, with a total mass of $\approx10^{5}\unit{M}_{\odot}$.
This is motivated by observations by for example \citet{2011ApJ...729..133M}, who
found GMCs ranging between 5 to 200 pc in
size, with masses from $10^{3}$ to $10^{7}$ $M_{\odot}$.
Realistic turbulence in our GMC model is driven by feedback from supernovae,
which are evolved self-consistently by allowing massive sink-particles to
explode and enrich the surrounding medium. The numerical method and
the model is discussed in Section 2. In Section 3 we present the stellar accretion and
disk formation processes, as obtained by our simulations.
In Section 4 we discuss the results.
Finally, we summarize in Section 5.

\section{Methods}

\begin{figure}[!htbp]
\subfigure{\includegraphics[width=\linewidth]{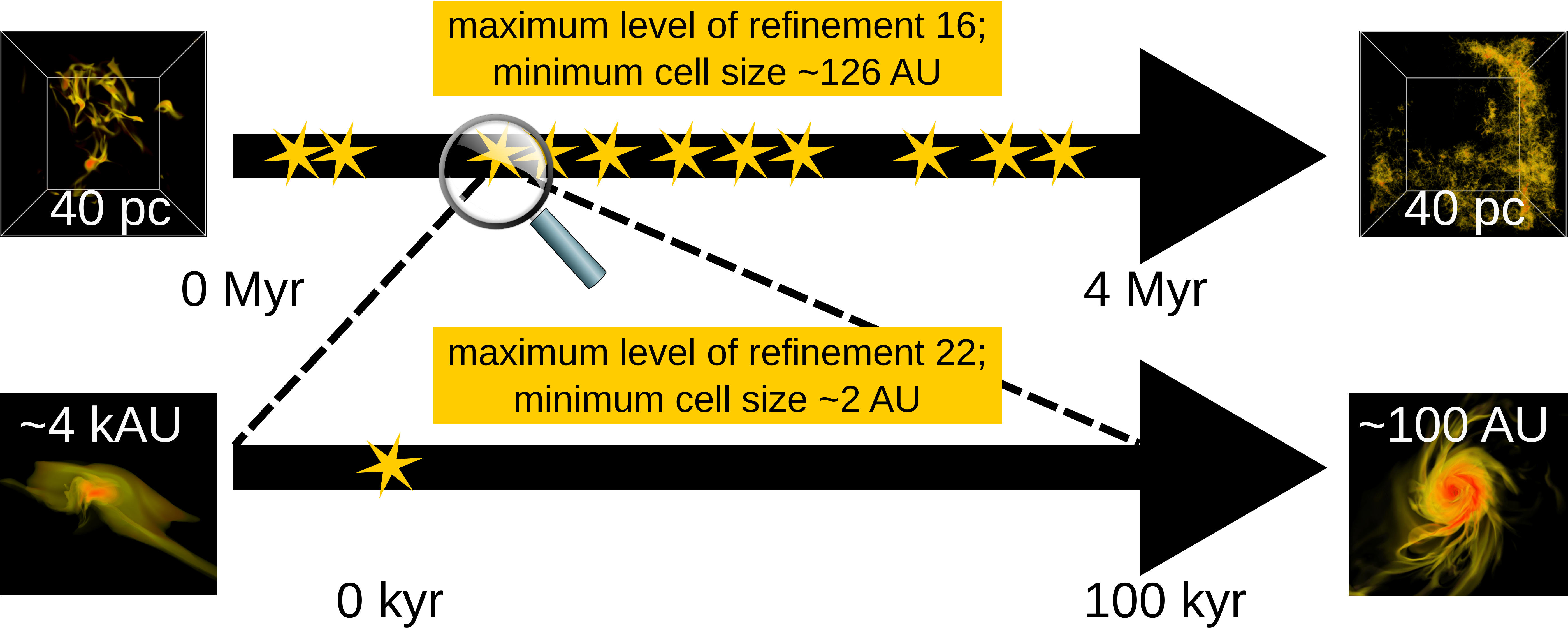}} \quad
\protect\caption{\label{fig:time-sketch} Sketch of the zoom-in procedure.
First we evolve a snapshot adopted from a previous simulation of a
GMC of $(40 \unit{pc})^3$ in size (upper left image) for about 4 Myr (upper right image).
During the evolution multiple sinks are created;
at the current time more than 500 stars have formed, and several supernovae have exploded.
We zoom in on selected 1-2 solar mass pre-stellar cores (lower left image) to resolve the
formation process with higher resolution for up to about 100 kyr after sink creation (lower right image).
This procedure is applied to altogether nine protostars.}
\end{figure}

In this section we describe the setup of our zoom simulations, with
focus on the refinement and accretion criteria, since these are the
key technical parameters of our model. Further descriptions of the
implementations of sink particles can be found in \citet{2014ApJ...797...32P} and in \citet{2016ApJ...826...22K}.
Our approach is feasible due to extensive application
of adaptive mesh refinement (AMR), using a heavily modified
version of the \ramses\ code \citep{2002A&A...385..337T,2006A&A...457..371F},
which can handle up to 29 levels of refinement relative to the box size \citep{2014IAUS..299..131N}.
In \Fig{time-sketch}, we sketch the procedure for our simulations.

\subsection{Evolving the initial conditions}
The starting point of the entire simulation is a GMC model, which is represented in \ramses\  by a cubic box of size (40 pc)$^3$,
with periodic boundary conditions. The box is filled with a self-gravitating, magnetized gas.
The average $H_{2}$ number density is initially $30$ cm$^{-3}$,
corresponding to a total mass in the box of approximately $10^{5}$M$_{\odot}$.
The initial magnetic field strength is 3.5 $\mu$G.
Initially, the box was evolved with driven turbulence and star cluster particles,
including supernovae, using the unigrid MHD {\stagger} code \citep{2011ApJ...737...13K}.
After an initial period of burn-in, a snapshot was restarted with \ramses, and refinement added
\citep[cf.][]{2013ApJ...769L...8V}.
Potentially a real GMC fragment might be broken
up by supernovae explosions or other forms of feedback. In the experiment,
this is prevented by the periodic boundary conditions, which can be interpreted as
corresponding to embedment in a similar or larger amount of nearby gas. The
assumed GMC lifetimes are in agreement with the 'star formation in
a crossing time' paradigm \citep{2000ApJ...530..277E,2003MNRAS.338..817E},
with recent numerical modeling \citep{2016ApJ...822...11P},
and with observational estimates \citep{2011ApJ...729..133M}.
To initialize a turbulent state solenoidal forcing in a shell of wave numbers in
the range of $1/L\le k/2\pi\le2/L$ was initially applied \citep{2002ApJ...576..870P},
but this forcing was turned off when sufficient forcing from exploding supernovae had
developed; no artificial forcing was active during the time interval reported on here.
The induced velocity dispersion of the cold and dense
medium is consistent with Larson's velocity law \citep{1969MNRAS.145..271L,1981MNRAS.194..809L}
\begin{equation}
\sigma_{u}\propto 1.2 L^{0.4}
\end{equation}
with $\sigma_{u}$ being the velocity dispersion in $\kms$ and $L$ the size in pc.
In reference to the recipe of \citep{1986PASP...98.1076F} for heating
due to UV-photons \citep{2006agna.book.....O} heating
is quenched in dense gas, and by using an optically thin cooling function
motivated by \citet{2012ApJS..202...13G} the heating
and cooling processes are otherwise modeled schematically as optically thin.

Based on lifetimes interpolated from \citet{1992A&AS...96..269S},
the evolution of stars with more than $8M_{\odot}$
is followed until they explode as core-collapse supernovae. According
to observations \citep{2002AJ....123..745R} supernovae release approximately 1 foe,
corresponding to $10^{51}$ erg of thermal energy. The integrated energy
yields from the stellar winds are expected to be smaller, and we therefore
omit their modeling (modeling winds from especially early type stars
in the current connection would also be extremely costly with current computer
codes, since the time step must be kept correspondingly short in the entire
model, forcing reduced spatial resolution and / or other compromises in
other parts of the model).
Initial temperatures of supernovae
are of the order of $10^{8}$ K and they expand with initial velocities of
the order of 1000 to 3000 km s$^{-1}$, thus requiring calculations
with time steps of only a few weeks for a brief period after each supernova
explosion.

\subsection{Zoom-in procedure}
\begin{table}[!htbp]
\centering
{
\begin{tabular}{r|r|r|r|r|r|r|r}

\begin{tabular}{@{}c@{}} \# of \\ sink \end{tabular} & \begin{tabular}{@{}c@{}} $\Delta x_{\rm min}$ \\ in AU \end{tabular} &
\begin{tabular}{@{}c@{}} $t_{\rm birth}$ \\ in kyr \end{tabular} & \begin{tabular}{@{}c@{}} x \\ in pc \end{tabular} &
\begin{tabular}{@{}c@{}} y \\ in pc \end{tabular} & \begin{tabular}{@{}c@{}} z \\ in pc \end{tabular} &
\begin{tabular}{@{}c@{}} $t_{\rm acc}$ \\ in kyr \end{tabular} &
\begin{tabular}{@{}c@{}} $m_{\rm sink}$ \\ in M$_{\odot}$ \end{tabular} \\ \hline
1  & 2   & 631  & 33.2 & 30.8 &  7.8 & 200  & 1.5  \\
2  & 2   & 667  & 13.5 & 27.4 & 25.6 & 200  & 1.3  \\
3  & 8   & 2212 & 37.9 & 27.3 & 33.0 & 1000 & 1.9  \\
4  & 2   & 2471 & 3.2  & 9.2  &  3.2 & 700  & 1.4  \\
5  & 2   & 2576 & 3.5  & 8.9  &  2.6 & 700  & 1.6  \\
6  & 2   & 2653 & 10.2 & 12.3 &  3.4 & 600  & 1.3  \\
7  & 2   & 3157 & 9.3  & 12.0 & 32.3 & 600  & 1.7  \\
8  & 2/8 & 3271 & 26.1 & 29.3 &  2.6 & >600 & 2    \\
9  & 2   & 3389 & 3.3  & 4.6  &  2.2 & 500  & 1.5  \\
\end{tabular} }
\caption{Overview of the nine sinks selected for zoom-in.
First column: number of sink, second column: cell size at the highest resolution,
third column: time of creation of the sink in the parental run,
fourth to sixth column: $x$, $y$ and $z$-coordinate of the sink at the time of creation,
seventh column: accretion time in the parental run, and
eighth column: final mass in the parental run.
Star 8 has been resolved with a minimum cell size of 2 AU for the first approximately 15 kyr, before reducing the resolution to 8 AU.}
\label{run-overview}
\end{table}

During the evolution of the GMC with \ramses, structures of high
density form as a consequence of convergent flows and compression
due to shocks in cold dense molecular clouds, which are characterized by super-sonic turbulence.
If the compression is sufficient, self-gravity overcomes opposing pressure forces, resulting in collapse. When the gas
density exceeds a critical value, well above the densities reached by turbulence alone,
and conditions fulfill additional
criteria such as convergence of flow, local potential minimum, etc \citep[see e.g.~][for details]{2014ApJ...797...32P},
the collapsing mass is transferred to a so called sink particle. The sink particle
represents a star, which only influences the environment through its
gravity and by accreting surrounding gas.

The analysis of star formation and protoplanetary disk formation around
single stars with \ramses\ is performed in two steps, using adaptive mesh
refinement on top of a uniform grid with $128^3$ cells (corresponding to 7 levels).
First, star formation is followed during 4 Myr of evolution, with 9 levels of refinement.
I.e.~the highest level of refinement is $\ell=16$, corresponding to a dynamic range
of $2^{16}$ relative to the entire box size of (40 pc)$^3$.  The minimum cell size
is thus $\Delta x_\textrm{min}= 2^{-\ell} \times 40 \unit{pc} \approx 126 \unit{AU}$.

In the next step, the environment of a single sink with a final mass of 1 to 2 M$_{\odot}$ is followed in
more detail, by zooming in on the particular object with an increased
maximum refinement level of $\ell=22$, corresponding to a minimum cell size of $\approx2$
AU, and by introducing an accordingly higher time cadence of output from
the simulation (in the following called snapshots).
The cadence is chosen to lie in the range 0.1 to 0.5 kyr. This second
stage is the key aspect of this study, given that it provides information
about protostellar accretion, including the
subsequent formation of protoplanetary disks.
To minimize numerical diffusion, and to simplify subsequent analysis, we perform velocity
transformations that keep the chosen sink particle stationary in the models frame of reference
during the second simulation step.
To avoid any tendency of coarse grid imprint on smaller scales that a sudden increase in resolution might have on
an already initiated collapse, we start the zoom-in simulations at a time shortly before the sink particle formed in the
parental run. We also allow the adaptive mesh refinement to adjust continuously and the AMR levels to appear sequentially
according to the developing collapse.

In total, we follow the evolution of the environment of nine sinks with higher resolution.
In Table \ref{run-overview}, we show the properties of the selected sinks.
We refine cells if they fulfill certain criteria, e.g. on mass density and distance from the sink particle.
When the protostellar system evolves most of the mass in the initial pre-stellar core is consumed by the star, and the
average density in the remaining envelope thus decreases. To maintain sufficient resolution in the inner part of the system
of the order of $10^6$ cells on highest level of refinement, we slowly change the density
refinement ladder during the run.
We illustrate the resulting distribution of cells refined to the highest level for the different sinks in \Fig{lmax_distr}.
For a more detailed description of the refinement strategy, please see the appendix.

\section{Results}
In this section, we present and discuss the influence of the large-scale environment on
the evolution of solar mass sinks.

\begin{figure}[!htbp]
\subfigure{\includegraphics[width=\linewidth]{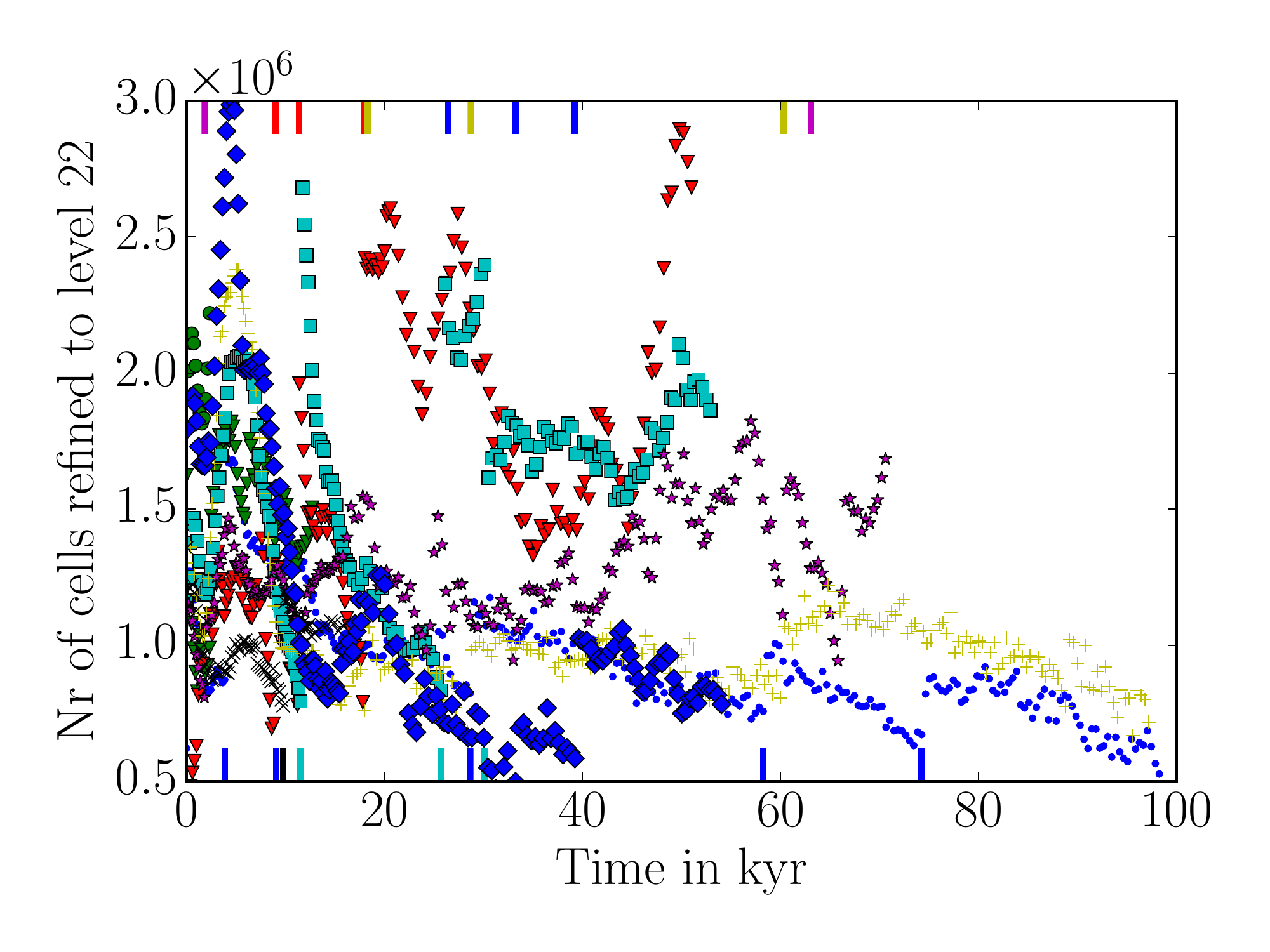}} \quad
\protect\caption{\label{fig:lmax_distr} Distribution of number of cells that are refined to level 22 (corresponding to a cell size of $\approx 2$ AU).
Blue dots correspond to sink 1, green triangles to sink 2, green dots to sink 3, red triangles to sink 4, cyan squares to sink 5,
magenta asterisks to sink 6, yellow pluses to sink 7, black crosses to sink 8 and blue diamonds to sink 9.
The colored tick marks on the x-axes show the time, when the refinement ladder was changed.
Tick marks on lower axis: Blue corresponds to sink 1, cyan to sink 5 and black to sink 8.
Tick marks on the upper axis: red to sink 4, magenta to sink 6, yellow to sink 7, blue to sink 9.  }
\end{figure}

\subsection{Structure of Stellar Environments}

\begin{figure*}[!htbp]
\subfigure{\includegraphics[width=\linewidth]{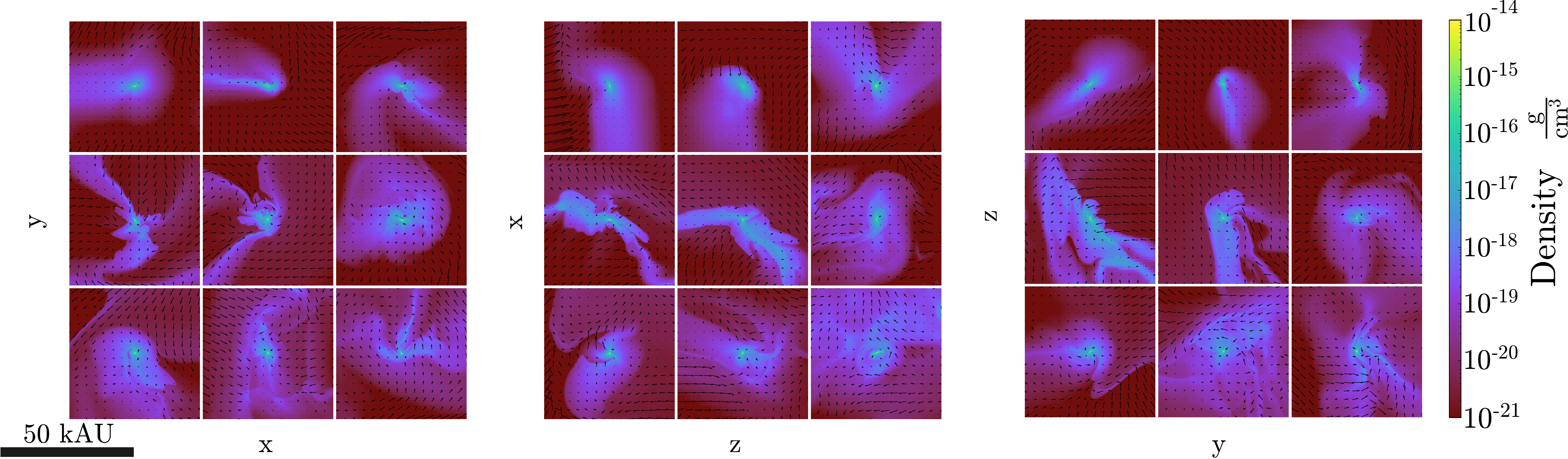}} \quad
\protect\caption{\label{fig:z-large-scale-env} Density distribution in the xy-plane around nine sinks evolving to masses
in-between 1 to 2 M$_\odot$. The black arrows indicate the orientation and strength of the velocity field in the plane.}
\end{figure*}

To illustrate the structure around the different sinks in detail,
we plot the density distribution in the three different planes of the coordinate system, together with the velocity field,
in rectangular slices of size $l=$50 kAU in \Fig{z-large-scale-env}.
The slices illustrate the shape of the pre-stellar cores and their environments at the time of sink creation.
At large distances from the sink (beyond $\approx 10$ kAU), the slices show very low densities --- less than $10^{-21}$ g$/$cm$^{-3}$ for sink 1 and 2.
Especially compared to sinks 4 to 9, the density is distributed more homogeneously around sinks 1 and 2 than for the other sinks,
where the pre-stellar cores are more perturbed.
In contrast to what is assumed in many local molecular cloud collapse models,
none of the pre-stellar cores appear as even roughly spherically symmetric.
The core around sink 1 is closest to resembling a classical pre-stellar core, flattened
due to the underlying magnetic field \citep{2003ApJ...599..363A}.
However, even in the least disturbed cases the cores show strong velocity differences with respect to their low density environment.
Moreover, some of the sinks appear to be at the end-points of filamentary arms about $\sim$100 AU to $\sim$1000 AU
in widths, which reach out to distances beyond several tens of kAU (cf.\ sink 2, 3, 4, 5, 6, 7 and 8 in xy-plane;
4, 5, 6 and 8 in xz-plane as well as 4, 5 and 9 in yz-plane). This may be related to the gravitational collapse, which
tends to concentrate mass at the end of filaments \citep{2015MNRAS.452.2410S}.

\begin{figure}[!htbp]
\subfigure{\includegraphics[width=\linewidth]{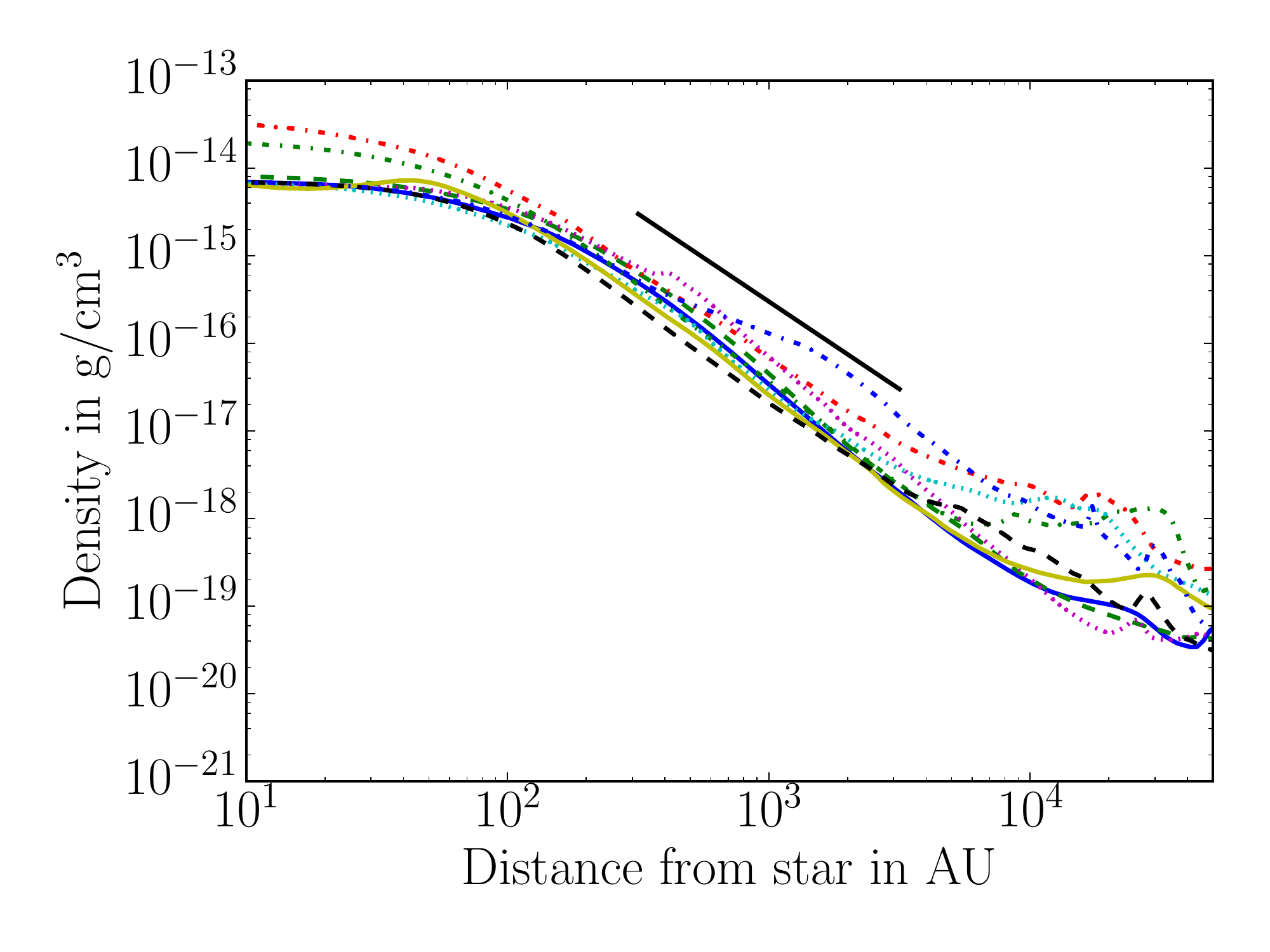}} \quad
\protect\caption{\label{fig:r-dens-0} Radial density profile for the gas around nine sinks in our simulation,
at times within a few hundred years of sink creation.
Blue solid corresponds to sink 1, green dash-dot to sink 2, green dash-dash to sink 3, red dash-dot to sink 4,
cyan dot-dot to sink 5, magenta dot-dot to sink 6, yellow solid to sink 7, black dash-dash to sink 8 and blue dash-dot to sink 9.
The black solid line between 300 and 3000 AU indicates a $\rho \propto r^{-2}$ profile
corresponding to a hypothetical spherical, isothermal density profile.}
\end{figure}

In \Fig{r-dens-0} we show the density distribution (average densities in spherical shells)
in the radial direction for the snapshots closest to (generally within a few hundred years of) sink creation.
We can see that depending on the sink, the densities can vary by several orders of magnitudes at
distances beyond $10^3$ AU to $10^4$ AU.
For comparison, we also plot the density profile of a singular isothermal sphere (SIS)
\begin{equation}
\rho_{\rm SIS} = \frac{\sigma_v^2}{2\pi Gr^2}.
\end{equation}

The averaged density profiles do not differ dramatically from a spherical isothermal Bonnor-Ebert density profile
with $\rho \propto r^{-2}$ and a central flat core.
Although the azimuthally averaged density profiles around different protostars may be similar,
taking into account the perturbations of and around the pre-stellar cores
(\Fig{z-large-scale-env}),
the actual stellar environment can nonetheless be significantly different due to the filamentary structure.
The densities at larger distances from sink 1 and 2 are lower than $10^{-21} \unit{g}\unit{cm}^{-3}$,
whereas the other sinks show higher densities beyond distances of about $10^4$ AU.
At radii smaller than $\approx 10^3$ AU the sinks have a similar generic profile, with a flat density distribution
close to the sink that decreases beyond a certain radius.
We require a threshold value for the density to trigger sink creation, which corresponds to having the Jeans length resolved
by a number of cells at the highest level of refinement. Therefore the inner flat core in the profile is approximately at the same
level of several $10^{-15} \rhu$ just after sink formation for all sinks, except for sink 4, where we required slightly different conditions
for sink creation, namely a higher threshold density, corresponding to 25 instead of 50 cells per Jeans length.
By using our method of sink creation, the lower the maximum level of refinement, the earlier the sink forms.
Therefore, the time of sink creation forestalls the actual birth of the protostar (with an amount of time similar
to the free-fall time at the central densities, which is $< 1000$ yr).
Such a time delay is short compared to the evolution around
young protostars for time intervals of the order of $100$ kyr, and therefore negligible.

\subsection{Accretion profiles}
In this section, we investigate the evolution of the sinks for the first 1 Myr, based on the parental run,
before we focus on the zoom-ins for a more detailed study of their evolution during the first 100 kyr.

\subsubsection{Following the evolution during the first 1 Myr at low resolution}
\begin{figure}[!htbp]
\subfigure{\includegraphics[width=\linewidth]{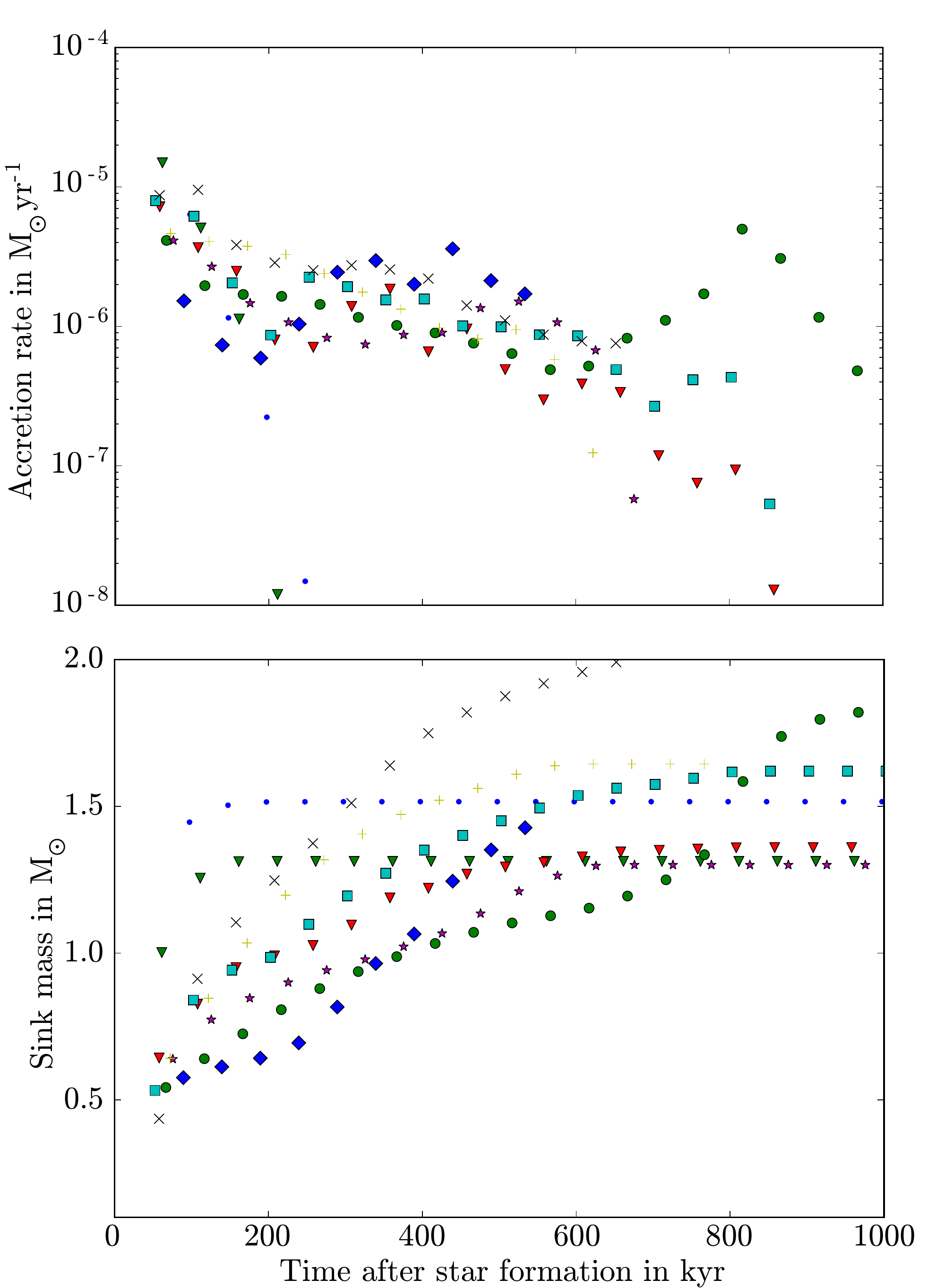} }
\protect\caption{\label{fig:acc_mass_run9} Accretion profiles (top) and mass evolution (bottom)
of the nine sinks that are selected for zoom-ins during the first 1 Myr after sink creation.
Blue dots correspond to sink 1, green triangles to sink 2, green dots to sink 3, red triangles to sink 4, cyan squares to sink 5, magenta asterisks to sink 6,
yellow pluses to sink 7, black crosses to sink 8 and blue diamonds to sink 9.}
\end{figure}

\Figure{acc_mass_run9} shows the mass accretion and the corresponding accretion rates for the different sinks.
Intentionally, we select sinks that accrete to masses higher than 1 M$_\odot$ to constrain the formation of a solar-mass star.
The selected sinks are representative for stars that accrete to about 1 M$_\odot$, because in the low resolution step
we neither resolve mass outflows
around the sinks, nor do we artificially remove a certain fraction of the accreting mass.
Taking into account that protostellar mass loss is expected to be about 50 \% of the accreted mass \citep{2007A&A...469..811Z}, the mass range is appropriate for our purposes.
Given that we use an output cadence of 50 kyr in the low-resolution parental run,
we can only roughly estimate the accretion onto the sink, but it is striking that the sinks accrete their masses on very different time-scales.
While sink 1 and 2 accrete most of their mass within only about 100 kyr,
some other sinks accrete only a fraction of the mass that they have after 1 Myr in that time,
and in some cases undergo long periods with accretion rates exceeding $10^{-6}$ M$_\odot$/yr at times later than 800 kyr after sink creation.
Considering the different accretion profiles together with the fact that some of the selected sinks have only evolved for less than 1 Myr in the parental run,
we are aware that some of the sinks have not yet finished their accretion process or may be fed at later times after quiet periods with in-falling mass from large scales.

Comparing the different accretion profiles in the context of the stellar environment, we conclude that more isolated,
less distorted pre-stellar cores (star 1 and 2) correlate with a shorter accretion time of the protostar. Essentially, they
consume the gas reservoir of their pre-stellar core, and the accretion process stops. This is in accordance
with a classical picture of star formation from spherically symmetric isolated pre-stellar cores.
The differences in duration of the accretion process are a consequence of the spatial extent of the pre-stellar core or
its possible feeding filament(s). Taking into account that the free-fall time scales as $t_{\rm ff} \propto R^{3/2}$,
infalling mass located at larger distances from the sink accretes later onto the sink.
As seen in \Fig{z-large-scale-env}, several pre-stellar cores are strongly perturbed,
and many sinks lie along filaments of gas that extend to distances beyond $10^4$ AU from the sink.
These filaments can feed the sink during its accretion phase with material initially located far away from the sink,
thus prolonging the accretion time scale compared to the classical scenario of a collapsing Bonnor-Ebert sphere of about $10^4$ AU in size ---
even though the collapsing cores may in fact initially be smaller. As pointed out by \citet{2014ApJ...797...32P},
mass clumps at distances beyond $10^4$ AU may eventually fall onto the system and thus cause accretion bursts,
which potentially explain the luminosity problem observed for YSOs \citep{1990AJ.....99..869K}.
We investigate the effect of the accretion profiles on the stellar evolution including a detailed comparison with observations in a follow-up paper.

\subsubsection{Accretion during the first 100 kyr of evolution}

\begin{figure}[!htbp]
\subfigure{\includegraphics[width=\linewidth]{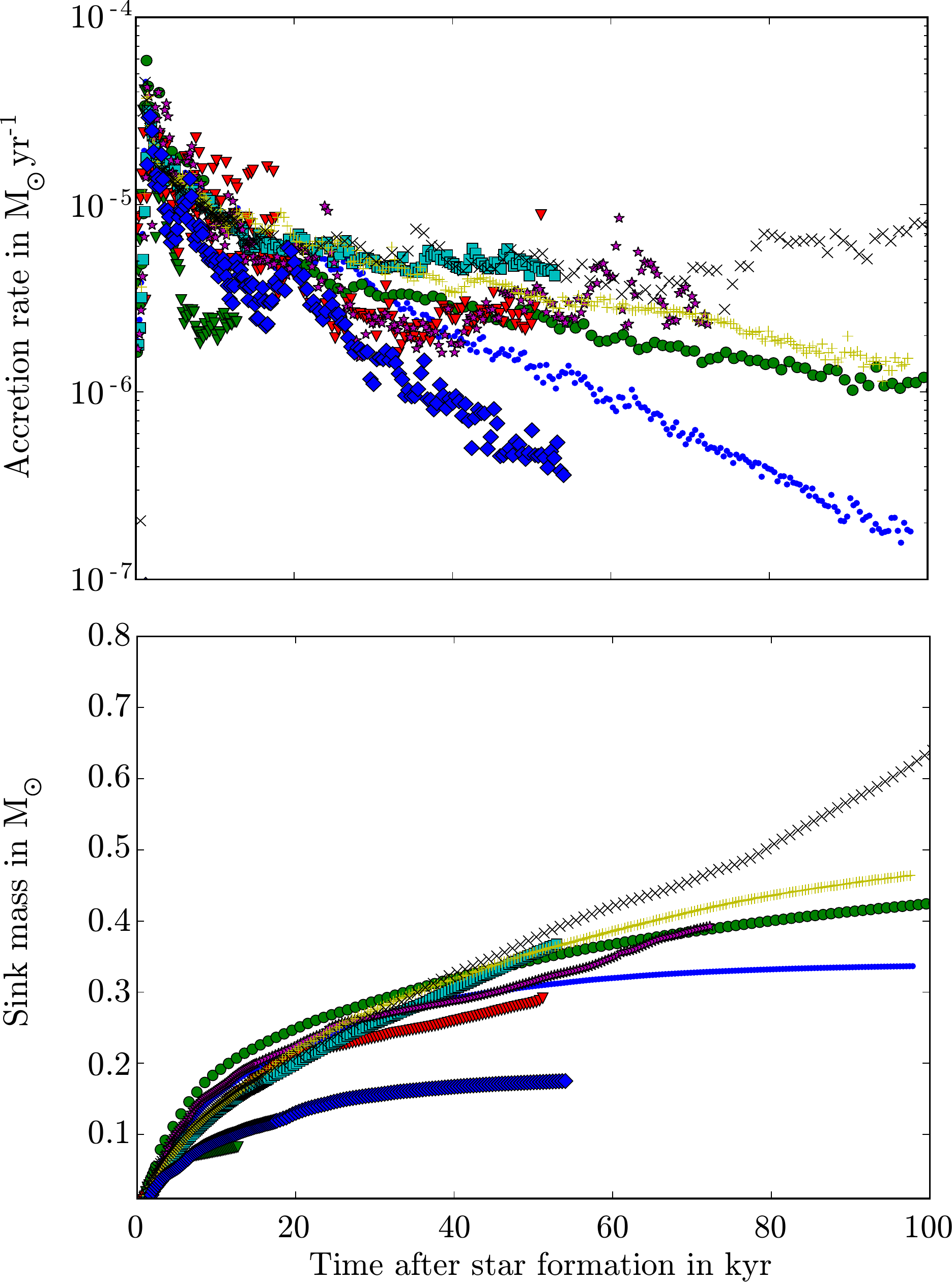} }
\protect\caption{\label{fig:acc-prof} Accretion profile (top) and sink mass evolution (bottom) for the 6 sinks created in zoom-ins started with increased resolution before sink creation.
The symbols correspond to the same sinks as in \Fig{acc_mass_run9}.
}
\end{figure}

\Fig{acc-prof} shows the evolution of the accretion rates during the first 100 kyr based on the zoom simulations.
We start by presenting the similarities for the different sinks before we focus on the differences in their accretion histories.
All accretion profiles quickly peak to values in the range of $3 \times 10^{-5}$ M$_\odot$/yr for sink 4
and $6 \times 10^{-5}$ M$_\odot$/yr for sink 3, before the accretion rates decrease during the subsequent evolution.
The quick rise within at most a few kyr seen for the sinks indicates the accuracy of the selected time of sink creation as discussed above.
The accretion rate peaks with a small offset after t=0, due to the flat density profiles around the different sinks at the time of their creation
(\Fig{r-dens-0}).

After the initial peak, the accretion rates of the sinks decrease in ways that differ between individual sinks.
Sink 1 (blue dots) shows the most consistent continuously decreasing accretion rate,
while other sinks settle to a nearly stable average accretion rate or have only slowly decaying profiles, after an initial drop in the rate.
After 100 kyr, sink 1 has an accretion rate of only $\sim 10^{-7}$ M$_\odot$/yr, and is therefore transitioning to a state that observationally
would be classified as either Class I or early Class II.
Sink 3 and sink 7 have very similar profiles and, as for sink 1, they also show a continuously decreasing accretion rate.
However, the change is more modest, starting at peak values of about a few $10^{-5}$ M$_\odot$/yr.
They still show accretion rates of about $10^{-6}$ M$_\odot$/yr after $t=100$ kyr.
Sink 7 also shows a  drop in the accretion, from a peak value of $4 \times 10^{-5}$ to about $6 \times 10^{-6}$ M$_\odot$/yr at approximately 20 kyr after sink creation.
However, the accretion rate then only decreases slightly, to a minimum of $3 \times 10^{-6}$ M$_\odot$/yr after $t=75$ kyr,
after which the accretion profile starts to rise again, to about $8 \times 10^{-6}$ M$_\odot$/yr at the end of the simulation at $t=100$ kyr.
Although we do not account for the later accretion profile of sink 5 due to its shorter simulated evolution,
we can see that the profile is almost identical to the one for star 8 and also similar to the evolution of sink 7 and sink 9.
Such a period of re-increasing accretion rate is also seen for sink 6 in the time interval between $t=27$ kyr and $t=62$ kyr.
The accretion profiles for this sink as well as for sink 9 and (especially) for sink 4 are more intermittent and episodic.

\begin{figure*}[!htbp]
\subfigure{\includegraphics[width=\linewidth]{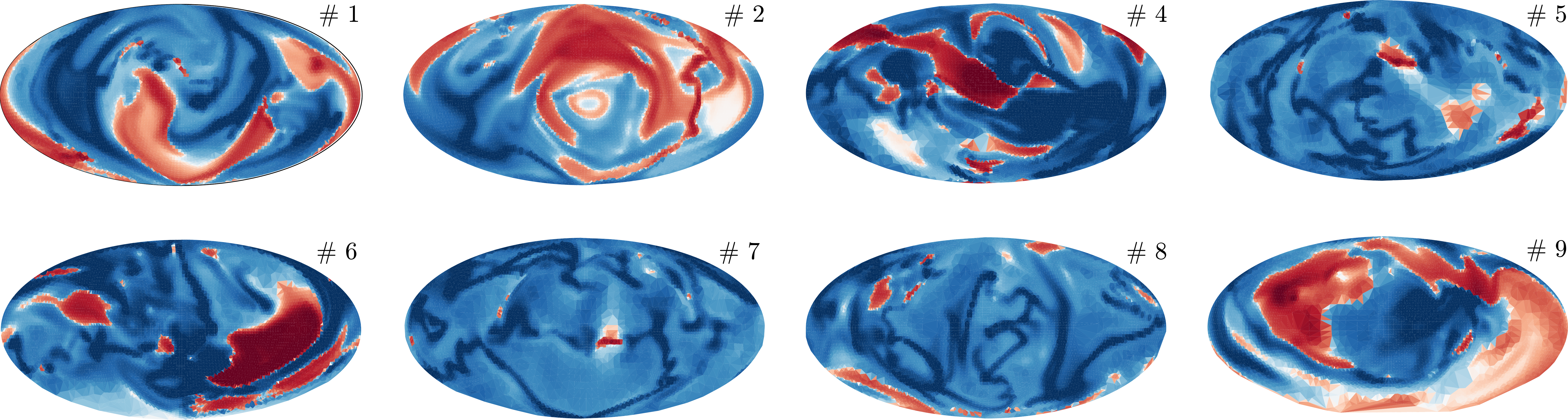} }
\begin{center}
\subfigure{\includegraphics[scale=0.4]{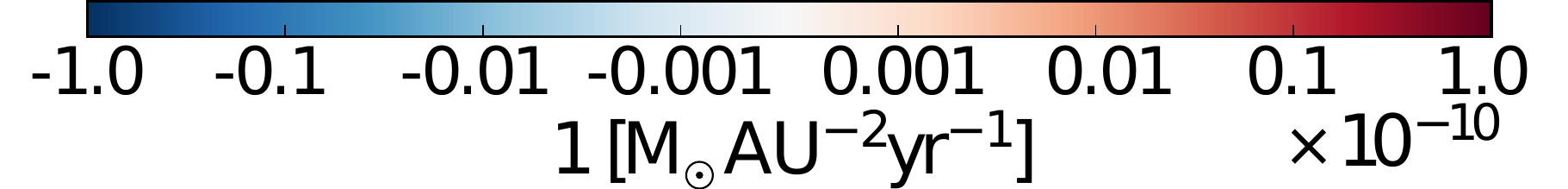}} \quad
\end{center}
\protect\caption{\label{fig:hammer-10a} Hammer projection of the accretion at a radius of 50 AU from the sink at time $t=10$ kyr
for sinks 1, 2, 4, 5, 6, 7, 8 and 9 that were evolved with a minimum cell size of 2 AU.
The colors represent inflow and outward motion according to the color bar.
In order to illustrate both positive and negative values,
we decided to use a linear range in-between $\pm 10^{-13}$ M$_{\odot}$ AU$^{-2}$yr$^{-1}$ and a logarithmic scale beyond.}
\end{figure*}
We caution the reader that we are generally underestimating the amplitudes and extents of the bursts for two reasons.
First, even our highest resolution of 2 AU is coarse compared to the actual size of a protostar, and
second, the snapshot cadences we used were at least 200 years, with the effect that shorter bursts are often missed in this plot.
Therefore, we postpone a detailed analysis and comparison of the accretion histories with observations to future work.
Nevertheless, the fact that some profiles show intermittent profiles may help to understand observations such as the ones by \citet{2015ApJ...800L...5S} of a
class 0 YSO (HOPS 383) with fluctuations of more than 30 in the mid-infrared.
We stress that these early accretion bursts that occur within a few tens of thousands of years after stellar birth are different
from the large scale in-falls, previously suggested by \citep{2014ApJ...797...32P}, explaining the luminosity variations of class I YSOs.

Compared to the mass evolution of the sinks in the parental run,
the mass evolution of the sinks in the zoom runs show significant differences.
First of all the sinks accrete much less mass than in the parental run,
which is mainly a consequence of the accretion efficiency parameter \acceff\ being 0.5 instead of 1 as in the parental run.
In retrospect, a lower \acceff\ could have been used also in the parental run,
to compensate for the missing outflow due to the unresolved winds and jets.
Simply multiplying the sink mass in the zoom runs
according to our choice of the accretion efficiency by a factor of 2
is not sufficient, because a higher \acceff\ would cause a larger sink mass,
and thus a deeper gravitational potential. This would cause additional accretion,
as the sink would be able to gravitationally attract mass from larger distances
that is gravitationally unbound for a lower sink mass. From smaller test runs evolved
with both recipes we have found that \acceff=0.5 corresponds to reducing
the sink mass by 1/3 in an identical run with \acceff=1.
Sink 1 is the most massive of the plotted sinks after 100 kyr in the parental run,
and is almost three times as massive as sink 3, and about twice as massive as sink 7 and 9,
although it has accreted less mass than these sinks in the zoom runs.

By tracing the history of the gas with tracer particles that are passively advected with the gas motion,
we showed in \citet{2016ApJ...826...22K} that a small fraction of the gas located within 100 AU
from the protostar at time $t$ after sink creation was located beyond $10^4$ AU at the time of sink creation.
Given that local core-collapse models usually use a homogeneous mass distribution similar to the theoretical
profile of a Bonnor-Ebert sphere,
we expect that the majority of low-mass sinks accrete on time-scales that are longer than predicted by these models.
Furthermore, the diversity shows the limitations of an isolated core collapse model as an initial condition,
when aiming to comprehend the entire process involved in sink creation, especially the occurrence of late mass infall through filaments.

\subsubsection{Diversity in the angular distribution of accretion}
In contrast to the case of a non-magnetized, rotation-less collapsing sphere, we find that the gas does not accrete uniformly.
Instead, gas accretes along accretion channels and accretion sheets, and some parts of the gas even flow in the outward direction.

In \Fig{hammer-10a}
, we illustrate this spatial heterogeneity of the accretion process by showing the angular distribution of the accretion rates
at a radius of 50 AU, 10 kyr after sink creation,
for sinks 1, 2, 4, 5, 6, 7, 8 and 9.
In order to highlight the infall and outward motion of material as well as the differences in
magnitude, we use a symmetric log color scale.
That means values between $\pm 10^{-13}$ \MAacc\ are plotted according to a linear
color scale, while we use a logarithmic scale for values below or above the threshold values.
Furthermore, we choose cut-off values of $\pm10^{-10}$ \MAacc\ to emphasize that red corresponds to outward motion
of gas and blue to infall.
When comparing the projections of the sinks with each other,
we find differences in the extent of the accretion channels and sheets as well as the amount of
gas moving away from the sink.
Sink 5, 7 and 8 only show weak outward motions of gas in contrast to the other four sinks,
where a significant surface area is dominated by the outward motion of gas.
In contrast to what is found in models of collapsing envelopes with uniform density, we do not see a pure bipolar outflow
at 50 AU at this stage, and distribution of outflowing and infalling gas does not show a clear morphology.
Nevertheless, we occasionally see outward motions
approximately perpendicular to the disk plane with speeds similar to the Kepler speed at distances of a few AU.
These speeds are consistent with magnetocentrifugally driven winds launched on scales of our highest resolution of 2 AU.

\begin{figure*}[!htbp]
\subfigure{\includegraphics[width=\linewidth]{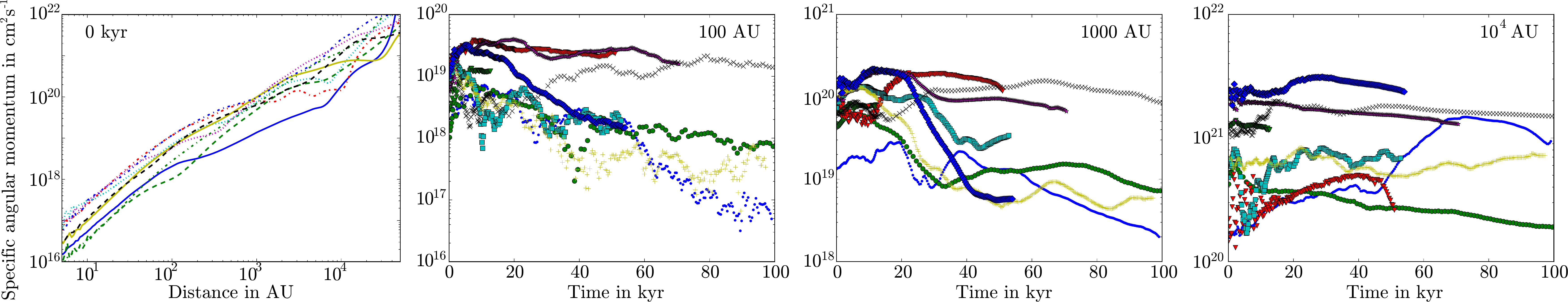} }
\protect\caption{\label{fig:ang_mom_gas} From left to right:
Specific angular momentum as a function of radius from the nine host sinks at t=0,
and evolution of total specific angular momentum of the gas located within a distance of 100 AU, 1000 AU and 10000 AU
from the eight sinks of the advanced zoom-ins.
The symbols belong to the same sinks as in \Fig{acc_mass_run9}.}
\end{figure*}

The strength of the magnetic fields around the protostar is probably overestimated in our models due
to the absence of non-ideal MHD effects \citep{2010ApJ...716.1541K,2011ApJ...738..180L,%
2011ApJ...733...54K,2016MNRAS.457.1037W,2015MNRAS.452..278T,2016ApJ...830L...8H}, and a significant part
of the gas initially moving outward may be caused
by magnetic interchange instabilities that occur because of high magnetic pressure close to the protostar, resulting in the emergence of
strong and expanding magnetic loops \citep{2014ApJ...793..130L}.
Considering the high densities at the center, we expect Ohmic dissipation to have the strongest effect in reducing the field strength.
Nevertheless we speculate that a build-up of high magnetic pressure, such
as seen in our simulations, might occur also to some extent during the real formation phase of protostars.
Other groups carrying out ideal MHD simulations of a local cloud collapse see this effect
as well for their turbulence-free simulations \citep{2011MNRAS.417.1054S,2012A&A...543A.128J}.
\citet{2013MNRAS.432.3320S} do not detect the occurrence of the magnetic interchange instability,
when including turbulence in their simulations.
We think that the instability does not occur in their case because
the resolution is lower than in our study (8 AU compared to 2 AU in our simulations).
The lower resolution has the effect that the magnetic pressure does not pile up as much as seen in our case.
Recent simulations by \citet{2016A&A...587A..32M}, accounting for dissipation effects from ambipolar diffusion and Ohmic resistivity,
show that non-ideal effects can circumvent the occurrence of the magnetic interchange instability.
However, the strengths of the non-ideal MHD coefficients depend on the degree of ionization of the gas,
which may be highly variable between protostars for several reasons.
First, the efficiency of shielding
by dust depends crucially on how the dust is distributed in space.
In cases where the dust distribution is very filamentary, with region in-between with much lower dust
densities, the effective absorption will be much less than if the distribution were uniform.
Secondly,
we point out that the efficiency of shielding even for a uniform dust-to-gas ratio strongly
depends on the grain size distribution.
Moreover, the cosmic ray intensity may depend on the location \citep{2009A&A...501..619P,2013A&A...560A.114P,2014A&A...571A..33P}, e.g.\ because of `magnetic bottle' effects,
and because of the distance in space-time from acceleration regions such as SN shock fronts.
Finally, if the cosmic ray intensity is efficiently reduced in dense regions,
the main ionization source becomes short-lived radionuclide $^{26}Al$ \citep{2009ApJ...690...69U,2014ApJ...794..123C,2015ApJ...801..117T},
and given that the abundance of $^{26}Al$ can differ among stars by orders of magnitude \citep{2013ApJ...769L...8V},
this can also contribute significantly to variations of the ionization degree in space and time.

In the case of sink 7 we find that, even at $t=75$ kyr, the gas falls towards the sink through sheets from all spatial directions.
However, the infall rates relax to more modest values and accretion appears to happen mainly
through two sheets with a small distance from each other at $t=100$ kyr.
In contrast, the Hammer-projections for sink 6 (not shown) show slightly broader sheets, with
higher accretion rates as well as a significant contribution of gas moving in the outward direction.
We point out that the enhanced infall rates for sink 6 are not surprising, taking into account that the density
over several hundred AU in the xy-plane from the sink is enhanced by up to a factor of 10 (\Fig{r-dens-0}) at the time of stellar birth,
while the density is about the same or only slightly higher in the other coordinate-axis planes.
Consequently, there is more mass available that can accrete onto the star-disk system.
Nevertheless, the pure difference in the absolute amount of mass cannot account for the significant outward motions
observed for sink 4 compared to the nearly pure infall morphology around sink 7.

\begin{figure*}[!htbp]
\subfigure{\includegraphics[angle=-90,origin=c,clip,trim={7cm 0cm 7cm 0},
width=\linewidth]{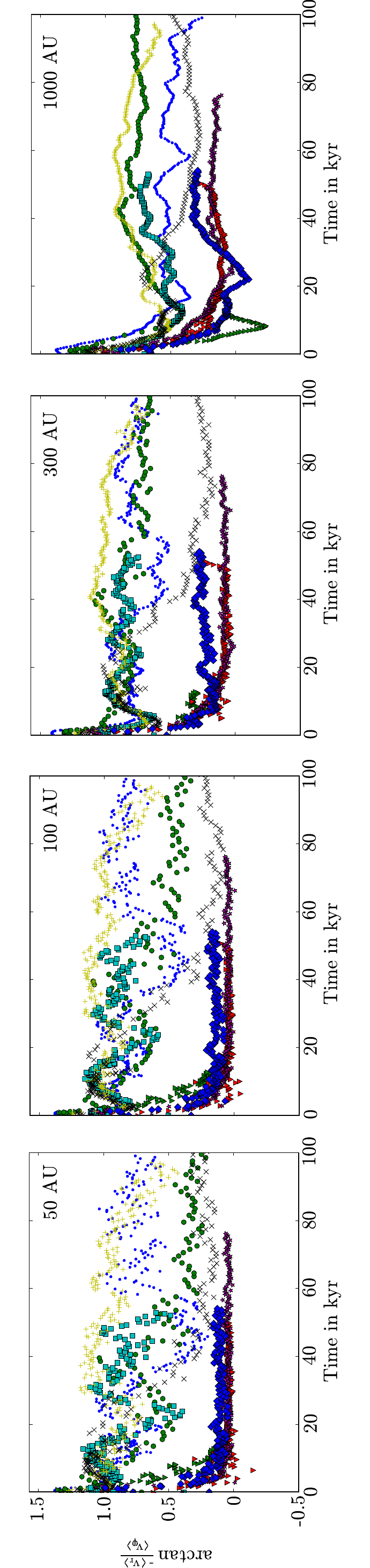}} 
\vspace{-26em}
\protect\caption{\label{fig:Brinch-prof} Evolution of the $\alpha$ angle
for four spheres of radii 50 AU, 100 AU, 300 AU and 1000 AU.
The symbols belong to the same sinks as in \Fig{acc_mass_run9}.}
\end{figure*}

\subsection{Disk formation and evolution}
An increasing number of observations of class 0 objects reveal the existence of circumstellar disks in the midst of their formation process.
They are not perfectly symmetric and their profiles differ from static thin standard accretion disks \citep{2013Sci...340.1199V}.
Due to the violent processes during protostar formation, the identification of a disk depends on the definition of the term disk.
The fundamental property that causes disk formation is the conservation of angular momentum during the infall, which determines
the size of the disk.
The left panel of \Figure{ang_mom_gas} shows the average specific angular momenta in the stellar surroundings
of the nine sinks close to their individual $t=0$.
For a one solar mass star, the specific angular momentum at 1 AU is about
$4.5 10^{19}$ cm$^2$ s$^{-1}$, and the plot illustrates that at this time the average specific angular momentum of
essentially all mass inside about $1000$ AU has low enough angular momentum to reach orbits inside 1 AU.  Remarkably,
the distribution over radius is similar for all sinks, scaling roughly linearly with radius.  Since we are resolving
each level of the refinement ladder with a large number of cells, and we have of the order of 50 cells per Jeans'
length, it is unlikely that this scaling is significantly influenced by numerical resolution effects.

To compare the evolution of the level of rotation around the
different sinks in our simulation, we plot the sum of the specific angular
momentum of the gas within a distance of $100$($1000$,$10^4$) AU in the right panels of \Fig{ang_mom_gas}.

All sinks show an increase of specific angular momentum in their vicinity during the
evolution. However, the strength of rotation and the rate at which the rotation increases
differs from sink to sink. Comparing the evolution of the specific angular momentum within 100 AU
with the accretion profile (upper panel in \Fig{acc-prof}) shows similarities.
The accretion profile of sink 1 drops very quickly, as does the amount of specific angular momentum within 100 AU.
The specific angular momentum around sink 1 peaks at a value of about $5\times 10^{18}$ cm$^2$s$^{-1}$ at $t\approx 10$ kyr
and then drops below $10^{17}$ cm$^2$s$^{-1}$ after $t\approx 100$ kyr.
In contrast, sink 8 has a much flatter accretion profile and the specific angular momentum within 100 AU is roughly constant
of the order of $10^{19}$ cm$^2$s$^{-1}$ since $t\approx 40$ kyr.
Also, the initial strength of specific angular momentum at distances beyond 1000 AU can differ by more than an order of magnitude.

The differences in extent of protoplanetary disks and the initial variation in the rotational velocity profiles
around the sinks at larger radii raises the question to what extent these two properties are connected.
Considering that gas falls towards the sink, potentially from larger distances,
the specific angular momentum at larger distances determines the rotational structure of the accreting gas.
The specific angular momentum within a spherical shell of 100 AU
around the sinks quickly increases during the first $\sim$5 kyr for sinks 4, 6 and 9,
as may be seen in the initial rise at the very left in the second panel of \Fig{ang_mom_gas}.
Below, we describe in more detail that circumstellar disks quickly form around these sinks.

Similar to \citet{2016A&A...587A..59F}, we quantify the rotational extent of the gas motion by investigating
the angle $\alpha$ \citep{2007A&A...461.1037B,2008A&A...489..607B}
\begin{equation}
\alpha = \arctan{\frac{- \langle v_r \rangle }{\langle v_{\phi} \rangle}}.
\label{Brinch-alpha}
\end{equation}
Theoretically, the angle can vary between $\frac{\pi}{2}$ for pure infall and $-\frac{\pi}{2}$ for pure outward motion with a value close to 0 representing pure rotation.
As shown by \citet{2016A&A...587A..59F} it serves as an adequate measure of the rotational support of the gas around the sink, and thus as a valid first proxy of a possible disk.
In \Fig{Brinch-prof}, we show the evolution of $\alpha$ within spheres of radii of 50 AU, 100 AU, 300 AU and 1000 AU around the sinks.
The decrease of $\alpha$ with time for all of the sinks shows that the relative amount of rotation increases for all sinks during their evolution
(although sink 7 has not evolved long enough to show a clear decrease).
However, just as with the accretion profiles of the different sinks (see \Fig{acc-prof}), the curves evolve remarkably different.

We interpret low $\alpha$ values as an indicator for the early formation of rotationally supported disks around these sinks.
The gas around the remaining sinks do not show such low values.
In particular the velocity profile around sink 7 is significantly dominated by the infall of material,
even after 100 kyr, with a $\alpha$ of more than 0.5. This indicates
that no rotationally supported disks have formed around them.

\begin{figure}[!htbp]
\subfigure{\includegraphics[bb=90bp 0bp 602bp 580bp,clip,width=\linewidth]{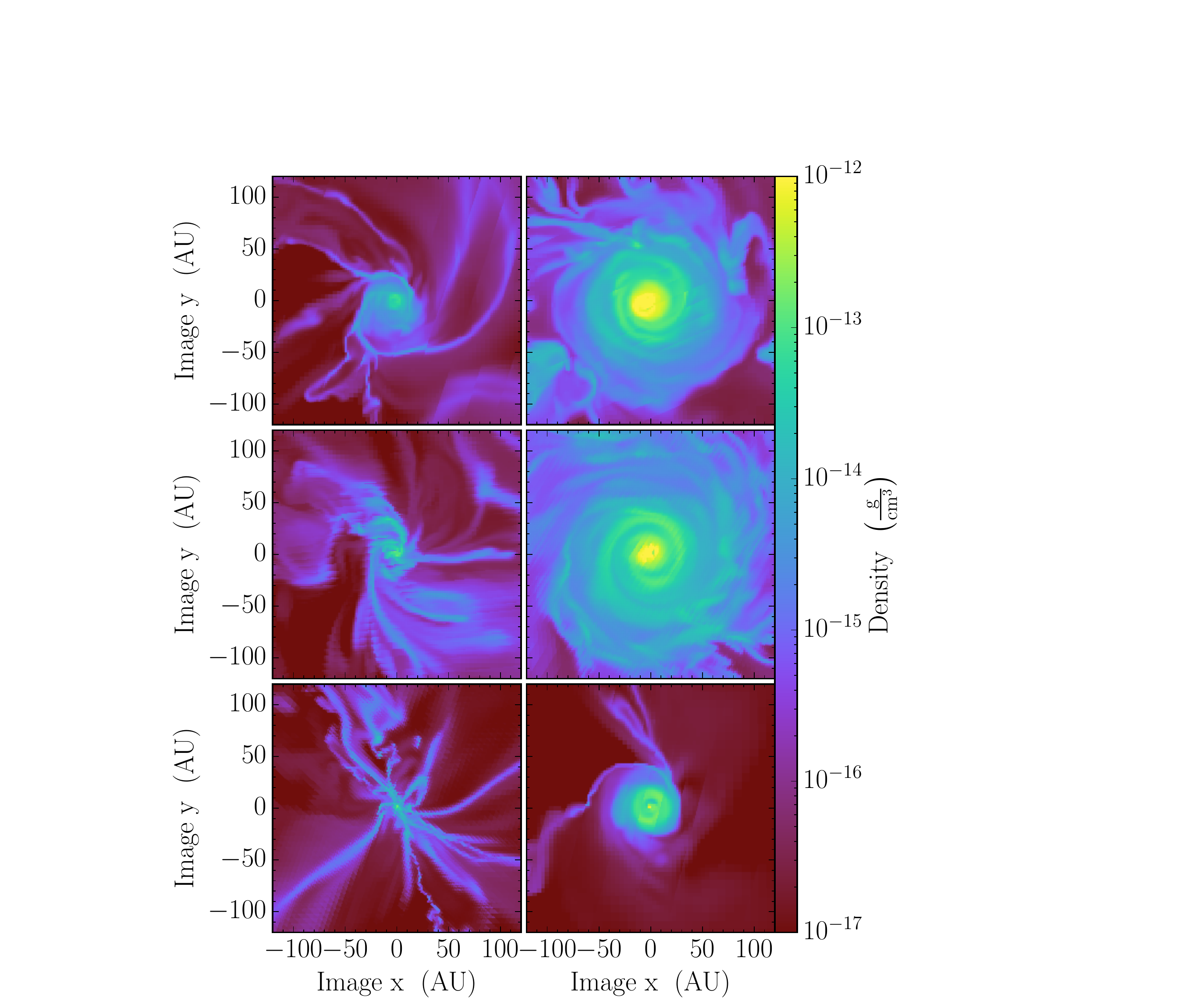} } \quad
\protect\caption{\label{fig:disk-slices} Slices in the plane vertical to the mean angular momentum vector
calculated for a sphere of 100 AU at $t=50$ kyr.
First row: sink 1 (left), sink 4 (right),
second row: sink 5 (left), sink 6 (right),
third row: sink 7 (left), sink 9 (right). }
\end{figure}

To illustrate the structure around the sinks in our study,
we show slices in the plane perpendicular to the mean angular momentum vector
at $t=50$ kyr for sink 1, 4, 5, 6, 7 and 9 in \Fig{disk-slices}.
The images reveal the variety in disk formation for the different stellar environments,
as already seen in the evolution of $\alpha$.
Similarly to the images of the pre-stellar cores
(\Fig{z-large-scale-env}),
the images show filamentary arms feeding the forming protoplanetary disk.
These filaments are not necessarily aligned with the disk plane,
but more often approach the disk from various angles.
Moreover, in agreement with recent observations from ALMA
and the Subaru Next Generation Adaptive Optics (HiCIAO)
the disks show evidence
of large-scale features, such as spiral arms or inflowing gas streams \citep{2014ApJ...793....1Y,Liue1500875}.

\subsection{Angular momentum transport}
An important aspect in protoplanetary disk studies is how the angular momentum is transported
from the environment to the star-disk system. To study the flow of angular momentum we consider a cylindrical test
volume with the height equal to the diameter ($h=2R$) and calculate the angular momentum flux through the cylinder
wall -- the ``radial direction'' -- and through the top and bottom of the cylinder -- the ``vertical direction''.
We refer the reader to appendix for the detailed calculation. Three terms contribute to the transport of angular momentum
in the vertical and radial directions. The magnetically induced transport is associated with the Maxwell stress $-\boldsymbol{B}\otimes\boldsymbol{B}$.
In the vertical direction it is
\begin{multline}
F_v^B(R) = \\
  \mp \int_0^{R} \mathrm{d}r \int_0^{2\pi} r\,\mathrm{d}\phi\, r \frac{B_{\phi}(r,\phi,\pm h/2) B_z(r,\phi,\pm h/2)}{4\pi} \,,
\end{multline}
while in the radial direction it is
\begin{multline}
F_r^B(R) = - \int_{-h/2}^{h/2} \mathrm{d}z \int_0^{2\pi} R\,\mathrm{d}\phi\, R \frac{B_{\phi}(R,\phi,z)B_r(R,\phi,z)}{4\pi}\,.
\end{multline}
The mechanical flux of angular momentum is associated with the Reynolds stress
$\rho\boldsymbol{v}\otimes\boldsymbol{v}$.  In the vertical directions it is
\begin{multline}
F_v^v(R) = \\
  \pm \int_0^{R} \mathrm{d}r \int_0^{2\pi} r\, \mathrm{d}\phi\, r\, \rho\, v_{\phi}(r,\phi,\pm h/2) v_z(r,\phi,\pm h/2)
\end{multline}
(where $\pm$ indicates the sings at top and bottom, respectively),
while in the radial direction it is
\begin{multline}
F_r^v(R) = \int_{-h/2}^{h/2} \mathrm{d}z \int_0^{2\pi} R\,\mathrm{d}\phi\, R\, \rho\, v_\phi(R,\phi,z)v_r(R,\phi,z) \,.
\end{multline}

Finally, we have the contribution associated with the gravitational potential ${\bf \nabla\Phi \nabla\Phi}$,
accounting for angular momentum transport through spiral arms and similar non-axisymmetric
structures.  In the vertical direction it is
\begin{multline}
F_v^g(R) = \\
 \pm \int_0^{R} \mathrm{d}r \int_0^{2\pi}r\,\mathrm{d}\phi\,r \frac{(\nabla \Phi)_{\phi}(r,\phi,\pm h/2) (\nabla \phi)_z(r,\phi,\pm h/2)}{4\pi G}\,,
\end{multline}
while in the radial direction it is
\begin{multline}
F_r^g(R) = \\
 \int_{-h/2}^{h/2} \mathrm{d}z \int_0^{2\pi} R\,\mathrm{d}\phi\,R \frac{(\nabla \Phi)_{\phi}(R,\phi,z)(\nabla \Phi)_r(R,\phi,z)}{4\pi G}\,.
\end{multline}

\begin{figure}[!htbp]
\subfigure{\includegraphics[width=\linewidth]{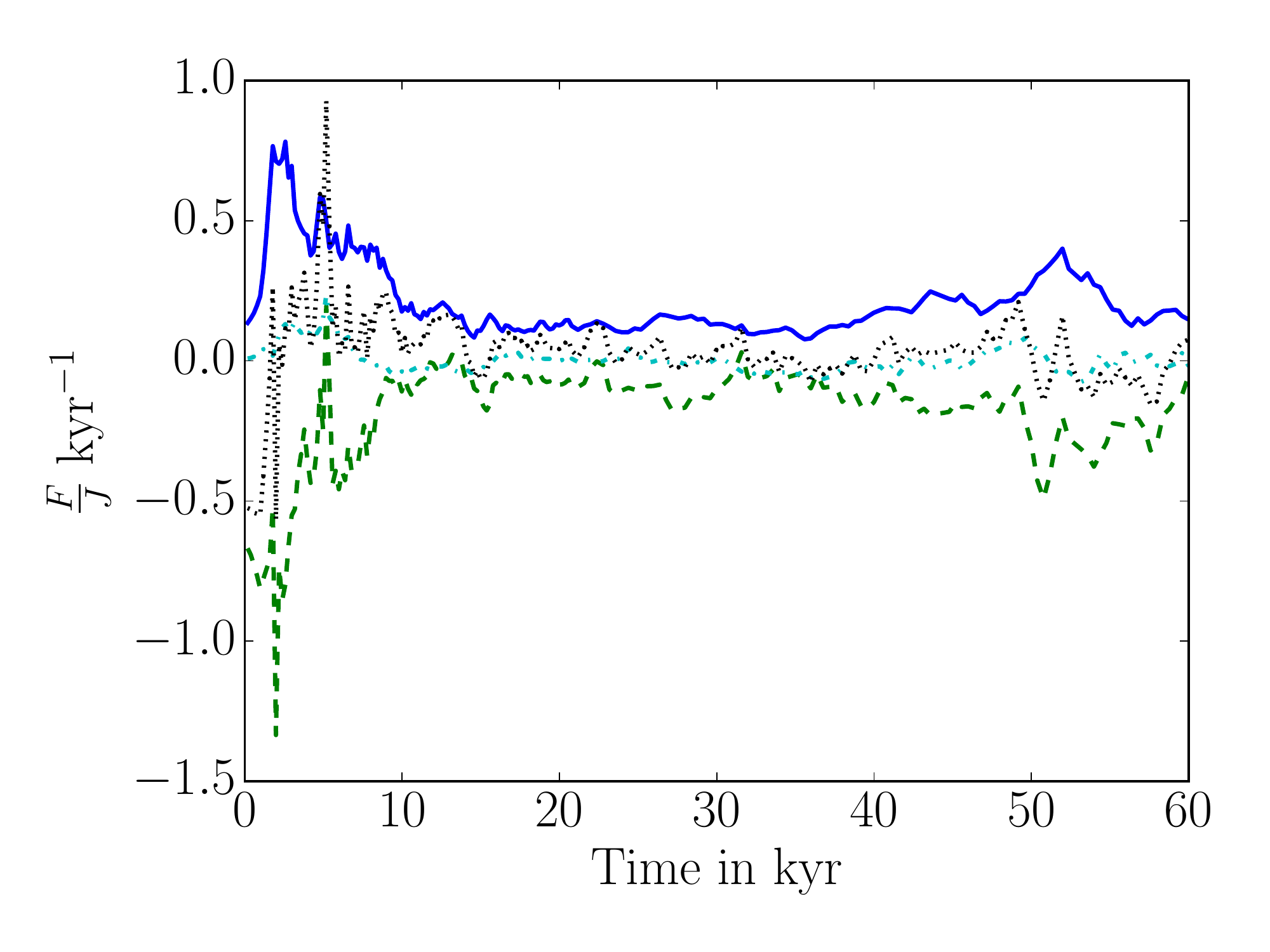} }
\protect\caption{\label{fig:transport_evo} Evolution of the angular momentum flux for sink 4
at the surfaces of a cylinder of size $h=2R=180$ AU.
Blue solid corresponds to the magnetic component, green dots to the mechanical component,
cyan dash-dot to the gravitational component, and the black dotted line represents the total flux.}
\end{figure}

\begin{figure*}[!htbp]
\subfigure{\includegraphics[width=\linewidth]{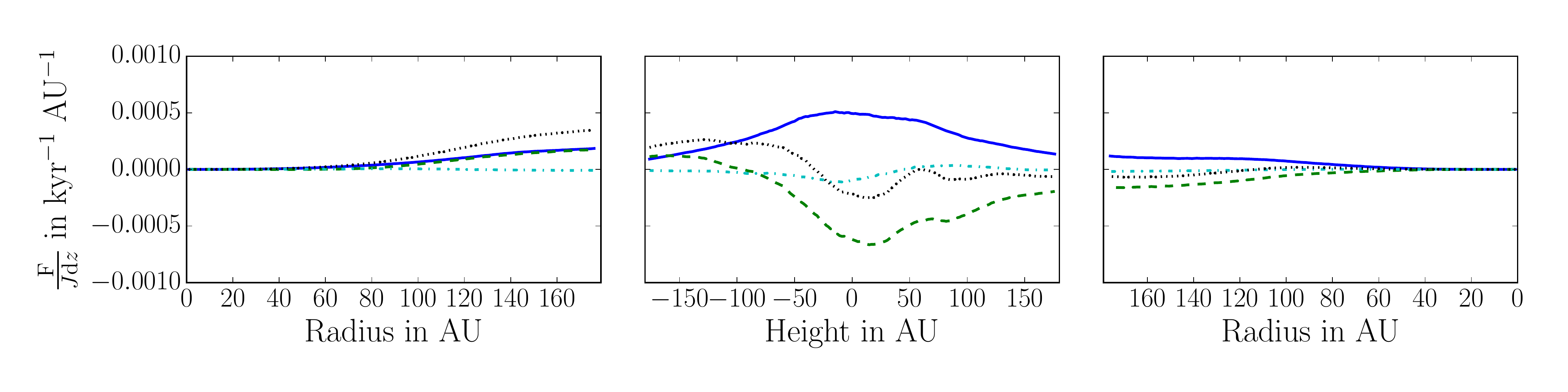} }
\protect\caption{\label{fig:transport_histo} Time averaged fluxes in the vertical (left and right panels)
and radial (middle panel) directions during the period from 15 to 45 kyr for sink 4.
The left panel shows the vertical flux as a function of radius at a constant height of 180 AU.
The middle panel shows the vertical flux as a function of height at a constant radius of 180 AU,
and the right panel shows the vertical flux as a function of radius at a constant height of -180 AU.
Blue solid corresponds to the magnetic component, green dots to the mechanical component,
cyan dash-dot to the gravitational component, and the black dotted line represents the total flux.
}
\end{figure*}

In contrast to \citet{2012A&A...543A.128J}, we consider the total value of the signed fluxes, rather than splitting
each contribution into negative and positive parts.
In this manner, we average out the natural spatial fluctuations induced by turbulence,
and obtain net fluxes (by definition positive in the outward direction).

We compute the fluxes of angular momentum within cylinders of different radii
($r=28$ AU, $r=80$ AU, $r=128$ AU and $r=180$ AU), with tops and bottoms at $+h$ and $-h$,
with in one case $h=r$, and in another case $h=2r$.
Using these cylindrical control volumes we compared the different contributions
relative to one another, and also compared their radial and vertical components.
In \Fig{transport_evo}, we illustrate the time evolution of the different components around sink 4.

As may be seen from Figures \ref{fig:transport_evo} and \ref{fig:transport_histo}, angular momentum is
--- consistent with expectations --- predominantly transported inwards by the mechanical Reynolds flux,
while the gravitational acceleration and the magnetic Maxwell stress generally account for transport in the outward direction.
Transport induced by the magnetic stress is typically stronger than the transport caused by gravity
by a factor of a few to several, though strong spiral arms can occasionally cause enhancements of the
gravitational component.
Comparing the contributions in the radial and vertical directions \Fig{transport_histo}, we find
that the strongest transport of angular momentum transport occurs in the vicinity of the disk midplane
in the radial direction, consistent with the fact that the fluctuating radial component of the
magnetic field, which tends to outline trailing spirals, is strongest near the disk midplane.
The vertical Maxwell component is stronger at larger radii.
This indicates that angular momentum
is not transported in a narrow jet.
However, this is a consequence of the resolution given that the minimum cell size of 2 AU is too coarse to resolve the launching of narrow jets.
The flux in the radial direction near the top and bottom is similar to
the vertical flux through the top and bottom, respectively. This
indicates that the angular momentum flux is oriented roughly at 45 degrees there.
Another interesting aspect concerns the asymmetry of the fluxes in the vertical direction.
In contrast to idealized core-collapse setups without turbulence, but in line with our results presented above,
the angular momentum is transported heterogeneously.
Even when averaging over periods of a few ten thousand years
--- corresponding to about 100 orbital times at 50 AU ---
we clearly see that more angular momentum is transported on one side than the other.
While the resolution is insufficient to study outflows via jets and winds quantitatively,
these asymmetries are nevertheless consistent with recent observations of asymmetric outflows (e.\,g.\ IRAS 03292+3039 \citep{2015ApJ...805..125T}).
We interpret them as a natural consequence of the underlying filamentary and chaotic
nature of the large scale accretion flows.

\subsection{Disk size evolution}
\begin{figure}[!htbp]
\subfigure{\includegraphics[width=\linewidth]{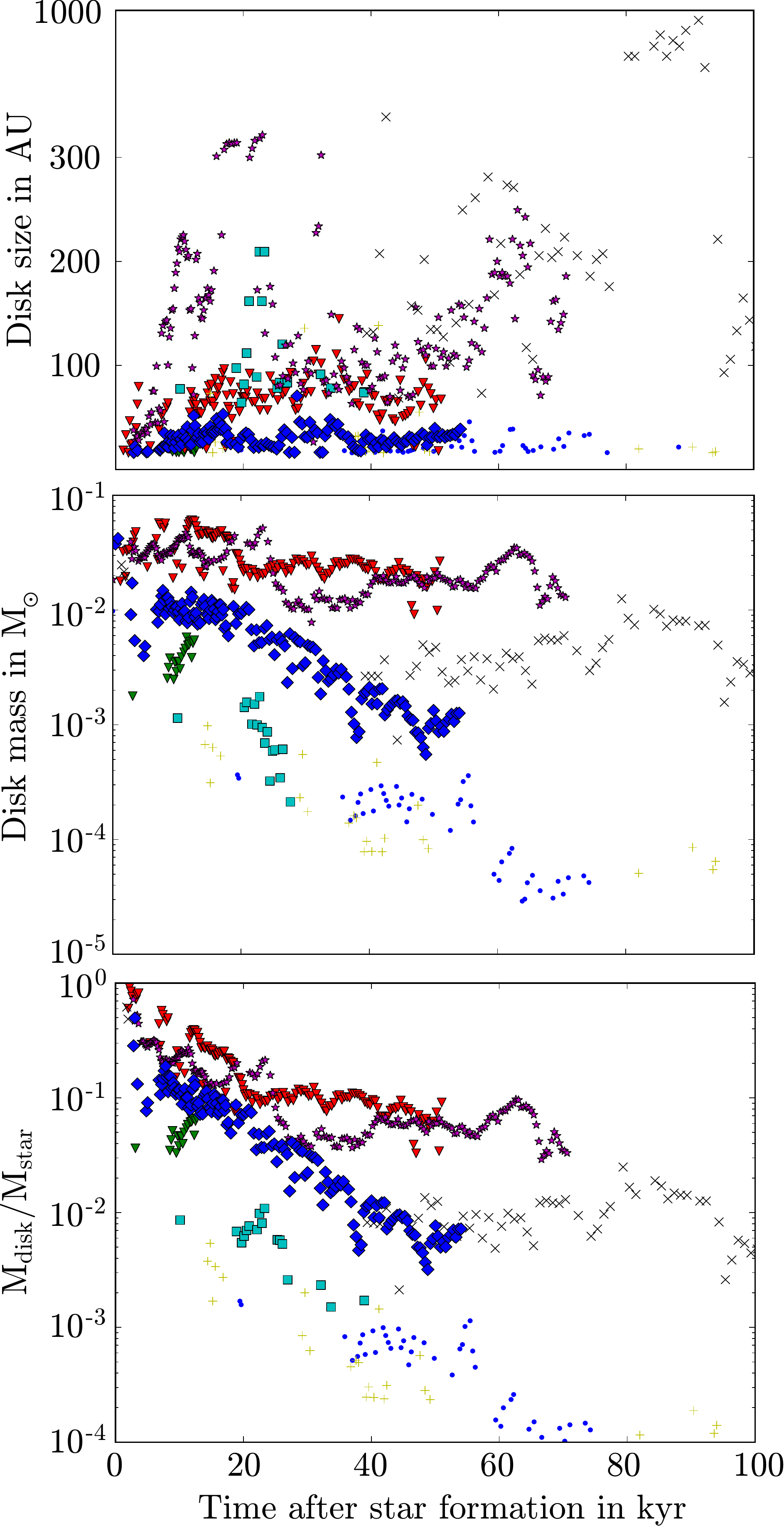} } \quad
\protect\caption{\label{fig:disk-size-mass-mratio} Disk size (upper panel), disk mass (middle panel)
and disk-to-stellar mass ratio (lower panel) as a function of time around the different protostars after sink creation.
The symbols belong to the same sinks as in \Fig{acc_mass_run9}.}
\end{figure}

An important quantity in understanding the disk evolution around the different protostars
is the time evolution of the disk sizes.
$\alpha$ is a good indicator of the relation of rotational to radial velocity,
but early disks might already form, although the gas has a strong radial velocity contribution.
We thus estimate the disk size in the following way.
We first calculate the total angular momentum vector $L_{100}$ inside a sphere of 100 AU around the protostar.
We then calculate the azimuthal velocity for all cells that are located inside a cylinder of 1000 AU in radius,
$\pm 8$ AU in height and with the radial direction being perpendicular to $L_{100}$.
Afterwards we estimate an average azimuthal velocity $v_{\phi}$ for all cells that are located between radii $r$ and $r+\mathrm{d}r$.
Altogether we consider 100 radial bins with $\mathrm{d}r$ increasing exponentially with increasing radius.
Finally, we determine the disk size as the radius where $v_{\phi}/v_K$ (with $v_K$ being the Kepler speed)
drops below a threshold value $a$,
though we do not take into account velocities inside a distance of 7 cells  (14 AU) from the sink
in order to avoid potentially low rotational velocities that are induced by the sink parameters.
Theoretically, a thin rotating gaseous disk with the radial structure described by power
laws has azimuthal velocities 
\begin{equation}
v_{\phi} = v_K \left(1-\mathcal{O}\left(\frac{h}{r}\right)^2\right)^{1/2}
\end{equation}
\citep{2007astro.ph..1485A}.
The small deviation of the rotational speed of the gas from the Kepler speed is induced
by the decrease of gas pressure with radius in the disk.
For thin disks, we thus expect $v_{\phi}$ to be nearly equal to $v_K$.
However, considering that early disks can be rather thick, and taking in to account the violent accretion process
from the outside, we relax the lower velocity limit somewhat, and choose a threshold value of $a=0.8$.
Based on this method, we plot in the upper panel of \Fig{disk-size-mass-mratio} the evolution of the disk sizes $r_{\rm disk}$ during
protostellar evolution for the different protostars.
In general, we can see that the disk sizes increase during the evolution.
However, similar to the time evolution of $\alpha$, we see strong differences among the different protostars.
While some disks extend to several hundred AU, others only extend to at most a few tens of AU.
This is in agreement with differing disk sizes depending on the initial density profile
in local core collapse models \citep{2014MNRAS.438.2278M}.
We find a significant time variation in the disk size during the evolution,
indicating that disk formation is an intermittent process, such as expected when accounting for turbulence.
Part of this intermittency is most probably the result of some of the disks being marginally Toomre unstable, as discussed below,
but this is of course ultimately related to the time-variation and strength of mass infall from the environment on to the disk.
We expect this intermittency to decline with time, when the infall rate decreases and the replenishment time increases in the disks.

\begin{figure}[!htbp]
\subfigure{\includegraphics[scale=0.43]{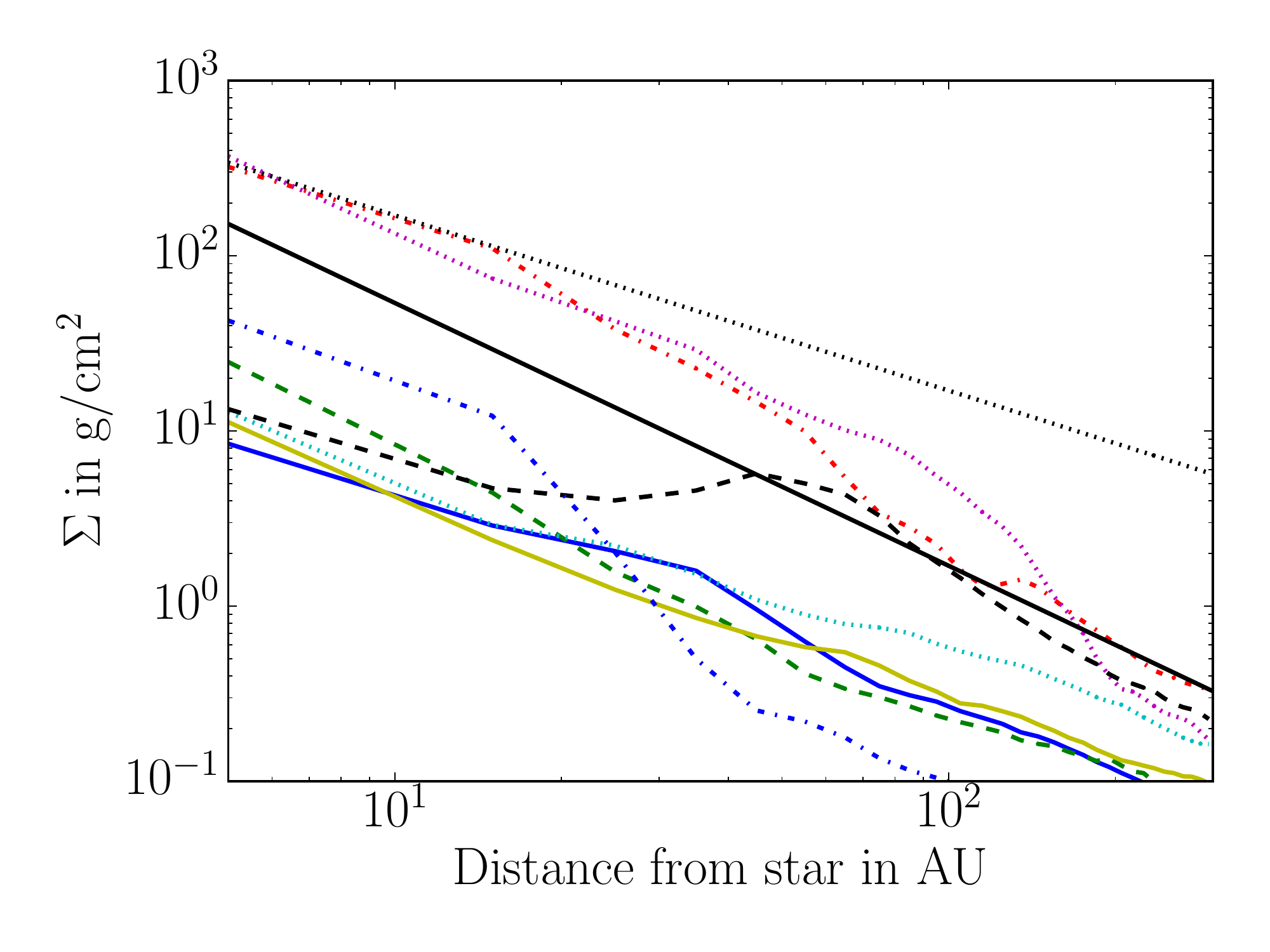} } \quad
\protect\caption{\label{fig:Sigma_50} Column density around the different sinks at $t=50$ kyr.
The lines belong to the same sinks as in \Fig{r-dens-0}.
The black solid line illustrates the
Minimum Mass Solar Nebula relation of $\Sigma = 1700 \frac{\unit{g}}{\unit{cm}^2} \cdot \left(r/\unit{AU}\right)^{-3/2}$,
and the black dotted line shows a surface density profile of $\Sigma = 1700 \frac{\unit{g}}{\unit{cm}^2} \cdot \left(r/\unit{AU}\right)^{-1}$.
}
\end{figure}
\begin{figure*}[!htbp]
\subfigure{\includegraphics[scale=0.43]{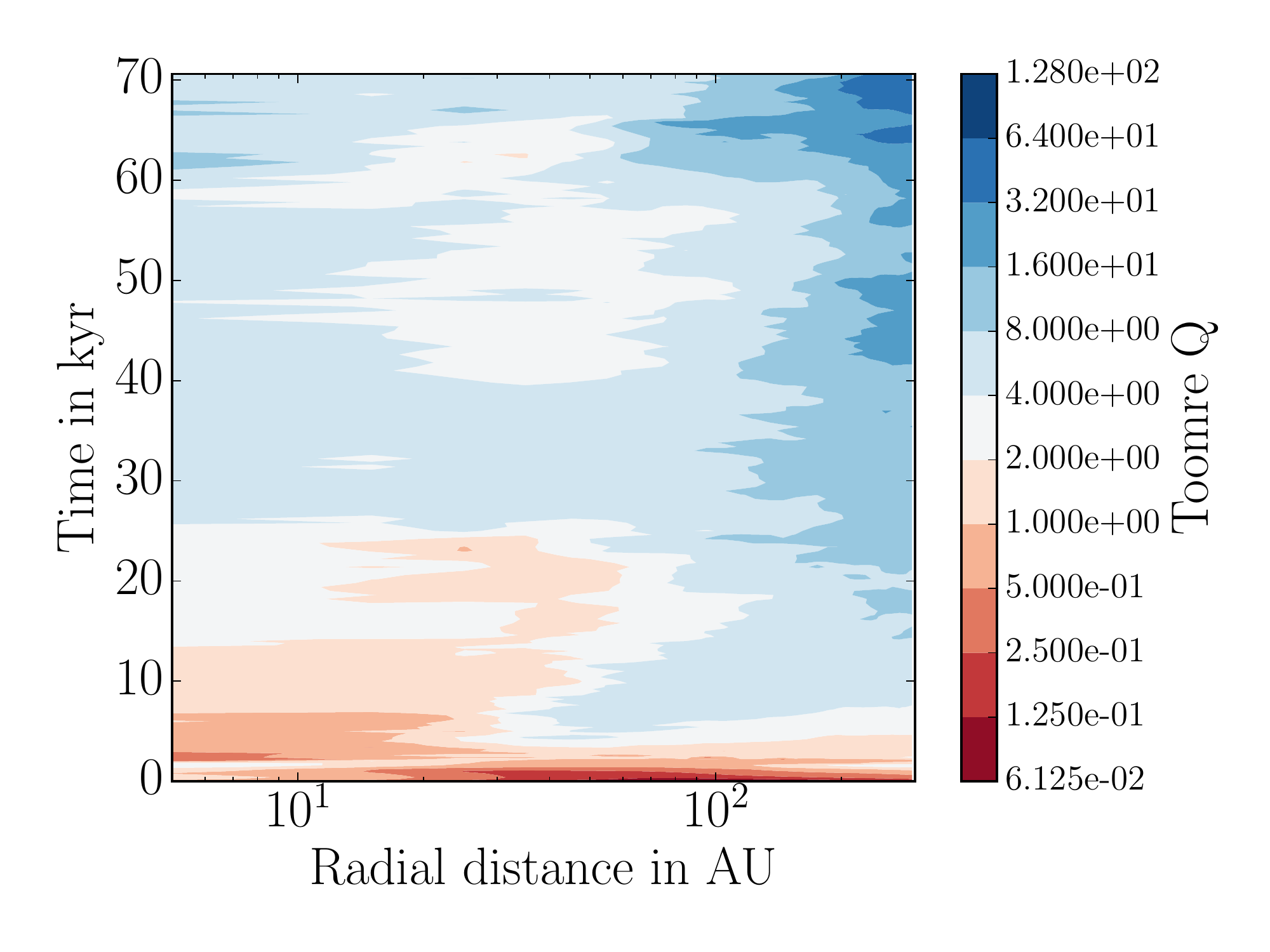} }
\subfigure{\includegraphics[scale=0.43]{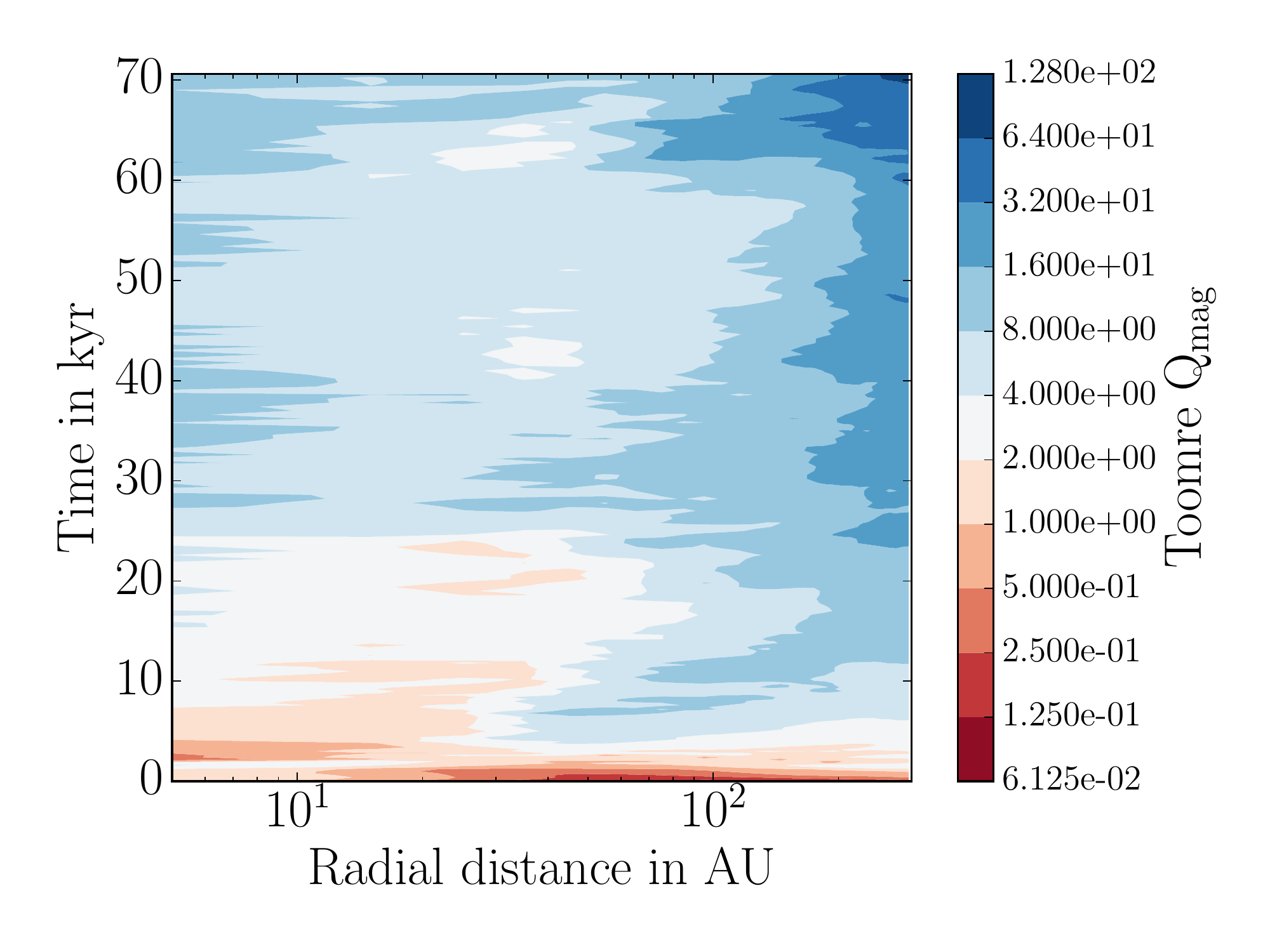} } \quad
\protect\caption{\label{fig:ToomreQ_contour} Azimuthally averaged Toomre parameter $Q$
(left panel) and magnetic Toomre parameter $Q_{\rm mag}$ (right panel) around sink 6 during disk evolution.
The color bar indicates the strength of $Q$ and $Q_{\rm mag}$, respectively.}
\end{figure*}

The differences are also reflected in the diversity of disk masses for the different disks during their evolution.
In order to compare disk masses, we add up the mass of the cells that are located within $r_{\rm disk}$,
within a maximum $\frac{h}{r}$ ratio of 0.2, and within a maximum vertical distance from the mid plane of $h=8$ AU (middle panel in \Fig{disk-size-mass-mratio}).
We find that the disk masses are highest ($\sim$ 1 \% M$_\odot$) around the sinks
where $\alpha$ approaches 0 most quickly and remains close to $\alpha=0$ (sink 4 and 6).
Moreover, we compare the disk masses with the mass of the host sink in the bottom panel of \Fig{disk-size-mass-mratio}.
We notice two trends.
First, early disks tend to have large disk-to-stellar mass ratios of up to $\sim 1$, because of the short duration of mass accretion by that time.
Second, after about 50 kyr the ratios vary between $0.1 \%$ and $10 \%$ for the stable disks in our simulation.
Such differences of three orders of magnitude in disk-to-star ratios are remarkable
and stress the influence of the stellar environments on star-disk properties --
even more so, when accounting for the cases where disks do not even form.

In order to constrain the surface density profiles around the sinks,
we compute the mass located in cylindrical shells of size $\Delta r = 10$ AU and $h=500 AU$ in height
with the vertical z-axis being aligned with the total angular momentum vector in a sphere of $r=100$ AU.
Afterwards we divide the massed by the corresponding disk surfaces to estimate the column densities.
When looking at the surface density profiles around the sinks after 50 kyr (\Fig{Sigma_50}),
we find that the profiles of the most evident disks (sink 4, sink 6 and sink 9),
show a dependence in-between $\Sigma \propto r^{-1}$ and $\Sigma \propto r^{-1.5}$,
with a profile closer to $\Sigma \propto r^{-1}$ in the inner part.
The overall surface densities of the two more massive disks (sink 4 and sink 6)
are larger by about a factor of two compared to the
Minimum Mass Solar Nebula (MMSN) \citep{1977MNRAS.180...57W,1981PThPS..70...35H}.
However, in the case of the weaker disk around sink 9,
the surface density profile is 10 to 20 times lower than the MMSN.
According to our rotational velocity criterion, sink 8 also hosts a disk-like structure.
Its surface density profile is much flatter and more perturbed than for the other sinks.
However, we caution that the run around this sink was carried out with lower resolution --
a minimum cell size of 8 AU instead of 2 AU.
Also, even though the surface density profile for the higher resolved disks are rather smooth,
we stress that abundant fluctuations (see \Fig{disk-slices}) are by construction averaged out in a radial profile.

Considering that the surface densities for two of our disks are above MMSN values,
we investigate whether the disks are gravitationally unstable.
To constrain that, we estimate the azimuthally averaged Toomre parameter \citep{1964ApJ...139.1217T} as based on cylindrical shells of height $100$ AU,
\begin{equation}
Q = \frac{c_s\Omega}{\pi G\Sigma},
\end{equation}
where $c_s$ is the sound speed, $\Omega$ is the orbital frequency and $G$ is the gravitational constant.
For reasons of simplicity, we assume perfect rotation inside the shell such that
\begin{equation}
\Omega = \sqrt{\frac{GM}{r^3}},
\end{equation}
where $M$ is the mass of the sink
and $r$ is the radius of the shell.
\begin{figure*}[!htbp]
\subfigure{\includegraphics[scale=0.4]{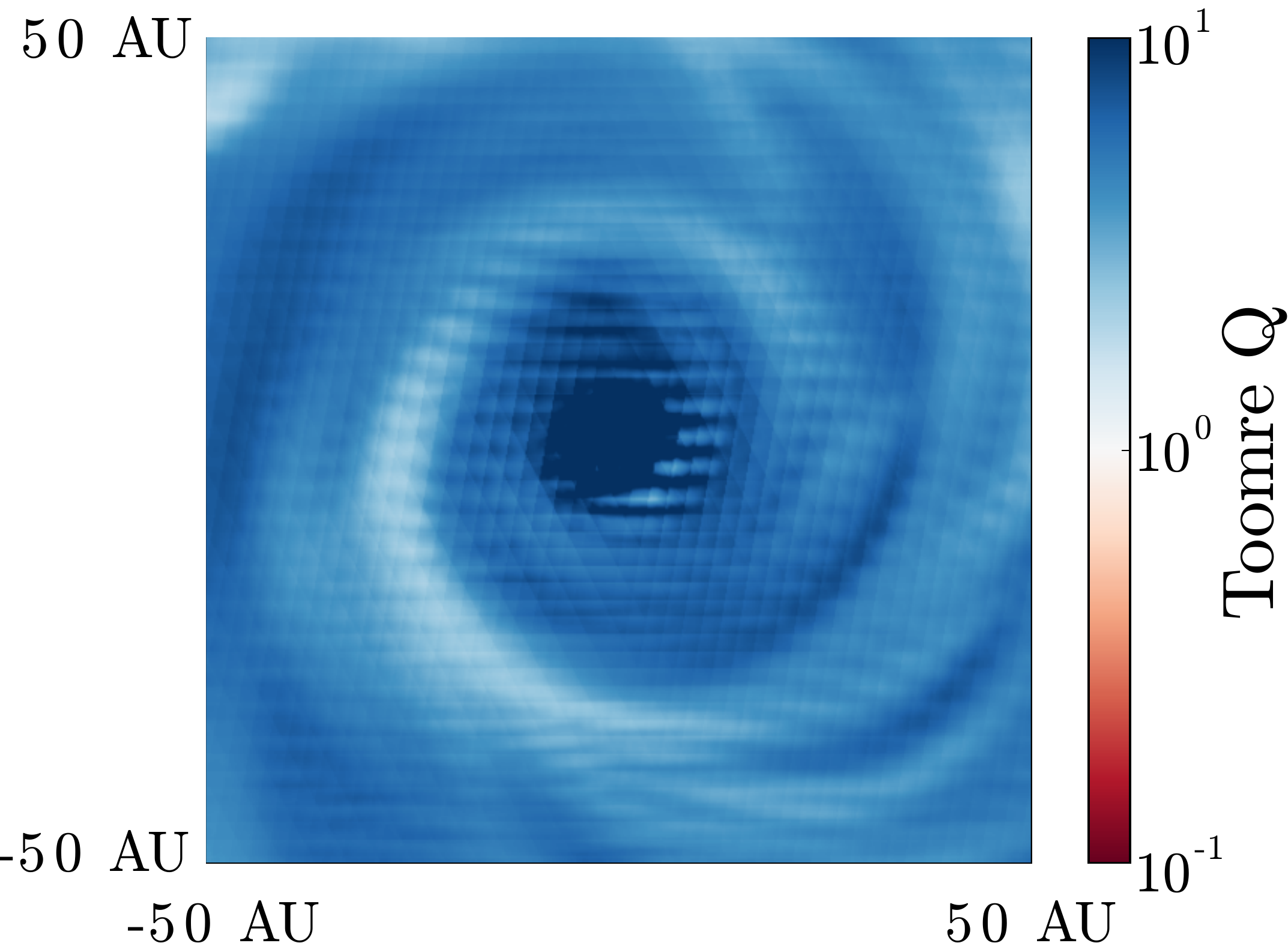} }
\subfigure{\includegraphics[scale=0.4]{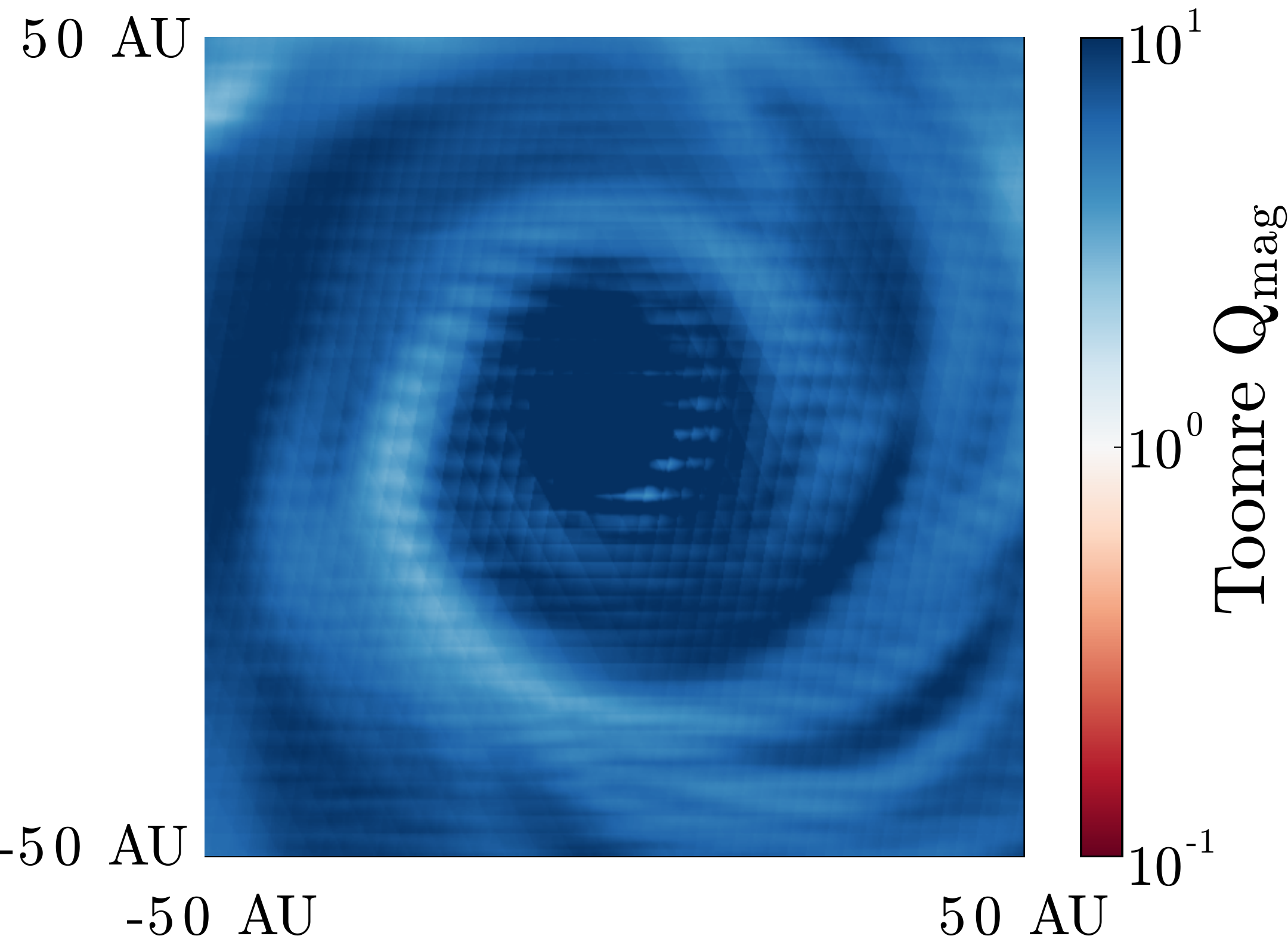} } \quad
\protect\caption{\label{fig:ToomreQ_proj} Toomre $Q$ parameter without (left panel) and including (right panel)
the magnetic support in form of the Alfv\'{e}n speed in the disk plane around sink 6 at $t=50$ kyr.
A value of $Q$ of about 1 or less means that the disk is gravitationally unstable.
}
\end{figure*}

We illustrate Toomre's $Q$ inside the disk around sink 6,
which is one of the two massive disks,
during its evolution in a contour plot (left panel of \Fig{ToomreQ_contour}).
We find that $Q$ drops below 1 at early times,
but we stress that at such early times no disk is present,
and hence the Toomre criterion should not be applied due to its dependence on the
Keplerian orbital angular velocity $\Omega$.
At later times, Q is larger than 1, but occasionally very close to 1.
If we applied more refinement and were able to resolve the disk inside $2$ AU,
we might expect $Q$ to be even lower at some locations in the disk.
However, considering only the pure hydrodynamical $Q$ parameter
is an oversimplification.
Instead we also have to take into account potential magnetic support of the disk.
In \Fig{ToomreQ_contour}\ is also shown the magnetic Toomre parameter \citep{2001ApJ...559...70K}
\begin{equation}
Q_{\rm mag} = \frac{\sqrt{(c_s^2 + v_A^2)} \Omega}{\pi G\Sigma},
\end{equation}
where $\mathbf{v_A}=\mathbf{B} / \sqrt{4\pi \rho}$ is the Alfv\'{e}n-velocity.
As for the sound speed, we take the mass-weighted average value
inside the column for $v_A$.
The toroidal magnetic fields inside the disk
contribute to the pressure support of the disk, partly stabilizing it 
against gravitational collapse.
As seen in the right panel of \Fig{ToomreQ_contour},
compared to $Q$, $Q_{\rm mag}$ is generally slightly enhanced in the disk,
but still only marginally larger than 1 at some locations at times when a disk is present.

Moreover, these values are spatially averaged,
and there may be denser clumps within the shell that lead to $Q$ values lower than 1,
and thus to gravitational collapse inside the disk.
In \Fig{ToomreQ_proj} we show the distribution of both $Q$ parameters
inside the disk at $t=50$ kyr around sink 6.
Although we find spiral structures with low $Q$ values (left panel in \Fig{ToomreQ_proj}),
we clearly see the additional support provided by the magnetic field
(right panel in \Fig{ToomreQ_proj}) preventing collapse at least at this point in time.

Nevertheless, the results show that $Q_{\rm mag}$ can well be in the marginal range.
Taking into account non-ideal MHD effects inside the disk may further
reduce the magnetic field strength and thus the magnetic disk support.
Several works investigated the effects
of non-ideal MHD in idealized symmetrical disks in detail
\citep[e.g][]{2014A&A...566A..56L,2015ApJ...801...84G,2015ApJ...798...84B}.
The importance of non-ideal effects remains to be investigated for our disks.
For computational reasons, we leave both tasks, the study of disks with higher resolution
as well as the investigation of non-ideal MHD effects to future work,
but already at this point we can conclude that some stars may host disks
that become massive enough to be gravitational unstable, in particular in the early embedded
phase of the evolution.

\subsection{Mass-to-flux ratio}
An often considered quantity in analytical and numerical core-collapse models is the ratio between mass and
magnetic flux threading a sphere of uniform density, in short the
mass-to-flux ratio \citep[e.g.][]{2003ApJ...599..363A,2008A&A...477....9H}.
It is commonly given as
\begin{equation}
\label{MtoFl_eq}
\mu = \frac{M_{\rm core}}{\Phi_{\rm core}} / \left( \frac{M}{\Phi}\right)_{\rm crit} = \frac{M_{\rm core}}{\int_A B_{\perp} \mathrm{d}A}/\left( \frac{0.13}{\sqrt{G}}\right),
\end{equation}
with $M_{\rm core}$ being the mass enclosed in the core,
$\Phi_{\rm core}$ the magnetic flux through the sphere with cross section $A$,
$\left( \frac{M}{\Phi}\right)_{\rm crit}$ the critical mass-to-flux ratio \citep{1976ApJ...210..326M},
and with $G$ being the gravitational constant.
However, as shown above,
stars do not form from pre-stellar cores of uniform density.
Instead, the cores
are significantly distorted, with feeding filaments
such as seen in \Fig{z-large-scale-env}.
Therefore,
the assumption of an isolated spherical core as the initial condition for modeling star formation
is an oversimplification.
Accounting for these asymmetries also shows the difficulty of the concept of an initial mass-to-flux ratio.
For cores that are highly distorted and continuously fed by accreting filaments
the mass budget is not fixed, and hence a changing mass-to-flux ratio does not necessarily imply
diffusion of mass relative to the magnetic field.
\begin{figure}[!htbp]
\subfigure{\includegraphics[width=\linewidth]{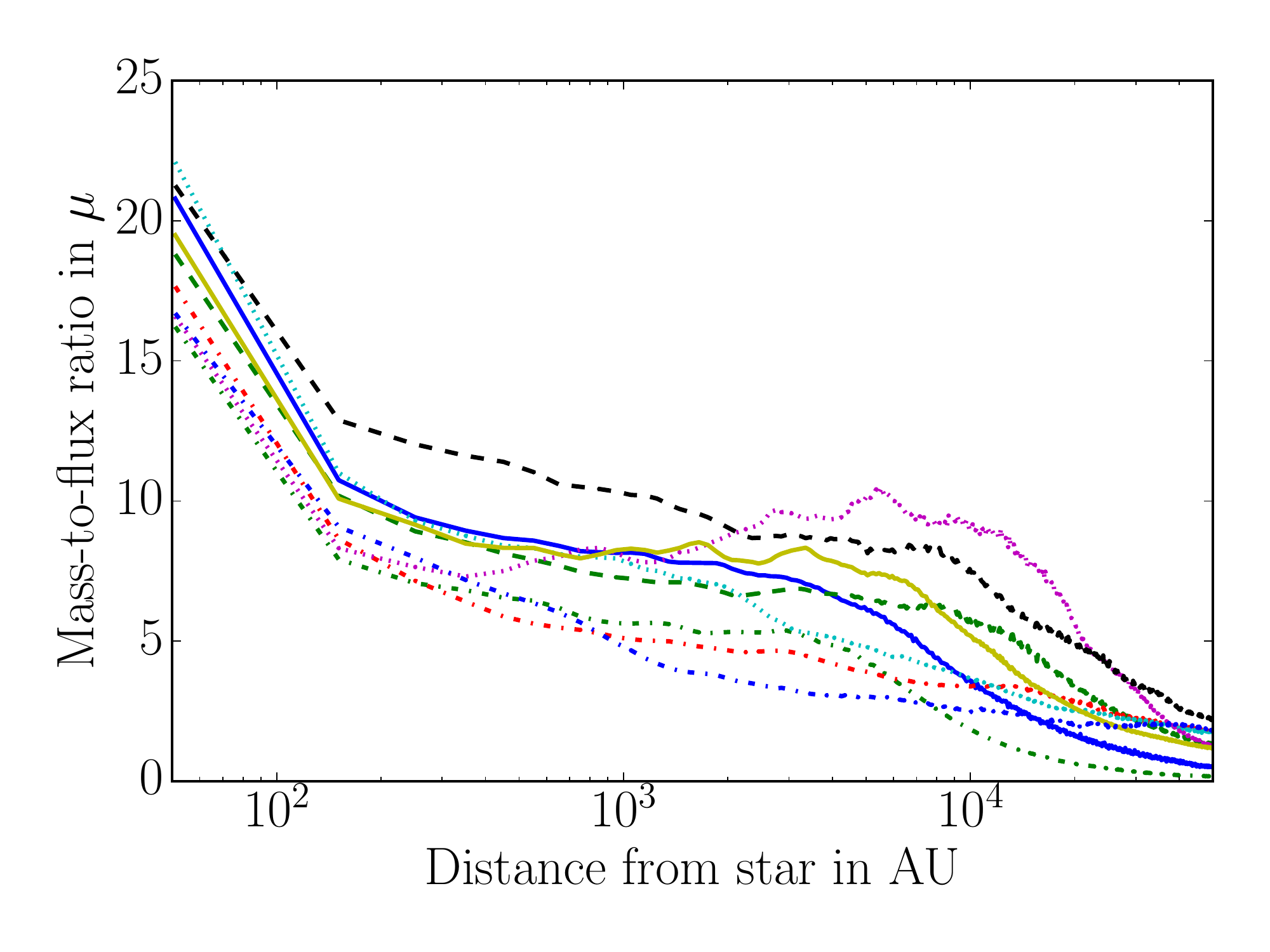} }
\protect\caption{\label{fig:MtoFl} Radial profile of mass-to-flux ratio at $t=0$ kyr for the different protostars.
The lines belong to the same sinks as in \Fig{r-dens-0}.
}
\end{figure}

Nevertheless, we estimate a proxy of the mass-to-flux ratio
to compare the relative magnetization around the different sinks at the time of their creation,
and to investigate how they might influence the formation of disks and their properties(see \Fig{MtoFl}).
We compute the mass-to-flux ratio according to eq. \ref{MtoFl_eq}
at different distances and $t=0$ by accounting for all the mass within the given distance from the forming sink
and by integrating over the absolute magnetic flux through the spherical test surface.
The plot shows a general trend of decreasing mass-to-flux ratio with increasing radius.
This correlation is caused by the fact that gravitational collapse already started at $t=0$,
with the effect that mass already has piled up close to the center of the collapsing core.
We are aware that the implementation of accretion onto sink particles causes diffusion 
of mass across field lines, deviating from the assumption of ideal MHD.
Therefore, we plot the mass-to-flux ratios at $t\approx 0$, when the sink mass is almost zero and the diffusion effect is negligible.
We point out that one cannot avoid some magnetic diffusion in numerical MHD simulations,
but that the magnetic diffusivity is certainly underestimated in our case, 
compared to models that explicitly account for non-ideal MHD effects.
Considering the mass-to-flux ratios only on scales of the pre-stellar core
in the range of $10^3$ to $10^4$ AU, we find supercritical mass-to-flux ratios that lie in the
range between $\approx 2$ and $\approx 10$, in agreement with observations
\citep{2008A&A...487..247F,2009Sci...324.1408G,2010ApJ...724L.113B}.
With respect to disk formation, we find that mass-to-flux ratios at a few to several AU
distances are largest around sink 6 and sink 8, which also are the sinks that show the most
significant disk formation.
This sounds intuitively reasonable, given that a higher-mass-to flux ratio implies less magnetic field strength
and therefore potentially weaker magnetic braking.
However, the mass-to-flux-ratios around sink 4 and sink 9 are in contrast the lowest,
but as seen in \Fig{disk-size-mass-mratio}, disks form around both sinks.
We interpret this result to mean that a spherical mass-to-flux ratios is not a very precise proxy
for the formation of disks from pre-stellar cores, due to the effects of turbulent motions.
As discussed above, the pre-stellar-cores are not spherical in shape,
and apparently the differences in the distribution of specific angular momentum inside the cores
has a stronger effect on disk formation than the mass-to-flux ratios.

\subsection{Angular momentum and magnetic field misalignment}
The fact that the Sun contains more than 99.9 \% of the mass in the solar system,
but less than 1 \% of the total angular momentum of the entire system
has been one of the major puzzles in models of solar system formation and thus of star formation in general.
In classical core collapse models of star formation, it was thus suggested that angular momentum can be efficiently transported by magnetic fields through magnetic braking.
However, studies and simulations of magnetized spherical core collapse revealed that this mechanism is in fact effective enough to suppress the formation of circumstellar disks altogether.
One mechanism that could help prevent the so called `magnetic braking catastrophe' is ambipolar diffusion, as suggested by \cite{1977ApJ...211..147M}.
However, more detailed studies by \cite{2009ApJ...698..922M} and \cite{2011ApJ...738..180L} showed that ambipolar diffusion will
rather cause a strong magnetic field at small circumstellar radii leading to efficient accretion shocks.

Another suggestion of how to reduce the effect of magnetic braking is changing the angle between the total angular momentum vector and the total magnetic field vector of the collapsing core.
Theoretical analysis indicated that magnetic braking is most efficient when the angular momentum and the magnetic field vectors are
perpendicular, and comparatively weaker for a parallel configuration \citep{1979ApJ...230..204M}.
More recent results \citep{2012A&A...543A.128J,2013ApJ...767L..11K} confirmed the fundamental idea that disk formation can be suppressed for certain angles, but in contrast to the theoretical prediction by \cite{1979ApJ...230..204M}, found that
magnetic braking acts strongest in case of a parallel alignment of the two vectors and weakest in a perpendicular configuration.
As pointed out by \cite{2012A&A...543A.128J}, the reason is that magnetic field lines are dragged towards
the center of the collapse, such that an initial parallel field develops sections that are nearly perpendicular to the angular
momentum vector, that act as efficient ``lever arms''.
Simulations of the collapse of spherical magnetized cores including the effect of turbulence
by \cite{2013MNRAS.432.3320S} also showed that circumstellar disks can form around young protostars
without relying on increased magnetic diffusion due to the non-ideal MHD effects.
Although they generally agree with the result that a misalignment of the magnetic field with respect to the angular momentum vector
facilitates disk formation, their conclusion is that this misalignment is rather the consequence of a more fundamental quantity present in GMCs, namely turbulence.
In our zoom simulations, we have the unique possibility to investigate the angular momentum transport in a consistent setup,
rather than imposing \textit{ad hoc} conditions on idealized core collapse models.

\begin{figure}[!htbp]
\subfigure{\includegraphics[clip, width=\linewidth]{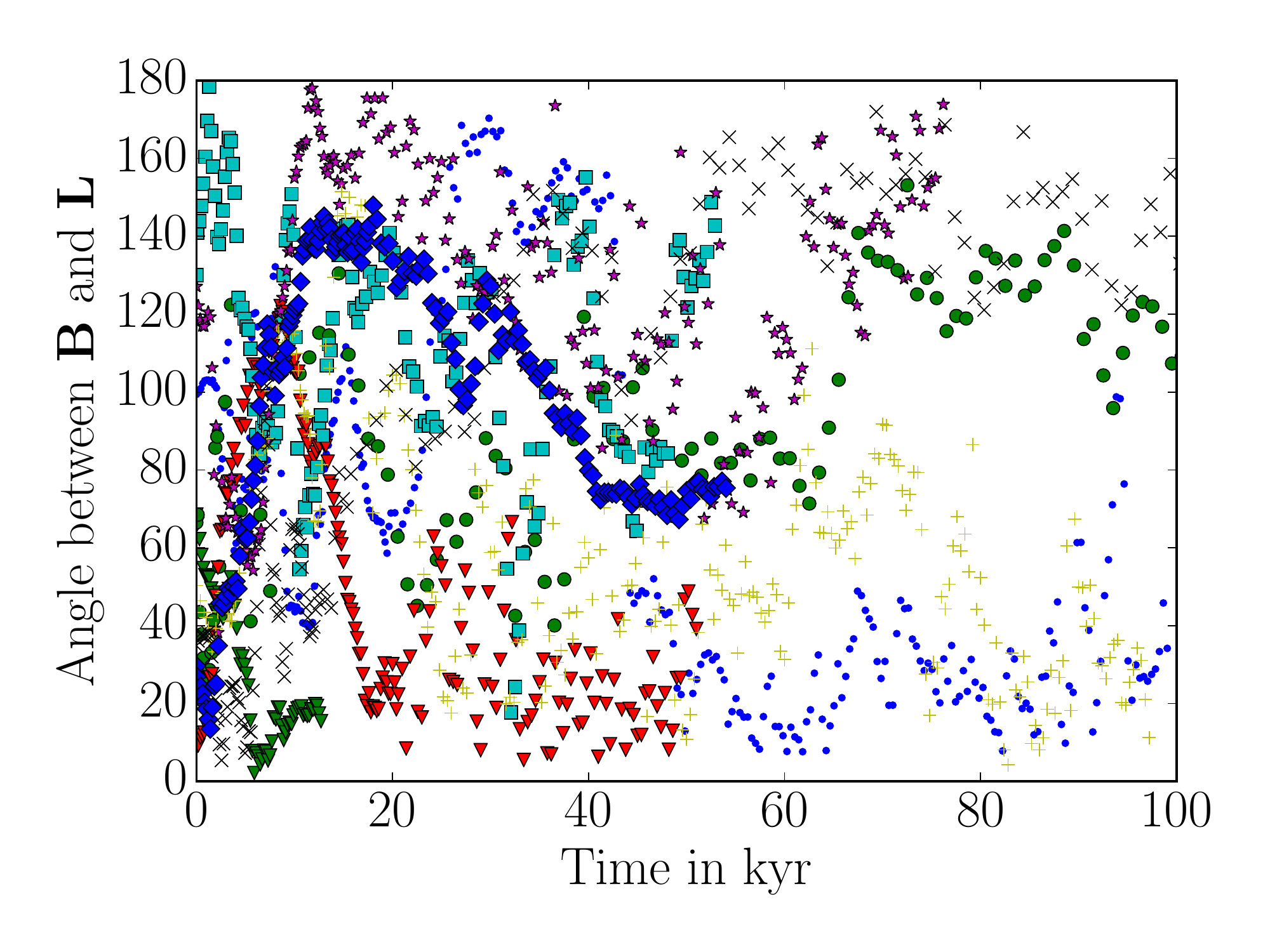} }
\protect\caption{\label{fig:L-B} Evolution of the angle between total angular momentum vector and total magnetic field direction
within a sphere of 100 AU around the eight different sinks.
The symbols belong to the same sinks as in \Fig{acc_mass_run9}.}
\end{figure}

In \Fig{L-B} we show the evolution of the angle between the angular momentum vector and the total magnetic field orientation within a radial distance of 100 AU from the sinks.
The plot illustrates the significant fluctuations of the angle around all sinks during their evolution,
and thus we conclude that in the early phases of protostellar evolution, where the majority of the protostar is assembled, the relative
orientation of the two quantities is highly dominated by the underlying turbulence,
and reflects the infall of gas clumps from different locations and with different angular momentum.
We also see that in the case of sinks that quickly evolve a strong rotationally supported circumstellar
velocity profile, the angle typically varies around $90$ degrees with $\pm \approx50$ degrees during the formation period.

In some cases, the angle between the angular momentum vector and magnetic-field vector can still vary significantly even during late phases, such as seen for
sink 8 around 60 kyr. This particular change from an angle of about $20$ degrees to nearly $180$ degrees is most probably related
to the peak in the accretion profile shortly after 60 kyr.
Considering that the rise in the accretion profile is caused by an infalling clump of mass,
it is not surprising that this clump can account for a significant shift in the alignment between magnetic field
and angular momentum.
Comparing the angle between the total angular momentum vector and
the total magnetic field vector of the gas around sinks with $\alpha$,
does not show a clear correlation between the level of rotation and the angle between the mean magnetic field vector
and the total angular momentum vector, indicating that in realistic situations it is an oversimplification to focus
on alignment / misalignment.

\subsection{Replenishment time}

\begin{figure*}[!htbp]
\subfigure{\includegraphics[width=\linewidth]{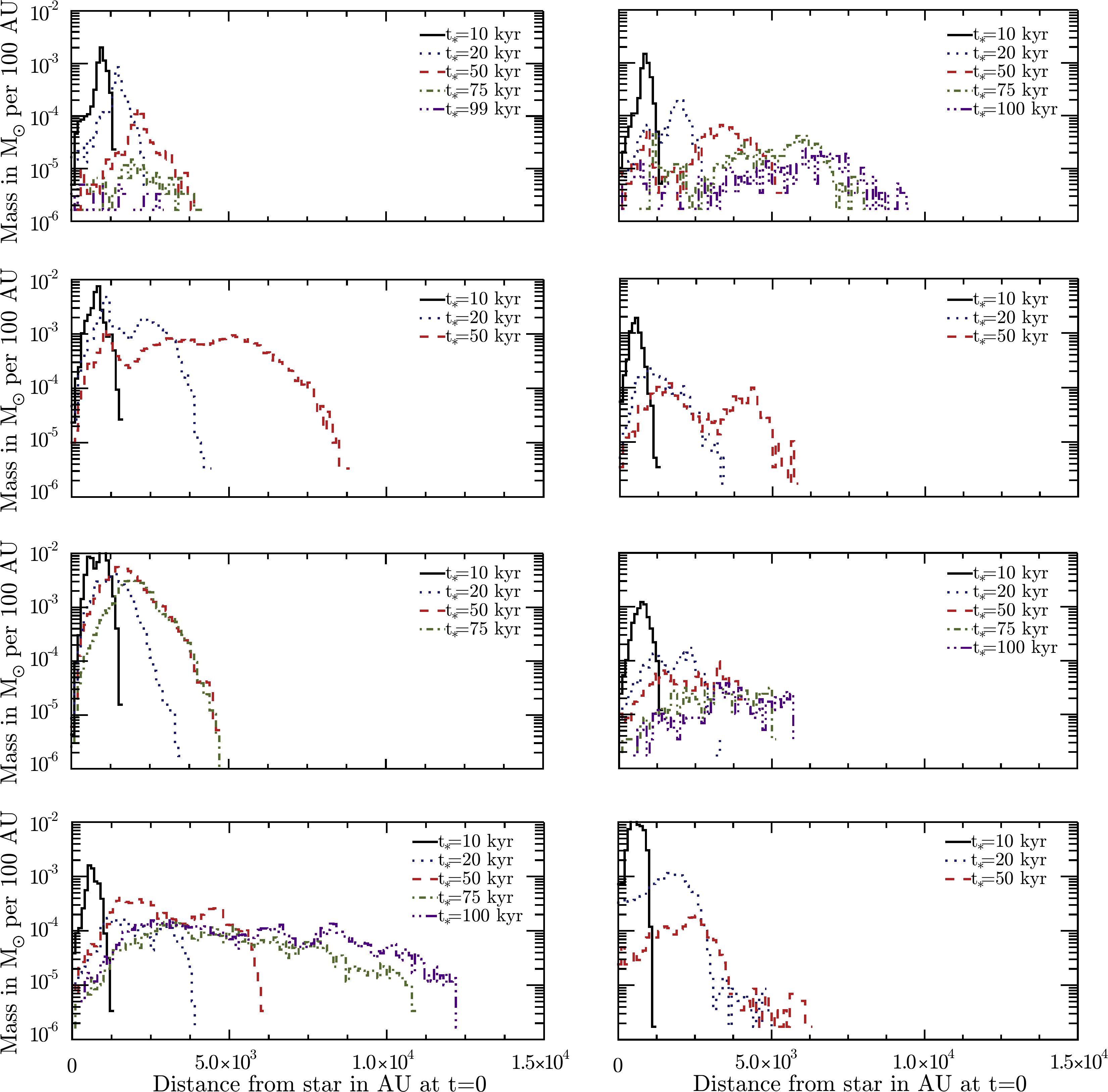} }
\protect\caption{\label{fig:origin_9} Distribution of where the gas that has accreted onto the different sinks
was located after times indicated in the legends at t=0 kyr.
The panels show from left to right and top to bottom the distribution for sink 1, 3, 4, 5, 6, 7, 8 and 9. }
\end{figure*}

To improve our understanding of the dynamics around the sink, we plot
where the mass that is located within a radial distance of 100 AU from the sink and that has not accreted onto the sink,
was located at the time of stellar birth \Fig{origin_9}.
One can see that for increasing times the gas stems from regions that were initially further away from the sink,
consistent with the classical model of inside-out collapse.
However, comparing the different sinks at similar late times, a fraction of the gas may stem from very different distances
(e.g. less than 5000 AU for sink 1, but beyond $10^4$ AU for sink 10).
We interpret this as a sign of spatial variations of the pre-stellar cores and particularly as evidence of infall through
large-scale filaments feeding the cores.
Looking closer at the individual profiles, the origin of the gas around sink 6 is striking.
Initially, it shows the same trend of an increasing original distance with increasing time,
but when comparing t=50 kyr with t=100 kyr,
almost no difference can be seen except for a lower mass contribution
from the gas that was initially closely located to the sink.
This feature may be caused by the fact that the gas orbits inside the disk for a long time
without any addition of new gas.
However, recalling the significant accretion rates of more than $10^{-6}$ \Macc in the early phases of star formation (\Fig{acc-prof}),
and the substantial disk masses around some of the protostars (and of sink 6 in particular),
indicates that the gas inside the protoplanetary disk gets replenished rather quickly.

\begin{figure}[!htbp]
\subfigure{\includegraphics[width=\linewidth]{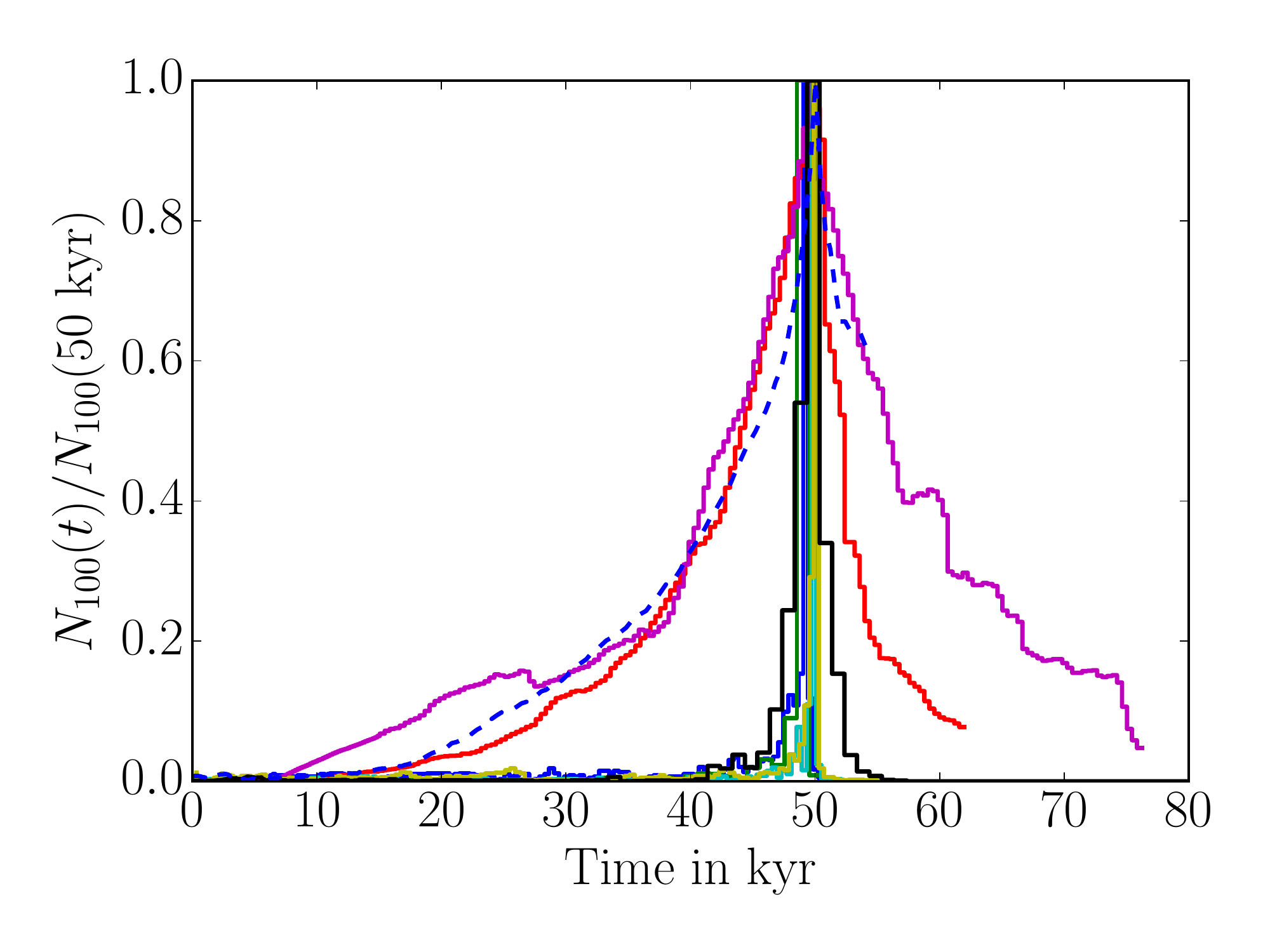} }
\protect\caption{\label{fig:N50} The number of tracer particles located within 100 AU distance
from the sink at t=50 kyr,compared to number of particles within 100 AU.
Blue corresponds to sink 1, green to sink 3, red to sink 4, cyan to sink 5,
magenta to sink 6, yellow to sink 7, black to sink 8 and blue dashed to sink 9.}
\end{figure}

To illustrate the short replenishment time, we plot in \Fig{N50} the fraction of tracer particles
that are located within 100 AU from the sink at t=50 kyr during the evolution of the stellar surrounding.
For the sinks that show no or only weak signs of disks,
the tracer particles move rapidly in radial direction
as indicated by the sharp peak in the plot.
The sinks with stronger signs of disks show broader peaks, but even in the case of the strongest disk (sink 6, magenta line),
most of the gas only remains in that region for at most a few ten thousand years.
This is in approximate agreement with the expected replenishment times of the disks defined as
\begin{equation}
t_{\rm repl} = \frac{m_{\rm disk}}{\dot{m}_{\rm accr}}.
\end{equation}

\subsection{Impact on Planet formation}

We have shown that circumstellar disks form at different times after stellar birth.
In some cases disks of significant size already form less than a few ten thousand years after stellar birth, while in other cases they have not formed within the first 100 kyr.
Assuming that these disks are indeed protoplanetary disks, the difference in disk formation times suggests that planet formation may occur at
significantly different times after protostellar birth.

Current models of planet formation assume power-law profiles for disk properties that depend on radius and height.
Motivated by the MMSN assumption,
these power-laws are time independent.
However, as shown in this study and in previous works, as well as in agreement with observations \citep{2015ApJ...805..125T},
protoplanetary disks may extend to radii of a few to several tens of AU already in the early phase of protostellar evolution.
Together with observations of gaps in the dust distribution of disks that are younger than 1 Myr (HL Tau) \citep{2014A&A...563L...2M,2015ApJ...808L...3A}
and commonly assumed average disk ages of about 2 Myr (for a critical analysis of potential underestimating disk ages due to selection biases
please refer to \citet{2014ApJ...793L..34P}), this raises the question whether planet formation already happens in the first few Myr,
perhaps starting already in the first few 100 kyr after protostellar formation.
At this stage, material moves rapidly in the radial direction through the disk and the protoplanetary disks might still be fed with new material as discussed above.
Therefore, the assumption of the disk mass as the mass reservoir for planet formation may be an oversimplification.
Instead, the mass reservoir for planet formation might in practice be significantly larger, even for low disk masses,
when accounting for the effects of rapid radial motions inside the disks as well as external infall onto the disk.
Hence, we emphasize that planets may have a much larger mass reservoir, when properly accounting for all the mass that
travels through the protoplanetary disk during the period of planet formation.

\section{Conclusions}
In this study, we investigated the first $\sim100$ kyr of protostellar formation and evolution
with 2 AU minimum cell size in nine cases, where 1-2 solar mass stars (sink particles) formed when the
minimum cell size was 126 AU.
Accounting for the physical environments in which the stars were embedded, we found that the accretion
process onto protostars is heterogeneous in multiple ways, namely
\begin{enumerate}
\item in time,
\item in space,
\item among protostars.
\end{enumerate}

Accretion is heterogeneous in time, in the sense that accretion rate time profiles vary significantly.
Initially, accretion rates are of the order of $10^{-5}$ M$_{\odot}/$yr to $10^{-4}$ M$_{\odot}/$yr,
and generally decrease during the subsequent evolution.
However, the protostars in our simulations sometimes underwent periods of increased accretion,
in which the accretion rates were enhanced by factors of a few to several.  We saw evidence for
event amplitudes increasing with increasing spatial resolution, and we note that event maxima may be missed
unless the snapshot cadence is correspondingly increased.

Accretion is heterogeneous in space, in the sense that mass accretion onto the star-disk system is filamentary,
acting through accretion channels and accretion sheets, rather than in the form of a smooth, space-filling infall of mass.

Last, the accretion processes differ among protostars, both quantitatively and qualitatively,
dependent on physical properties such as density, magnetization, and the strength of turbulence.
These properties are typically determined by the dynamics of the environment on length scales
similar to or larger than $\sim10^4$ AU; i.e., on scales that characterize the dynamical fluctuations
in the general turbulence of the GMC.

We thus conclude that the diversity in the large-scale stellar environment profoundly
influences the formation and evolution of protoplanetary disks.
If the magnetization of the surrounding gas is not too large,
protoplanetary disks can form as early as a few thousand years after star formation.
In cases where the magnetization of the collapsing gas is sufficiently large (low mass-to-flux ratios),
no disk of more than a few AU in size may form around the star.
We suggest that the main reason why the magnetic braking catastrophe is avoided in many cases is the reduction of magnetic
braking caused by turbulence \citep[cf. also][]{2013MNRAS.432.3320S}.
The fact that protoplanetary disks form in our simulations,
even though we do not account for non-ideal MHD effects,
suggests that either non-ideal MHD effects are not important in situations with realistic turbulence,
or disks form even more frequently than seen in our study.
Limited numerical resolution (cf.\ the Appendix) is not likely to
play a significant role, since accretion rates and frequencies of disk formation were
similar in early versions of these zoom simulations, where the number of cells per Jeans' length
was  significantly smaller \citep{2014IAUS..299..131N}.

We also studied the setting of parameters in our sink particle recipe. On the one hand the average
mass accretion profiles of the sinks are rather robust to a broad range of settings, but on the
other hand changes in the settings can have significant effects on the exact conditions under
which protoplanetary disks form.
Choosing settings that favor disk formation has the side effect that massive clumps of gas may
accrete onto the sink once in a while, thus causing periods with significant accretion bursts.
We thus conclude that, at least when the accretion process is modeled with transition from flows
to sink particles occurring at scales of a few AU, there is a significant uncertainty from parameter
dependencies. However, these uncertainties do not change the overall physical conclusion, that the
outcome of star and disk formation strongly depends on the pre-stellar environment.

We conclude that protoplanetary disk formation is a ubiquitous process in GMCs, with
rotationally supported disks forming due to the specific angular momentum of the collapsing gas,
even though the process is counteracted by magnetic braking.
An important factor ensuring that the formation of disks is not entirely suppressed is the presence
of turbulence (in both flows and magnetic fields) inherited from the large-scale dynamics.

\acknowledgements
This research was supported by a grant from the Danish Council for Independent Research to {\AA}N,
and a Sapere Aude Starting Grant from the Danish Council for Independent Research to TH.
Research at Centre for Star and Planet Formation is funded by the Danish National Research Foundation (DNRF97).
We acknowledge PRACE for awarding us access to the computing resource CURIE based in France at CEA for
carrying out part of the simulations.
Archival storage and computing nodes at the University of Copenhagen HPC center, funded with a research
grant (VKR023406) from Villum Fonden, were used for carrying out part of the simulations and the post-processing.

\bibliographystyle{apj}

\begin{thebibliography}{}
\expandafter\ifx\csname natexlab\endcsname\relax\def\natexlab#1{#1}\fi

\bibitem[{{Allen} {et~al.}(2003){Allen}, {Li}, \& {Shu}}]{2003ApJ...599..363A}
{Allen}, A., {Li}, Z.-Y., \& {Shu}, F.~H. 2003, \apj, 599, 363

\bibitem[{{ALMA Partnership} {et~al.}(2015){ALMA Partnership}, {Brogan},
  {P{\'e}rez}, {Hunter}, {Dent}, {Hales}, {Hills}, {Corder}, {Fomalont},
  {Vlahakis}, {Asaki}, {Barkats}, {Hirota}, {Hodge}, {Impellizzeri}, {Kneissl},
  {Liuzzo}, {Lucas}, {Marcelino}, {Matsushita}, {Nakanishi}, {Phillips},
  {Richards}, {Toledo}, {Aladro}, {Broguiere}, {Cortes}, {Cortes}, {Espada},
  {Galarza}, {Garcia-Appadoo}, {Guzman-Ramirez}, {Humphreys}, {Jung}, {Kameno},
  {Laing}, {Leon}, {Marconi}, {Mignano}, {Nikolic}, {Nyman}, {Radiszcz},
  {Remijan}, {Rod{\'o}n}, {Sawada}, {Takahashi}, {Tilanus}, {Vila Vilaro},
  {Watson}, {Wiklind}, {Akiyama}, {Chapillon}, {de Gregorio-Monsalvo}, {Di
  Francesco}, {Gueth}, {Kawamura}, {Lee}, {Nguyen Luong}, {Mangum}, {Pietu},
  {Sanhueza}, {Saigo}, {Takakuwa}, {Ubach}, {van Kempen}, {Wootten},
  {Castro-Carrizo}, {Francke}, {Gallardo}, {Garcia}, {Gonzalez}, {Hill},
  {Kaminski}, {Kurono}, {Liu}, {Lopez}, {Morales}, {Plarre}, {Schieven},
  {Testi}, {Videla}, {Villard}, {Andreani}, {Hibbard}, \&
  {Tatematsu}}]{2015ApJ...808L...3A}
{ALMA Partnership}, {Brogan}, C.~L., {P{\'e}rez}, L.~M., {et~al.} 2015, \apjl,
  808, L3

\bibitem[{{Armitage}(2007)}]{2007astro.ph..1485A}
{Armitage}, P.~J. 2007, ArXiv Astrophysics e-prints, astro-ph/0701485

\bibitem[{{Bai}(2015)}]{2015ApJ...798...84B}
{Bai}, X.-N. 2015, \apj, 798, 84

\bibitem[{{Balsara}(2015)}]{2015JCoPh.295....1B}
{Balsara}, D.~S. 2015, Journal of Computational Physics, 295, 1

\bibitem[{{Bate} {et~al.}(2014){Bate}, {Tricco}, \&
  {Price}}]{2014MNRAS.437...77B}
{Bate}, M.~R., {Tricco}, T.~S., \& {Price}, D.~J. 2014, \mnras, 437, 77

\bibitem[{{Beuther} {et~al.}(2010){Beuther}, {Vlemmings}, {Rao}, \& {van der
  Tak}}]{2010ApJ...724L.113B}
{Beuther}, H., {Vlemmings}, W.~H.~T., {Rao}, R., \& {van der Tak}, F.~F.~S.
  2010, \apjl, 724, L113

\bibitem[{{Blitz}(1993)}]{1993prpl.conf..125B}
{Blitz}, L. 1993, in Protostars and Planets III, ed. E.~H. {Levy} \& J.~I.
  {Lunine}, 125--161

\bibitem[{{Brinch} {et~al.}(2007){Brinch}, {Crapsi}, {Hogerheijde}, \&
  {J{\o}rgensen}}]{2007A&A...461.1037B}
{Brinch}, C., {Crapsi}, A., {Hogerheijde}, M.~R., \& {J{\o}rgensen}, J.~K.
  2007, \aap, 461, 1037

\bibitem[{{Brinch} {et~al.}(2008){Brinch}, {Hogerheijde}, \&
  {Richling}}]{2008A&A...489..607B}
{Brinch}, C., {Hogerheijde}, M.~R., \& {Richling}, S. 2008, \aap, 489, 607

\bibitem[{{Cleeves} {et~al.}(2014){Cleeves}, {Bergin}, \&
  {Adams}}]{2014ApJ...794..123C}
{Cleeves}, L.~I., {Bergin}, E.~A., \& {Adams}, F.~C. 2014, \apj, 794, 123

\bibitem[{{Elmegreen}(2000)}]{2000ApJ...530..277E}
{Elmegreen}, B.~G. 2000, \apj, 530, 277

\bibitem[{{Elmegreen} \& {Shadmehri}(2003)}]{2003MNRAS.338..817E}
{Elmegreen}, B.~G., \& {Shadmehri}, M. 2003, \mnras, 338, 817

\bibitem[{{Falgarone} {et~al.}(2008){Falgarone}, {Troland}, {Crutcher}, \&
  {Paubert}}]{2008A&A...487..247F}
{Falgarone}, E., {Troland}, T.~H., {Crutcher}, R.~M., \& {Paubert}, G. 2008,
  \aap, 487, 247

\bibitem[{{Franco} \& {Cox}(1986)}]{1986PASP...98.1076F}
{Franco}, J., \& {Cox}, D.~P. 1986, \pasp, 98, 1076

\bibitem[{{Frimann} {et~al.}(2016){Frimann}, {J{\o}rgensen}, \&
  {Haugb{\o}lle}}]{2016A&A...587A..59F}
{Frimann}, S., {J{\o}rgensen}, J.~K., \& {Haugb{\o}lle}, T. 2016, \aap, 587,
  A59

\bibitem[{{Fromang} {et~al.}(2006){Fromang}, {Hennebelle}, \&
  {Teyssier}}]{2006A&A...457..371F}
{Fromang}, S., {Hennebelle}, P., \& {Teyssier}, R. 2006, \aap, 457, 371

\bibitem[{{Girart} {et~al.}(2009){Girart}, {Beltr{\'a}n}, {Zhang}, {Rao}, \&
  {Estalella}}]{2009Sci...324.1408G}
{Girart}, J.~M., {Beltr{\'a}n}, M.~T., {Zhang}, Q., {Rao}, R., \& {Estalella},
  R. 2009, Science, 324, 1408

\bibitem[{{Gnedin} \& {Hollon}(2012)}]{2012ApJS..202...13G}
{Gnedin}, N.~Y., \& {Hollon}, N. 2012, \apjs, 202, 13

\bibitem[{{Gressel} {et~al.}(2015){Gressel}, {Turner}, {Nelson}, \&
  {McNally}}]{2015ApJ...801...84G}
{Gressel}, O., {Turner}, N.~J., {Nelson}, R.~P., \& {McNally}, C.~P. 2015,
  \apj, 801, 84

\bibitem[{{Hayashi}(1981)}]{1981PThPS..70...35H}
{Hayashi}, C. 1981, Progress of Theoretical Physics Supplement, 70, 35

\bibitem[{{Hennebelle} {et~al.}(2016){Hennebelle}, {Commer{\c c}on},
  {Chabrier}, \& {Marchand}}]{2016ApJ...830L...8H}
{Hennebelle}, P., {Commer{\c c}on}, B., {Chabrier}, G., \& {Marchand}, P. 2016,
  \apjl, 830, L8

\bibitem[{{Hennebelle} \& {Fromang}(2008)}]{2008A&A...477....9H}
{Hennebelle}, P., \& {Fromang}, S. 2008, \aap, 477, 9

\bibitem[{{Hopkins}(2015)}]{2015MNRAS.450...53H}
{Hopkins}, P.~F. 2015, \mnras, 450, 53

\bibitem[{{Jiang} {et~al.}(2013){Jiang}, {Belyaev}, {Goodman}, \&
  {Stone}}]{2013NewA...19...48J}
{Jiang}, Y.-F., {Belyaev}, M., {Goodman}, J., \& {Stone}, J.~M. 2013, \na, 19,
  48

\bibitem[{{Joos} {et~al.}(2012){Joos}, {Hennebelle}, \&
  {Ciardi}}]{2012A&A...543A.128J}
{Joos}, M., {Hennebelle}, P., \& {Ciardi}, A. 2012, \aap, 543, A128

\bibitem[{{Joos} {et~al.}(2013){Joos}, {Hennebelle}, {Ciardi}, \&
  {Fromang}}]{2013A&A...554A..17J}
{Joos}, M., {Hennebelle}, P., {Ciardi}, A., \& {Fromang}, S. 2013, \aap, 554,
  A17

\bibitem[{{Kenyon} {et~al.}(1990){Kenyon}, {Hartmann}, {Strom}, \&
  {Strom}}]{1990AJ.....99..869K}
{Kenyon}, S.~J., {Hartmann}, L.~W., {Strom}, K.~M., \& {Strom}, S.~E. 1990,
  \aj, 99, 869

\bibitem[{{Kim} \& {Ostriker}(2001)}]{2001ApJ...559...70K}
{Kim}, W.-T., \& {Ostriker}, E.~C. 2001, \apj, 559, 70

\bibitem[{{Krasnopolsky} {et~al.}(2010){Krasnopolsky}, {Li}, \&
  {Shang}}]{2010ApJ...716.1541K}
{Krasnopolsky}, R., {Li}, Z.-Y., \& {Shang}, H. 2010, \apj, 716, 1541

\bibitem[{{Krasnopolsky} {et~al.}(2011){Krasnopolsky}, {Li}, \&
  {Shang}}]{2011ApJ...733...54K}
---. 2011, \apj, 733, 54

\bibitem[{{Kritsuk} {et~al.}(2011){Kritsuk}, {Nordlund}, {Collins}, {Padoan},
  {Norman}, {Abel}, {Banerjee}, {Federrath}, {Flock}, {Lee}, {Li},
  {M{\"u}ller}, {Teyssier}, {Ustyugov}, {Vogel}, \& {Xu}}]{2011ApJ...737...13K}
{Kritsuk}, A.~G., {Nordlund}, {\AA}., {Collins}, D., {et~al.} 2011, \apj, 737,
  13

\bibitem[{{Krumholz} {et~al.}(2013){Krumholz}, {Crutcher}, \&
  {Hull}}]{2013ApJ...767L..11K}
{Krumholz}, M.~R., {Crutcher}, R.~M., \& {Hull}, C.~L.~H. 2013, \apjl, 767, L11

\bibitem[{{Kuffmeier} {et~al.}(2016){Kuffmeier}, {Frostholm Mogensen},
  {Haugb{\o}lle}, {Bizzarro}, \& {Nordlund}}]{2016ApJ...826...22K}
{Kuffmeier}, M., {Frostholm Mogensen}, T., {Haugb{\o}lle}, T., {Bizzarro}, M.,
  \& {Nordlund}, {\AA}. 2016, \apj, 826, 22

\bibitem[{{Larson}(1969)}]{1969MNRAS.145..271L}
{Larson}, R.~B. 1969, \mnras, 145, 271

\bibitem[{{Larson}(1981)}]{1981MNRAS.194..809L}
---. 1981, \mnras, 194, 809

\bibitem[{{Lesur} {et~al.}(2014){Lesur}, {Kunz}, \&
  {Fromang}}]{2014A&A...566A..56L}
{Lesur}, G., {Kunz}, M.~W., \& {Fromang}, S. 2014, \aap, 566, A56

\bibitem[{{Li} {et~al.}(2011){Li}, {Krasnopolsky}, \&
  {Shang}}]{2011ApJ...738..180L}
{Li}, Z.-Y., {Krasnopolsky}, R., \& {Shang}, H. 2011, \apj, 738, 180

\bibitem[{{Li} {et~al.}(2014){Li}, {Krasnopolsky}, {Shang}, \&
  {Zhao}}]{2014ApJ...793..130L}
{Li}, Z.-Y., {Krasnopolsky}, R., {Shang}, H., \& {Zhao}, B. 2014, \apj, 793,
  130

\bibitem[{Liu {et~al.}(2016)Liu, Takami, Kudo, Hashimoto, Dong, Vorobyov, Pyo,
  Fukagawa, Tamura, Henning, Dunham, Karr, Kusakabe, \& Tsuribe}]{Liue1500875}
Liu, H.~B., Takami, M., Kudo, T., {et~al.} 2016, Science Advances, 2,
  http://advances.sciencemag.org/content/2/2/e1500875.full.pdf

\bibitem[{{Machida} {et~al.}(2007){Machida}, {Inutsuka}, \&
  {Matsumoto}}]{2007ApJ...670.1198M}
{Machida}, M.~N., {Inutsuka}, S.-i., \& {Matsumoto}, T. 2007, \apj, 670, 1198

\bibitem[{{Machida} {et~al.}(2014){Machida}, {Inutsuka}, \&
  {Matsumoto}}]{2014MNRAS.438.2278M}
---. 2014, \mnras, 438, 2278

\bibitem[{{Machida} \& {Matsumoto}(2011)}]{2011MNRAS.413.2767M}
{Machida}, M.~N., \& {Matsumoto}, T. 2011, \mnras, 413, 2767

\bibitem[{{Machida} {et~al.}(2006){Machida}, {Matsumoto}, {Hanawa}, \&
  {Tomisaka}}]{2006ApJ...645.1227M}
{Machida}, M.~N., {Matsumoto}, T., {Hanawa}, T., \& {Tomisaka}, K. 2006, \apj,
  645, 1227

\bibitem[{{Machida} {et~al.}(2004){Machida}, {Tomisaka}, \&
  {Matsumoto}}]{2004MNRAS.348L...1M}
{Machida}, M.~N., {Tomisaka}, K., \& {Matsumoto}, T. 2004, \mnras, 348, L1

\bibitem[{{Masson} {et~al.}(2016){Masson}, {Chabrier}, {Hennebelle}, {Vaytet},
  \& {Commer{\c c}on}}]{2016A&A...587A..32M}
{Masson}, J., {Chabrier}, G., {Hennebelle}, P., {Vaytet}, N., \& {Commer{\c
  c}on}, B. 2016, \aap, 587, A32

\bibitem[{{Maury} {et~al.}(2014){Maury}, {Belloche}, {Andr{\'e}}, {Maret},
  {Gueth}, {Codella}, {Cabrit}, {Testi}, \& {Bontemps}}]{2014A&A...563L...2M}
{Maury}, A.~J., {Belloche}, A., {Andr{\'e}}, P., {et~al.} 2014, \aap, 563, L2

\bibitem[{{Mayor} \& {Queloz}(1995)}]{1995Natur.378..355M}
{Mayor}, M., \& {Queloz}, D. 1995, \nat, 378, 355

\bibitem[{{Mellon} \& {Li}(2009)}]{2009ApJ...698..922M}
{Mellon}, R.~R., \& {Li}, Z.-Y. 2009, \apj, 698, 922

\bibitem[{{Mouschovias}(1977)}]{1977ApJ...211..147M}
{Mouschovias}, T.~C. 1977, \apj, 211, 147

\bibitem[{{Mouschovias} \& {Paleologou}(1979)}]{1979ApJ...230..204M}
{Mouschovias}, T.~C., \& {Paleologou}, E.~V. 1979, \apj, 230, 204

\bibitem[{{Mouschovias} \& {Spitzer}(1976)}]{1976ApJ...210..326M}
{Mouschovias}, T.~C., \& {Spitzer}, Jr., L. 1976, \apj, 210, 326

\bibitem[{{Murray}(2011)}]{2011ApJ...729..133M}
{Murray}, N. 2011, \apj, 729, 133

\bibitem[{{Myers} {et~al.}(2013){Myers}, {McKee}, {Cunningham}, {Klein}, \&
  {Krumholz}}]{2013ApJ...766...97M}
{Myers}, A.~T., {McKee}, C.~F., {Cunningham}, A.~J., {Klein}, R.~I., \&
  {Krumholz}, M.~R. 2013, \apj, 766, 97

\bibitem[{{Nordlund} {et~al.}(2014){Nordlund}, {Haugb{\o}lle}, {K{\"u}ffmeier},
  {Padoan}, \& {Vasileiades}}]{2014IAUS..299..131N}
{Nordlund}, {\AA}., {Haugb{\o}lle}, T., {K{\"u}ffmeier}, M., {Padoan}, P., \&
  {Vasileiades}, A. 2014, in IAU Symposium, Vol. 299, IAU Symposium, ed.
  M.~{Booth}, B.~C. {Matthews}, \& J.~R. {Graham}, 131--135

\bibitem[{{Osterbrock} \& {Ferland}(2006)}]{2006agna.book.....O}
{Osterbrock}, D.~E., \& {Ferland}, G.~J. 2006, {Astrophysics of gaseous nebulae
  and active galactic nuclei}

\bibitem[{{Padoan} {et~al.}(2014){Padoan}, {Haugb{\o}lle}, \&
  {Nordlund}}]{2014ApJ...797...32P}
{Padoan}, P., {Haugb{\o}lle}, T., \& {Nordlund}, {\AA}. 2014, \apj, 797, 32

\bibitem[{{Padoan} \& {Nordlund}(2002)}]{2002ApJ...576..870P}
{Padoan}, P., \& {Nordlund}, {\AA}. 2002, \apj, 576, 870

\bibitem[{{Padoan} {et~al.}(2016){Padoan}, {Pan}, {Haugb{\o}lle}, \&
  {Nordlund}}]{2016ApJ...822...11P}
{Padoan}, P., {Pan}, L., {Haugb{\o}lle}, T., \& {Nordlund}, {\AA}. 2016, \apj,
  822, 11

\bibitem[{{Padovani} {et~al.}(2009){Padovani}, {Galli}, \&
  {Glassgold}}]{2009A&A...501..619P}
{Padovani}, M., {Galli}, D., \& {Glassgold}, A.~E. 2009, \aap, 501, 619

\bibitem[{{Padovani} {et~al.}(2014){Padovani}, {Galli}, {Hennebelle},
  {Commer{\c c}on}, \& {Joos}}]{2014A&A...571A..33P}
{Padovani}, M., {Galli}, D., {Hennebelle}, P., {Commer{\c c}on}, B., \& {Joos},
  M. 2014, \aap, 571, A33

\bibitem[{{Padovani} {et~al.}(2013){Padovani}, {Hennebelle}, \&
  {Galli}}]{2013A&A...560A.114P}
{Padovani}, M., {Hennebelle}, P., \& {Galli}, D. 2013, \aap, 560, A114

\bibitem[{{Pfalzner} {et~al.}(2014){Pfalzner}, {Steinhausen}, \&
  {Menten}}]{2014ApJ...793L..34P}
{Pfalzner}, S., {Steinhausen}, M., \& {Menten}, K. 2014, \apjl, 793, L34

\bibitem[{{Richardson} {et~al.}(2002){Richardson}, {Branch}, {Casebeer},
  {Millard}, {Thomas}, \& {Baron}}]{2002AJ....123..745R}
{Richardson}, D., {Branch}, D., {Casebeer}, D., {et~al.} 2002, \aj, 123, 745

\bibitem[{{Safron} {et~al.}(2015){Safron}, {Fischer}, {Megeath}, {Furlan},
  {Stutz}, {Stanke}, {Billot}, {Rebull}, {Tobin}, {Ali}, {Allen}, {Booker},
  {Watson}, \& {Wilson}}]{2015ApJ...800L...5S}
{Safron}, E.~J., {Fischer}, W.~J., {Megeath}, S.~T., {et~al.} 2015, \apjl, 800,
  L5

\bibitem[{{Schaller} {et~al.}(1992){Schaller}, {Schaerer}, {Meynet}, \&
  {Maeder}}]{1992A&AS...96..269S}
{Schaller}, G., {Schaerer}, D., {Meynet}, G., \& {Maeder}, A. 1992, \aaps, 96,
  269

\bibitem[{{Seifried} {et~al.}(2011){Seifried}, {Banerjee}, {Klessen}, {Duffin},
  \& {Pudritz}}]{2011MNRAS.417.1054S}
{Seifried}, D., {Banerjee}, R., {Klessen}, R.~S., {Duffin}, D., \& {Pudritz},
  R.~E. 2011, \mnras, 417, 1054

\bibitem[{{Seifried} {et~al.}(2013){Seifried}, {Banerjee}, {Pudritz}, \&
  {Klessen}}]{2013MNRAS.432.3320S}
{Seifried}, D., {Banerjee}, R., {Pudritz}, R.~E., \& {Klessen}, R.~S. 2013,
  \mnras, 432, 3320

\bibitem[{{Seifried} {et~al.}(2012){Seifried}, {Pudritz}, {Banerjee}, {Duffin},
  \& {Klessen}}]{2012MNRAS.422..347S}
{Seifried}, D., {Pudritz}, R.~E., {Banerjee}, R., {Duffin}, D., \& {Klessen},
  R.~S. 2012, \mnras, 422, 347

\bibitem[{{Seifried} \& {Walch}(2015)}]{2015MNRAS.452.2410S}
{Seifried}, D., \& {Walch}, S. 2015, \mnras, 452, 2410

\bibitem[{{Teyssier}(2002)}]{2002A&A...385..337T}
{Teyssier}, R. 2002, \aap, 385, 337

\bibitem[{{Tobin} {et~al.}(2015){Tobin}, {Looney}, {Wilner}, {Kwon},
  {Chandler}, {Bourke}, {Loinard}, {Chiang}, {Schnee}, \&
  {Chen}}]{2015ApJ...805..125T}
{Tobin}, J.~J., {Looney}, L.~W., {Wilner}, D.~J., {et~al.} 2015, \apj, 805, 125

\bibitem[{{Tomida} {et~al.}(2015){Tomida}, {Okuzumi}, \&
  {Machida}}]{2015ApJ...801..117T}
{Tomida}, K., {Okuzumi}, S., \& {Machida}, M.~N. 2015, \apj, 801, 117

\bibitem[{{Tomida} {et~al.}(2013){Tomida}, {Tomisaka}, {Matsumoto}, {Hori},
  {Okuzumi}, {Machida}, \& {Saigo}}]{2013ApJ...763....6T}
{Tomida}, K., {Tomisaka}, K., {Matsumoto}, T., {et~al.} 2013, \apj, 763, 6

\bibitem[{{Tomida} {et~al.}(2010){Tomida}, {Tomisaka}, {Matsumoto}, {Ohsuga},
  {Machida}, \& {Saigo}}]{2010ApJ...714L..58T}
---. 2010, \apjl, 714, L58

\bibitem[{{Toomre}(1964)}]{1964ApJ...139.1217T}
{Toomre}, A. 1964, \apj, 139, 1217

\bibitem[{{Tsukamoto} {et~al.}(2015){Tsukamoto}, {Iwasaki}, {Okuzumi},
  {Machida}, \& {Inutsuka}}]{2015MNRAS.452..278T}
{Tsukamoto}, Y., {Iwasaki}, K., {Okuzumi}, S., {Machida}, M.~N., \& {Inutsuka},
  S. 2015, \mnras, 452, 278

\bibitem[{{Umebayashi} \& {Nakano}(2009)}]{2009ApJ...690...69U}
{Umebayashi}, T., \& {Nakano}, T. 2009, \apj, 690, 69

\bibitem[{{van der Marel} {et~al.}(2013){van der Marel}, {van Dishoeck},
  {Bruderer}, {Birnstiel}, {Pinilla}, {Dullemond}, {van Kempen}, {Schmalzl},
  {Brown}, {Herczeg}, {Mathews}, \& {Geers}}]{2013Sci...340.1199V}
{van der Marel}, N., {van Dishoeck}, E.~F., {Bruderer}, S., {et~al.} 2013,
  Science, 340, 1199

\bibitem[{{Vasileiadis} {et~al.}(2013){Vasileiadis}, {Nordlund}, \&
  {Bizzarro}}]{2013ApJ...769L...8V}
{Vasileiadis}, A., {Nordlund}, {\AA}., \& {Bizzarro}, M. 2013, \apjl, 769, L8

\bibitem[{{Watson} {et~al.}(2016){Watson}, {Calvet}, {Fischer}, {Forrest},
  {Manoj}, {Megeath}, {Melnick}, {Najita}, {Neufeld}, {Sheehan}, {Stutz}, \&
  {Tobin}}]{2016ApJ...828...52W}
{Watson}, D.~M., {Calvet}, N.~P., {Fischer}, W.~J., {et~al.} 2016, \apj, 828,
  52

\bibitem[{{Weidenschilling}(1977)}]{1977MNRAS.180...57W}
{Weidenschilling}, S.~J. 1977, \mnras, 180, 57

\bibitem[{{Wolszczan} \& {Frail}(1992)}]{1992Natur.355..145W}
{Wolszczan}, A., \& {Frail}, D.~A. 1992, \nat, 355, 145

\bibitem[{{Wurster} {et~al.}(2016){Wurster}, {Price}, \&
  {Bate}}]{2016MNRAS.457.1037W}
{Wurster}, J., {Price}, D.~J., \& {Bate}, M.~R. 2016, \mnras, 457, 1037

\bibitem[{{Yen} {et~al.}(2014){Yen}, {Takakuwa}, {Ohashi}, {Aikawa}, {Aso},
  {Koyamatsu}, {Machida}, {Saigo}, {Saito}, {Tomida}, \&
  {Tomisaka}}]{2014ApJ...793....1Y}
{Yen}, H.-W., {Takakuwa}, S., {Ohashi}, N., {et~al.} 2014, \apj, 793, 1

\bibitem[{{Zanni} {et~al.}(2007){Zanni}, {Ferrari}, {Rosner}, {Bodo}, \&
  {Massaglia}}]{2007A&A...469..811Z}
{Zanni}, C., {Ferrari}, A., {Rosner}, R., {Bodo}, G., \& {Massaglia}, S. 2007,
  \aap, 469, 811

\end{thebibliography}

\appendix
\section{Refinement and sink accretion parameters}

\begin{figure*}[!htbp]
\subfigure{\includegraphics[width=0.5\linewidth]{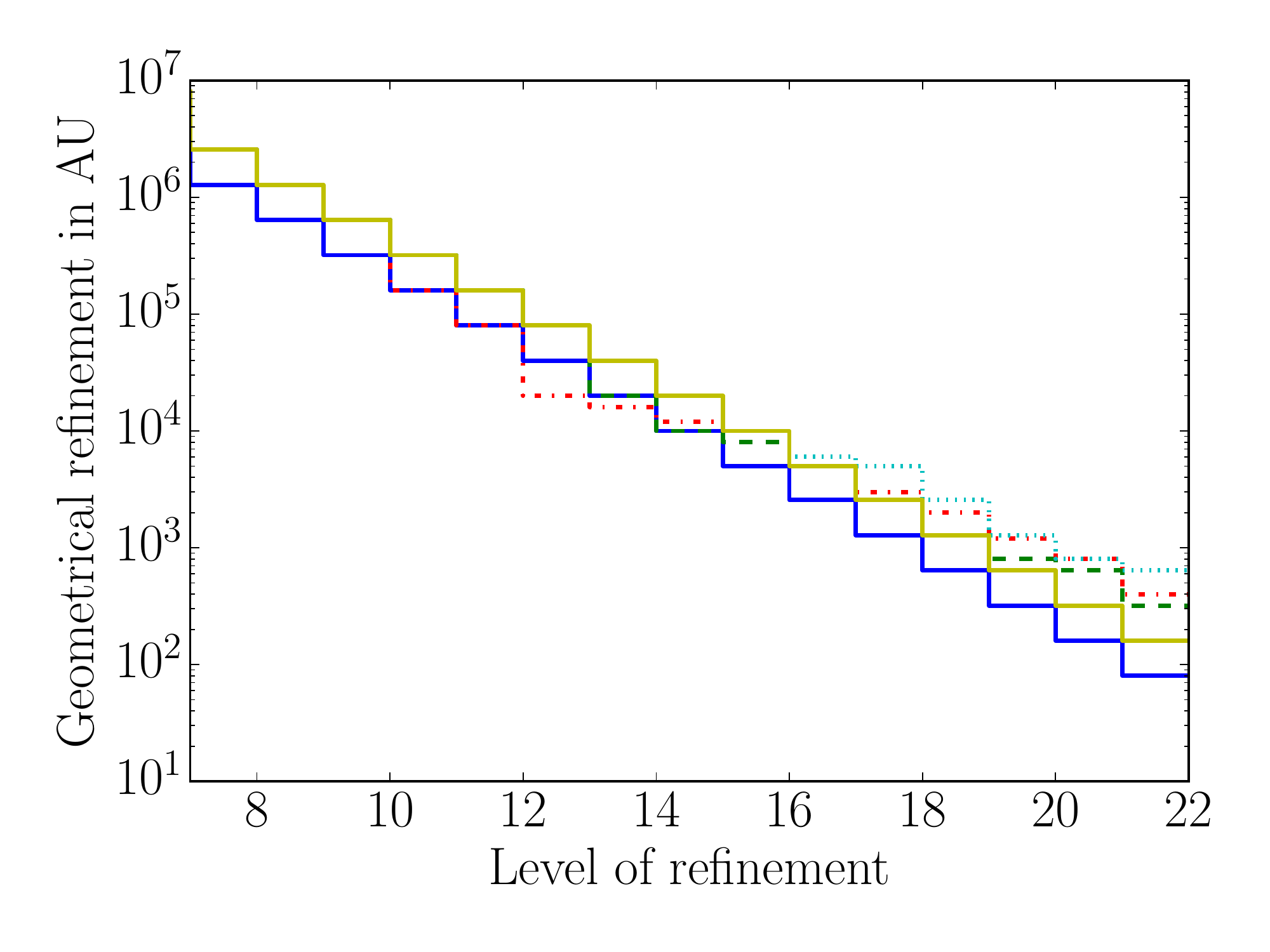}} \quad
\subfigure{\includegraphics[width=0.5\linewidth]{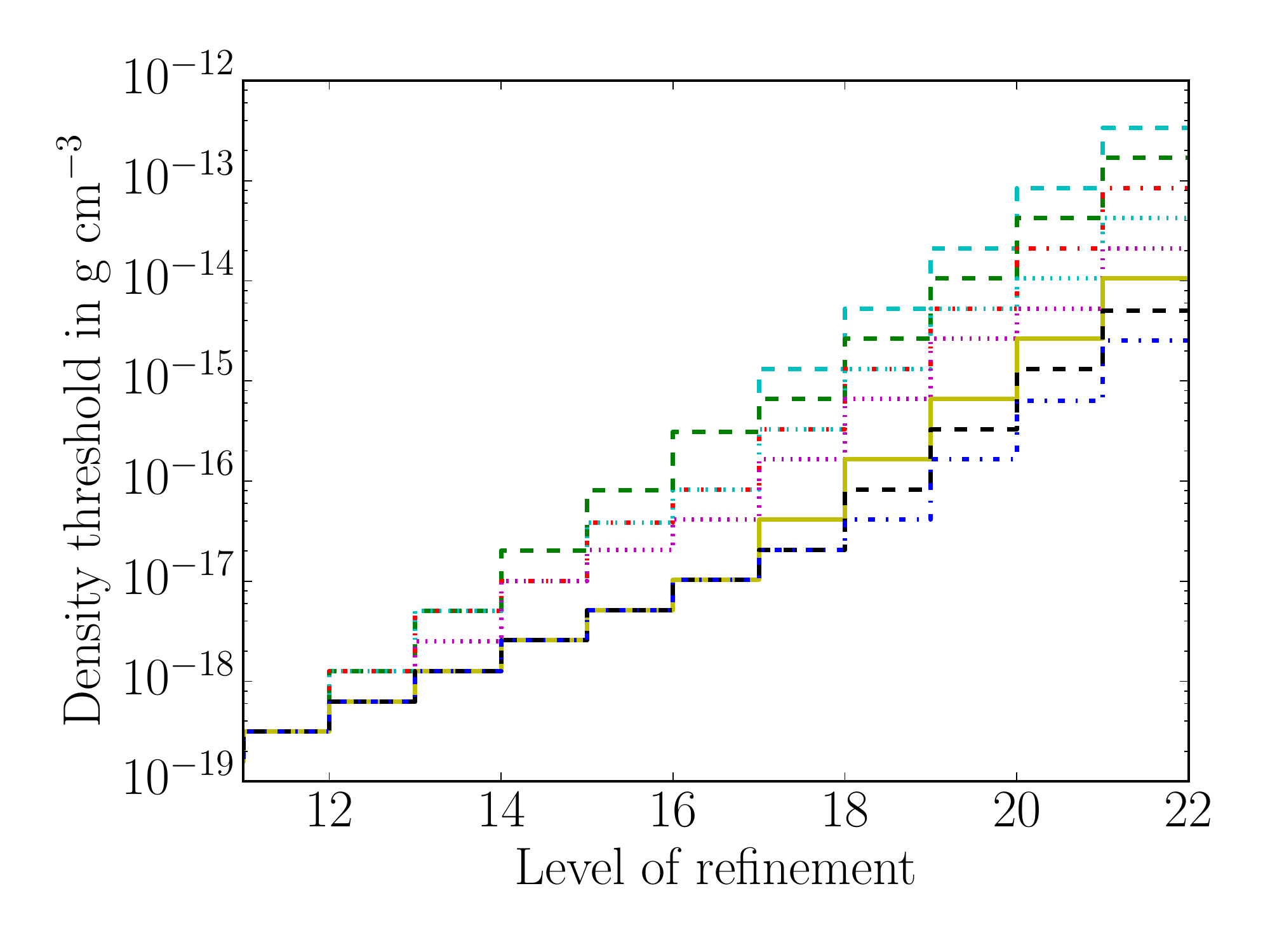}} \quad
\protect\caption{\label{fig:lref_distr} Left panel: the geometrical refinement ladder.
Blue shows the ladder around sink 1, sink 2, and sink 6 after $t=30$ kyr
and for the evolution around sink 7 after $t=18$ kyr.
Green dashed corresponds to the ladder around sink 3 and 9, red dash-dot around sink 4,
cyan dot-dot illustrates the initial ladder around sink 5 until $t=30$ kyr
and the yellow solid line corresponds to the initial ladder around sink 7 until $t=18$ kyr.
Right panel: refinement ladders for the gas around the different sinks at different periods.
In \Fig{lmax_distr} we show at what times the setting was changed.
The red dotted line corresponds to the initial density threshold for refinement
for sink 4, until its first change. The cyan dashed line corresponds to the second period for sink 4,
the green dashed line corresponds to the ladder during
third period for sink 4 and the initial ladder for the other sinks,
red dashed-dotted line corresponds to the fourth period of sink 4
and the second period for sink 5, 6, 7, 8 and 9,
cyan dot-dot shows the third period for sink 1, 5, 6, 7 and 9,
magenta dot-dot corresponds to the fourth period for sink 1, 7 and 9,
yellow solid to the fifth period for sink 1 and black dashed to the final period for sink 1.
}
\end{figure*}

To limit the computational costs only to the regions of interest,
we apply geometrical refinement in our zoom-runs, where a cell is only allowed to be resolved
to a location-dependent maximum level of $\ell=$\lmax(r).
We then mark cells for refinement if they exceed level-dependent threshold values for one of
the refinement criteria, such as number of cells per Jeans' length, steep gradients in velocity,
magnetic field magnitude, gas density, or gas pressure.
The maximum allowed refinement level decreases with increasing distance from the protostar
according to a refinement ladder (as shown in the left panel of \Fig{lref_distr}).
Such a limitation on the refinement (referred to as `geometrical refinement' in the
\ramses\ user guide) is thus only a necessary, but not a sufficient condition for cells to be refined.
Our most important sufficient condition for refinement is the density (Jeans' length) criterion.
A cell is marked for refinement to a certain level if the density exceeds a level dependent threshold value.
We illustrate the chosen refinement ladders for the different runs in the right hand side panel of \Fig{lref_distr}.

With respect to density gradients, cells are generally refined if the difference between neighboring cells
exceeds a factor of $20$.
However, this only applies up to level of 20, and steeper density gradients are allowed
for cells above level 20. Similarly, refinement is triggered by gradients in the speed
and the magnetic field strength. We have found that refining on gradients helps in maintaining
waves and features in the flow, and in resolving shocks, in particular in the vicinity of the protostars.
In Table \ref{run-params} the parameters for refinement due to gradients in
velocity and magnetic field strength in the different runs are listed.
The first column shows the number of the sink. The second column indicates the maximum level
at which cells are refined based on gradients in the velocity. The value in parentheses indicates the
time in kyr after sink creation at which the maximum level was increased from 20 to 22.
Refinement is triggered when the relative velocity gradient differs by more than a factor of $f=5$
compared to the sound speed using the following inequality to trigger refinement
\begin{equation}
\sum_{i \in x,y,z} \max\left(
\frac{2 \left|v^+_i - v^0_i\right|}{c_s^+ + c_s^0},
\frac{2 \left|v^-_i - v^0_i\right|}{c_s^- + c_s^0}
\right)^2 > f^2\,,
\end{equation}
where $c_s$ is the sound speed, and $-,0,+$ indicates cells at $-1,0,+1$ cells distance from the center cell.
The third column indicates whether cells are selected for refinement if the magnetic field strength between
neighboring cells differs by more than a factor of 3. We point out that although these settings differ slightly
between runs, their effect is only minor. The most important criterion for refinement is the density.

\begin{table}[!htbp]
\centering
{
\begin{tabular}{rrrr}

\begin{tabular}{@{}c@{}} \# of \\ sink \end{tabular} &
\begin{tabular}{@{}c@{}} \lumax \end{tabular} &
\begin{tabular}{@{}c@{}} $\Delta B_{\rm lim}$ \end{tabular} &
\begin{tabular}{@{}c@{}} \cB \end{tabular} \\ \hline
1a & 22      & No  & 100      \\
2  & 22      & No  & 100(9)   \\
3  & 22      & Yes & 100      \\
4  & 22      & Yes & 100      \\
5  & 20      & No  & No       \\
6  & 20      & No  & 100(7)   \\
7  & 22(16)  & No  & 100(29)  \\
8  & 22      & Yes & 100      \\
9  & 22      & Yes & 100      \\
\end{tabular} }
\caption{Overview of the refinement parameters for the nine sinks selected for zoom-in.
First column: number of sink,
second column: maximum level where refinement based on the speed gradient is applied,
third column: if refinement based on gradient of magnetic field is activated
fourth column: value of the fast magnetosonic speed at which the code starts using the LLF solver instead of the HLLD solver [km\ s$^{-1}$].
The number in parentheses in column two and four indicates the time after sink creation in kyr at which the value is changed.}
\label{run-params}
\end{table}

Besides the different refinement criteria, the remaining numerical parameter that is different in the runs
is listed in the fourth column: it is the threshold value of the fast magnetosonic velocity at which the code switches from
the HLLD solver to a more diffusive LLF (Lax-Friedrichs)
solver. This is to avoid developing exceedingly large magnetosonic speeds in otherwise uninteresting places,
which has aggravating effects on the time step.

\section{Sink parameter comparison}
To assess the robustness of our results we investigate the impact on the accretion process of setting different parameters
in our sink particle model. To do this, we carried out a parameter study with the sink settings given in Table
\ref{sink-params-070}, for zoom simulations around sink 1.
The table provides an overview of the parameters that are important for accretion onto the sink.
The third column displays the density limit that has to be exceeded by a cell in order to accrete mass onto the sink.
Column four gives the radius in cells inside which mass is allowed to be accreted onto the sink \racc.
In the fifth column, we give the accretion rate \accrate\ from a cell onto the
sink. \accrate\ is a prefactor, which is related to the assumed angle between the angular motion and the
radial motion of the gas very close to the sink, so that the amount of mass removed becomes proportional
to $\accrate\ \rho v_K$, where $v_K$ is the Keplerian speed.
The sixth column gives the fraction of the accreting mass that is added to the sink.
The remaining mass is removed from the box, to mimic the mass lost in outflows. According to observations -- in
particular the core mass function to initial mass function correspondence --
between half and two thirds of the envelope mass is lost in the outflows \citep{2016ApJ...828...52W}.
With a 2 AU cell resolution, we already
account for some of the mass loss in winds and due to envelope mass that is entrained in the wind, but it is non-trivial to
measure how much. We therefore experiment with setting the accreted fraction to both 50\% and 100\%, to gauge the importance.

\begin{figure}[!htbp]
\subfigure{\includegraphics[width=0.5\linewidth]{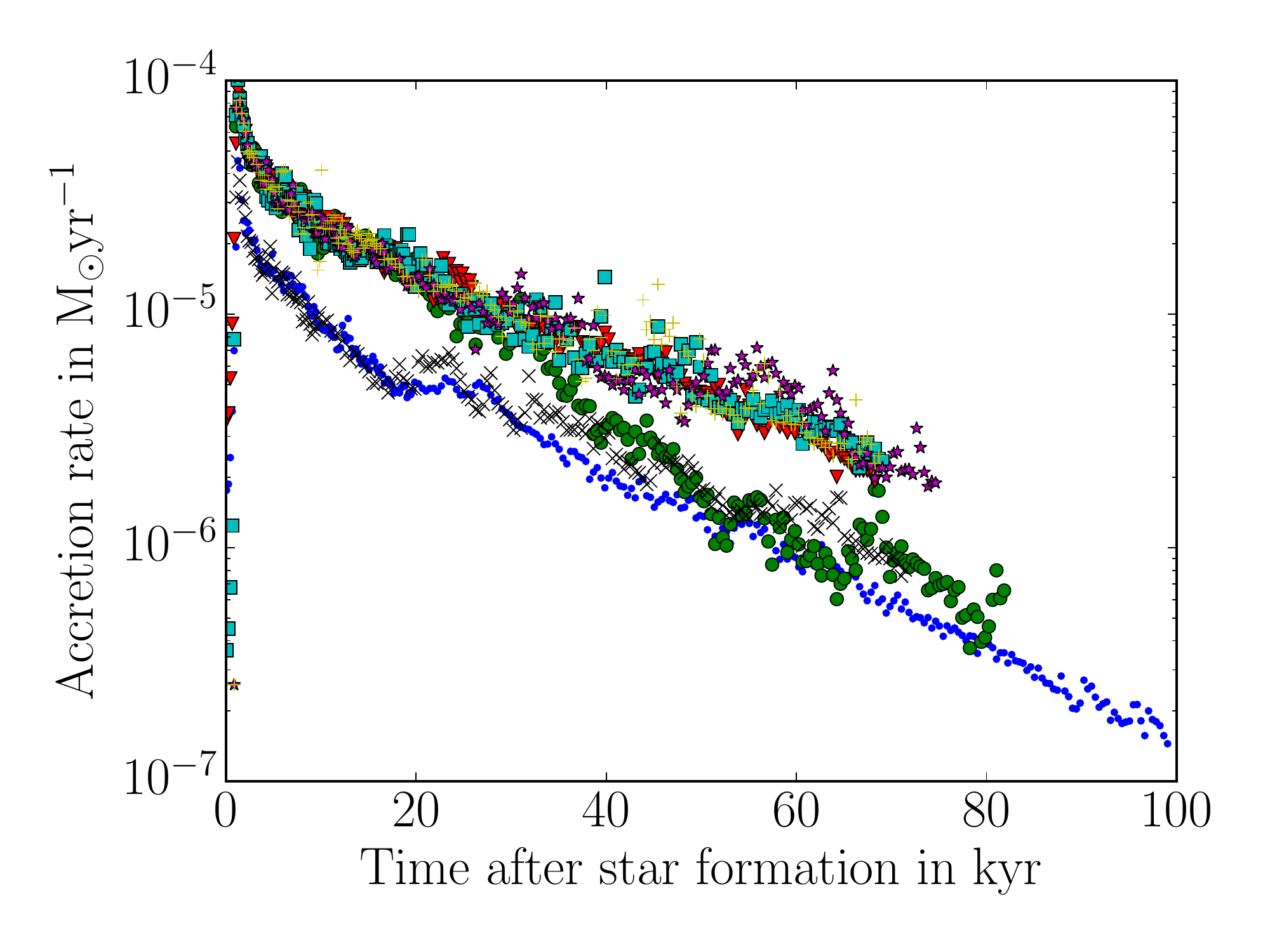} } \quad
\subfigure{\includegraphics[width=0.5\linewidth]{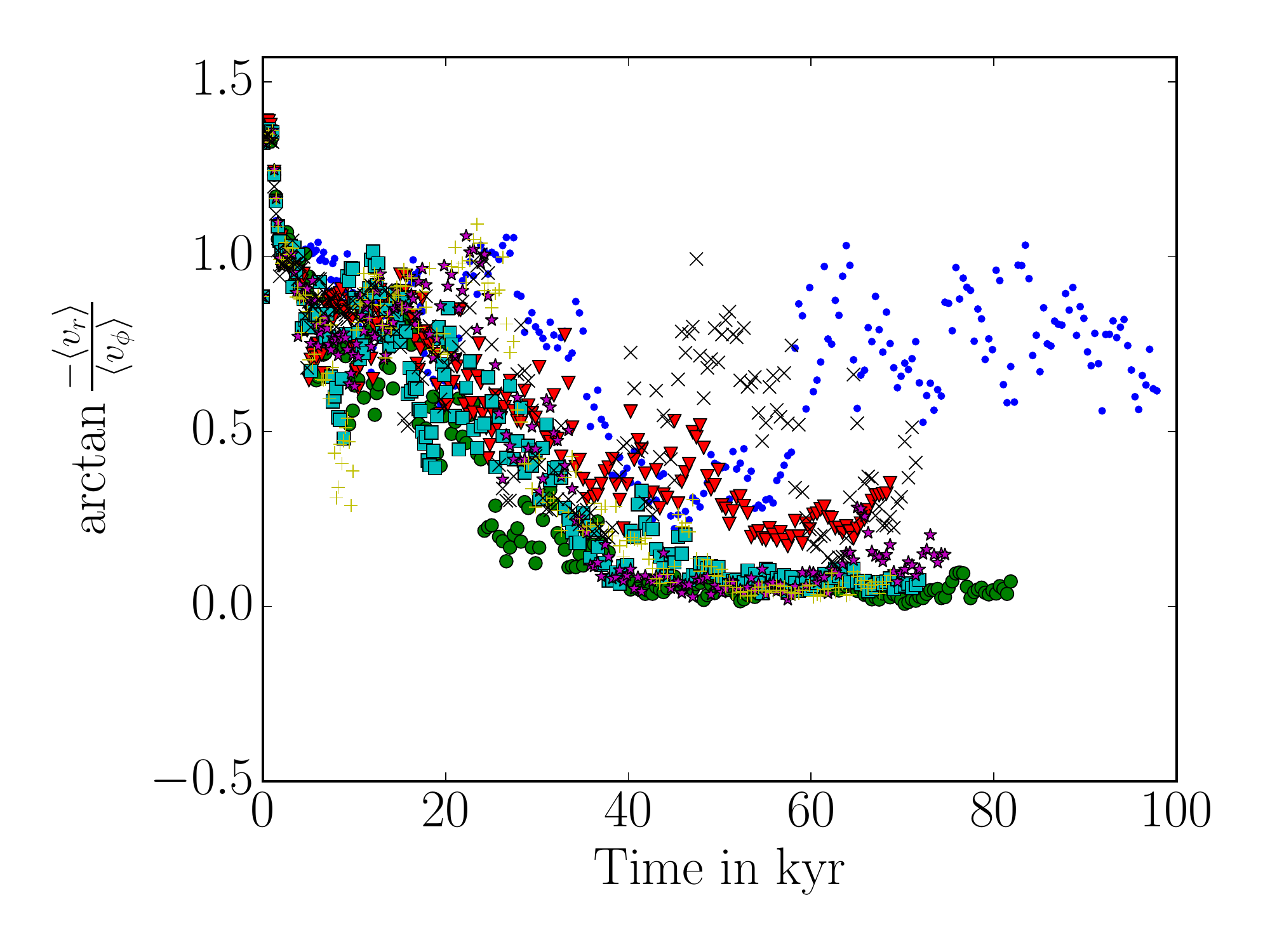} } \quad
\protect\caption{\label{fig:acc-prof_070} Accretion profile (left) and $\alpha = \arctan{\frac{- \langle v_r \rangle }{\langle v_{\phi} \rangle}}$ within 50 AU (right) for the zoom-ins with different sink settings of sink 1, from t=0 until 60 to 100 kyr after sink creation,
with the settings given in Table \ref{sink-params-070}, with maximum resolution of 2 AU, and with output cadence of 200 to 400 yrs.
Blue dots corresponds to run 1a, red triangles to run 1b, cyan squares to run 1c,
magenta asterisks to sink 1d, yellow pluses to run 1e, green circles to run 1f and black crosses to run 1g.}
\end{figure}

We started each run from shortly before the time of sink formation on level 22 and evolved the runs for up to 100 kyr.
In the left panel of \Fig{acc-prof_070} we illustrate the accretion profiles of the different runs.
In run 1a (blue dots), the accretion rate quickly increases to its maximum accretion rate of
several $10^{-5}$ $M_{\odot}$$yr^{-1}$ at the very beginning of the accretion process.
Afterwards, the accretion rate drops more or less continuously --- with a small shoulder
around 20 kyr --- to an accretion rate of $10^{-7}$ \Macc.
In run 1b (red triangles), we apply the same setting except for increasing \acceff\ from 50 \% to 100 \%.
The overall characteristic of a steep initial increase followed by a continuous relaxation is the same.
As expected, the higher accretion efficiency parameter causes a higher accretion rate during the evolution.
In run 1c (cyan squares), we decreased \accrate\ to $10^{-3}$ and used otherwise identical setting as for 1b.
The differences are very subtle, showing that lowering the instantaneous accretion rate just lets the sink
particle accrete more slowly or across more cells, even though sometimes the lower rate results in
a significant pileup.
Run 1c shows stronger fluctuations than 1a, including episodes of accretion bursts,
while generally following accretion rates of similar strength as 1b.
In run 1d (magenta asterisks), we decreased \racc\ to $7.5$ cells, with other parameters the same as in 1c.
Similar to 1c, the accretion rate shows fluctuations of up to a factor of 2.
The fluctuations -- especially at later times -- do not overlap with the ones in 1c, and seem to be somewhat stronger.
In run 1e (yellow pluses), we increased $\rhofrac$\ from $10^{-4}$ to $10^{-1}$ with respect to 1d and find a profile that shows a more smooth evolution.
In run 1f (green circles), we decrease \racc\ to 4 cells, with other parameters as in 1e. We again find fluctuations in the accretion of mass onto the sink.
Moreover, we see a striking difference, in that the accretion profile drops steeper after about 30 kyr than seen for the other sinks with \acceff\ $=1$.
After about 50 kyr and apart from the significant fluctuations, the strength of the accretion rate is close to the accretion rate for the run with \acceff\ $=0.5$ (1a).
Finally, we carry out a comparison run 1g, with the same settings as in 1f, but with \acceff\ $=0.5$ (black crosses).
We find that the profile is very similar to run 1a (both in absolute accretion rate and in the shape of the profile) for about the first 30 kyr and then shows stronger fluctuations.
Run 1g does not show the significant drop in the accretion rate that was seen in 1f.

The most striking difference is the effect of the efficiency parameter, which separates the accretion profiles in two groups.
The sinks that accumulate 100 $\%$ of the mass selected in their surroundings (1b to 1f)
show higher mass accretion rates than in the case where $50 \%$ of that mass is removed from the box (run 1a and run 1g).
Generally, the profiles then evolve in a similar manner, with some smaller fluctuations and more
or less strongly evolved brief burst periods (e.g.~run 1e at about $t=50$ kyr) until the end of the simulation.
However, there is one particular exception (run 1f), which evolves in a similar fashion as the other runs with \acceff\ $=100 \%$ until about $t=35$ kyr
before the accretion rate drops more steeply until at about $t=50$ kyr, and finally follows more the evolution of the sinks with low \acceff,
though with significantly stronger bursts than run 1a.
Both the stronger amount of fluctuations in the accretion profiles and the fall-off of run 1f can be understood by studying the velocity profiles around the sinks.

In the right panel of \Fig{acc-prof_070}, we illustrate the evolution of $\alpha$ for the different sinks
and find that for some settings a significant rotational gas motion evolves,
whereas the velocity profile for other settings is mostly infall dominated during the entire evolution.
The accretion profile of run 1a evolves rather smoothly and calmly,
because there is no sign of significant disk formation for this setting of the sink parameters.
The sinks with a surrounding disks show more intermittent accretion
profiles, because clumps of mass rotating in the disk may eventually fall into the sink.
Furthermore, the drop-off in the accretion profile in run 1f coincides with the drop in $\alpha$, and thus with the formation of a disk.
The fact that a disk starts to form means that infalling mass does not accrete directly onto the star, but starts to build up a disk first.
Therefore, the accretion profile drops more significantly compared to the runs where no disk or only a weak disk forms.
We interpret the fact that disk formation occurs more strongly for high \acceff\ as a direct consequence of the deeper gravitational potential induced by the more massive sinks.

\begin{table}[!htbp]
\centering
{
\begin{tabular}{r|r|r|r|r|r}

\begin{tabular}{@{}c@{}} \# of \\ sink \end{tabular} &
\begin{tabular}{@{}c@{}} \cB \end{tabular} &
\begin{tabular}{@{}c@{}} \acclim \end{tabular} &
\begin{tabular}{@{}c@{}} \racc \end{tabular} &
\begin{tabular}{@{}c@{}} \accrate \end{tabular} &
\begin{tabular}{@{}c@{}} \acceff \end{tabular} \\ \hline
1a & 100      & $1.7\times 10^{-19}$ & 22.5  & $10^{-2}$ & 0.5 \\
1b & 100      & $1.7\times 10^{-19}$ & 22.5  & $10^{-2}$ & 1.0 \\
1c & 100      & $1.7\times 10^{-19}$ & 22.5  & $10^{-3}$ & 1.0 \\
1d & 100      & $1.7\times 10^{-19}$ & 7.5  & $10^{-3}$ & 1.0 \\
1e & 100      & $1.7\times 10^{-16}$ & 7.5  & $10^{-3}$ & 1.0 \\
1f & 200      & $1.7\times 10^{-16}$ & 4  & $10^{-3}$ & 1.0 \\
1g & 100      & $1.7\times 10^{-16}$ & 4  & $10^{-3}$ & 0.5 \\
\end{tabular} }
\caption{Overview of the seven different settings for the study of the sink parameters.
All simulations were run with refinement based on density, gradients in density, and gradients in velocity
at all levels of refinement.
First column: number of sink,
second column: value of the fast magnetosonic speed at which the code starts using the LLF solver instead of the HLLD solver [km\ s$^{-1}$],
third column: limit above which gas is considered for accretion [g cm$^{-3}$],
fourth column: radius inside which gas is considered for accretion measured in cell widths at highest level,
fifth column: fraction of mass that is accreted,
sixth column: fraction of the accreted mass that is added to the sink (to account for outflows).}
\label{sink-params-070}
\end{table}

From the results we conclude that the general profile of the sink accretion is
robust to variations of the sink parameter settings over longer time scales.
However, changes (particularly of \acceff) affect the disk formation process,
and therefore also the accretion process on small time-scales.
If the sink forms a circumstellar disk, the sink accretes mass in occasional bursts, causing
a more intermittent accretion profile but affecting the long-term accretion profile only modestly.
The parameter study illustrates the difficulties in choosing physically correct parameter settings.
We selected the settings from run 1a, which shows the weakest sign of disk formation for
the zoom-in simulations, for the sinks that formed on highest level of refinement.
This was a conservative choice in terms of determining the efficiency of disk formation,
because choosing a higher accretion efficiency increases the probability of disk formation.
Therefore, the frequency of disk formation we find is likely to be a lower limit, and
would have been higher if we had chosen settings more favorable for disk formation.

In comparison to effects due to the choice of sink particle parameters, limited
numerical resolution is not likely to play a significant role, since accretion rates and
frequencies of disk formation were largely similar in early versions of these zoom simulations,
where the number of cells per Jeans' length was  significantly smaller \citep{2014IAUS..299..131N}.

\section{Angular Momentum Conservation}
Here we summarize how the change in angular momentum in a volume is related to flux densities through the surface
of the volume. Our derivation is similar to that given in
\citet{2012A&A...543A.128J}, but more compact, since we use a conservative formulation of the Euler equation. It is also
coordinate independent, and contains pressure terms, which canceled out for the specific geometry of the volume and
angular momentum component considered by \citet{2012A&A...543A.128J}.

The Euler equation in conserved form may be written
\begin{equation}
\frac{\partial \rho \boldsymbol{v}}{\partial t} = -\boldsymbol{\nabla}\cdot\left[\rho\boldsymbol{v}\otimes\boldsymbol{v}
-\frac{1}{4\pi}\boldsymbol{B}\otimes\boldsymbol{B}+\frac{1}{4\pi G}\boldsymbol{\nabla}\Phi\otimes\boldsymbol{\nabla}\Phi\right]
- \boldsymbol{\nabla} P_\textrm{tot}\,,
\end{equation}
where the total pressure is
\begin{equation}
P_\textrm{tot} = P + \frac{\boldsymbol{B}^2}{8\pi} + \frac{(\boldsymbol{\nabla}\Phi)^2}{8\pi G}
\end{equation}
and we have used a conserved formulation of the gravitational force \citep{2013NewA...19...48J}.
The total angular momentum inside a volume with surface $S$ is
\begin{equation}
\boldsymbol{L} = \int \boldsymbol{r} \times \rho \boldsymbol{v} \,\textrm{d}V
\end{equation}
The time evolution of the total angular momentum can be calculated by integrating the time evolution of the angular momentum density
$\boldsymbol{l} = \boldsymbol{r}\times \rho \boldsymbol{v}$, which is directly related to the Euler equation.
To transform from $\rho\boldsymbol{v}$ to $\boldsymbol{l}$ we use the tensor identity
\begin{align}\nonumber
\boldsymbol{r}\times\boldsymbol{\nabla}\cdot(\boldsymbol{X} \otimes \boldsymbol{Y}) & =
\epsilon_{ijk} \boldsymbol{r}_j \partial_l \left[\boldsymbol{X}_l \boldsymbol{Y}_k\right] \boldsymbol{e}_i \\
\label{eq:Tid}
& = \partial_l \left[ \boldsymbol{X}_l \epsilon_{ijk} \boldsymbol{r}_j \boldsymbol{Y}_k \right] \boldsymbol{e}_i
- \epsilon_{ijk} \boldsymbol{X}_j \boldsymbol{Y}_k \boldsymbol{e}_i \\ \nonumber
& = \boldsymbol{\nabla}\cdot \left[ \boldsymbol{X} \otimes (\boldsymbol{r} \times \boldsymbol{Y}) \right]
- \boldsymbol{X} \times \boldsymbol{Y}\,,
\end{align}
where we have written it out in component form too (using the Einstein summation convention),
to explicitly specify which tensor index the divergence applies to. The flux terms in the Euler equation are all symmetric tensors,
and as a consequence the last term in Eq.~(\ref{eq:Tid}) disappears, leaving a pure divergence. The pressure term may be written
\begin{equation}
\boldsymbol{r} \times \boldsymbol{\nabla} P_\textrm{tot} = - \boldsymbol{\nabla} \times (P_\textrm{tot} \boldsymbol{r})
\end{equation}
Transforming the volume integral into surface integrals we find the time evolution of the total angular momentum
expressed as a function of the flux through the surface
\begin{equation}\label{eq_angmom}
\frac{\textrm{d} \boldsymbol{L}}{\textrm{d} t} = - \int_S (\boldsymbol{r} \times \rho \boldsymbol{v}) (\boldsymbol{v} \cdot \boldsymbol{n})
-\frac{1}{4\pi} (\boldsymbol{r} \times \boldsymbol{B}) (\boldsymbol{B} \cdot \boldsymbol{n})
+\frac{1}{4\pi G} (\boldsymbol{r} \times \boldsymbol{\nabla}\Phi) (\boldsymbol{\nabla}\Phi \cdot \boldsymbol{n}) \,\textrm{d}A
- \int_S P_\textrm{tot}\, \boldsymbol{r} \times \boldsymbol{n} \,\textrm{d}A\,,
\end{equation}
where $\boldsymbol{n}$ is the normal vector to the surface element $\textrm{d}A$.

If our control volume is a sphere, the pressure term disappears, since $\boldsymbol{n} = \boldsymbol{e}_r$. By rotational
symmetry this is also the case for the $x$- and $y$-component.
In the case where the control volume is a cylinder the pressure term also disappears, but only for the $z$-component of the angular
momentum. If instead a cubical test volume is used, the pressure term has to be included when computing the time derivative of any component of the angular momentum.

In cylindrical coordinates, the change in the $z$-component of the angular momentum may be split up in a contribution from
the cylinder wall, assumed to be at $r=R$, and contributions from the top and bottom layer at $z=\pm h/2$.
The contribution from the cylinder wall is
\begin{equation}
\left.\frac{\textrm{d} L_z}{\textrm{d} t}\right|_\textrm{cyl,r} = - \int^{h/2}_{-h/2} \textrm{d}z \int^{2\pi}_0 R\, \textrm{d}\phi\, R \left[
 \rho v_\phi v_r  - \frac{1}{4\pi} B_\phi B_r + \frac{1}{4\pi G} (\boldsymbol{\nabla}\Phi)_\phi (\boldsymbol{\nabla}\Phi)_r \right]\,,
\end{equation}
where all variables are evaluated at the fixed radius $R$. The contributions from the top or bottom layer at fixed height $\pm h/2$ are
\begin{equation}
\left.\frac{\textrm{d} L_z}{\textrm{d} t}\right|_\textrm{cyl,top/bot} = \mp \int^R_0 \textrm{d}r \int^{2\pi}_0 r\, \textrm{d}\phi\, r \left[
 \rho v_\phi v_z  - \frac{1}{4\pi} B_\phi B_z + \frac{1}{4\pi G} (\boldsymbol{\nabla}\Phi)_\phi (\boldsymbol{\nabla}\Phi)_z \right]\,,
\end{equation}
where the sign differs according to if it is top or bottom layer respectively that is calculated.

In spherical coordinates the $z$-component of the angular momentum may be written
\begin{equation}
\left.\frac{\textrm{d} L_z}{\textrm{d} t}\right|_\textrm{sph} =
- \int^{\pi}_0 R\, \textrm{d}\theta \int^{2\pi}_0 R \sin\theta\, \textrm{d}\phi\, R \sin\theta
\left[ \rho v_\phi v_r  - \frac{1}{4\pi} B_\phi B_r + \frac{1}{4\pi G} (\boldsymbol{\nabla}\Phi)_\phi (\boldsymbol{\nabla}\Phi)_r \right]\,,
\end{equation}
where all variables are evaluated at the fixed radius $R$ and the coordinates are $(r, \phi, \theta)$, with $\theta$ being the polar angle.

\section{Grid alignment}
\begin{figure*}[!htbp]
\subfigure{\includegraphics[width=0.33\linewidth]{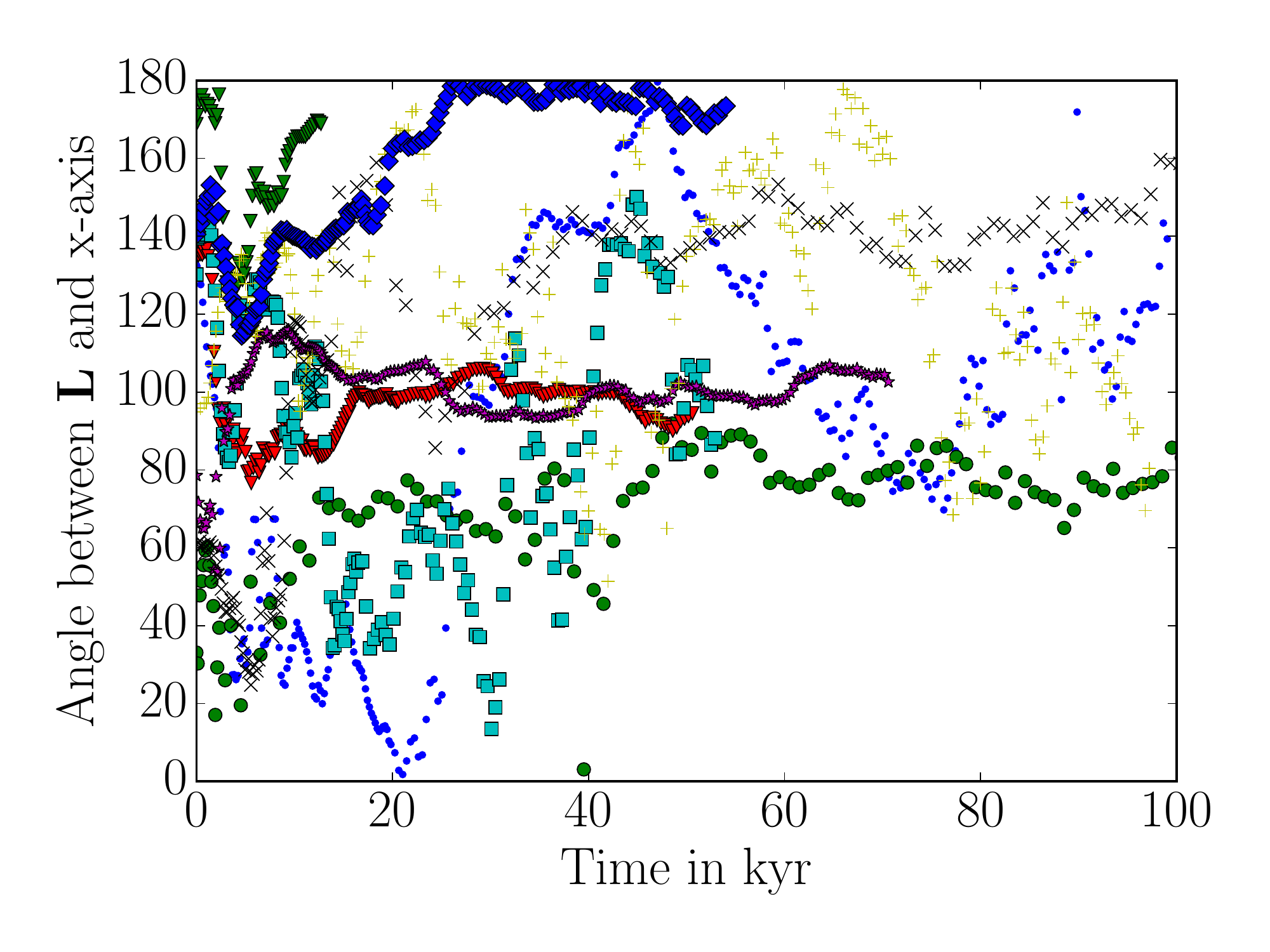} }
\subfigure{\includegraphics[width=0.33\linewidth]{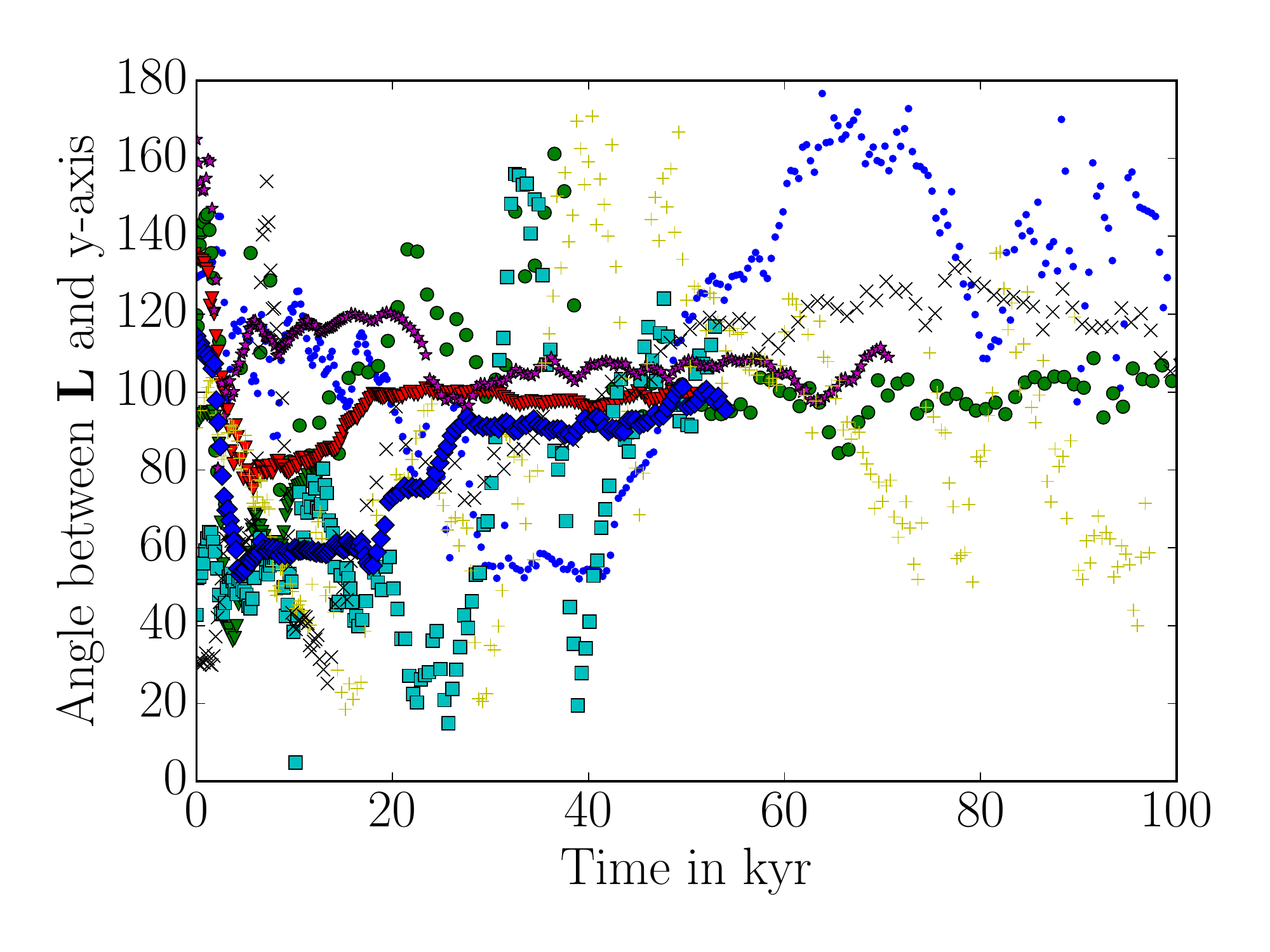} }
\subfigure{\includegraphics[width=0.33\linewidth]{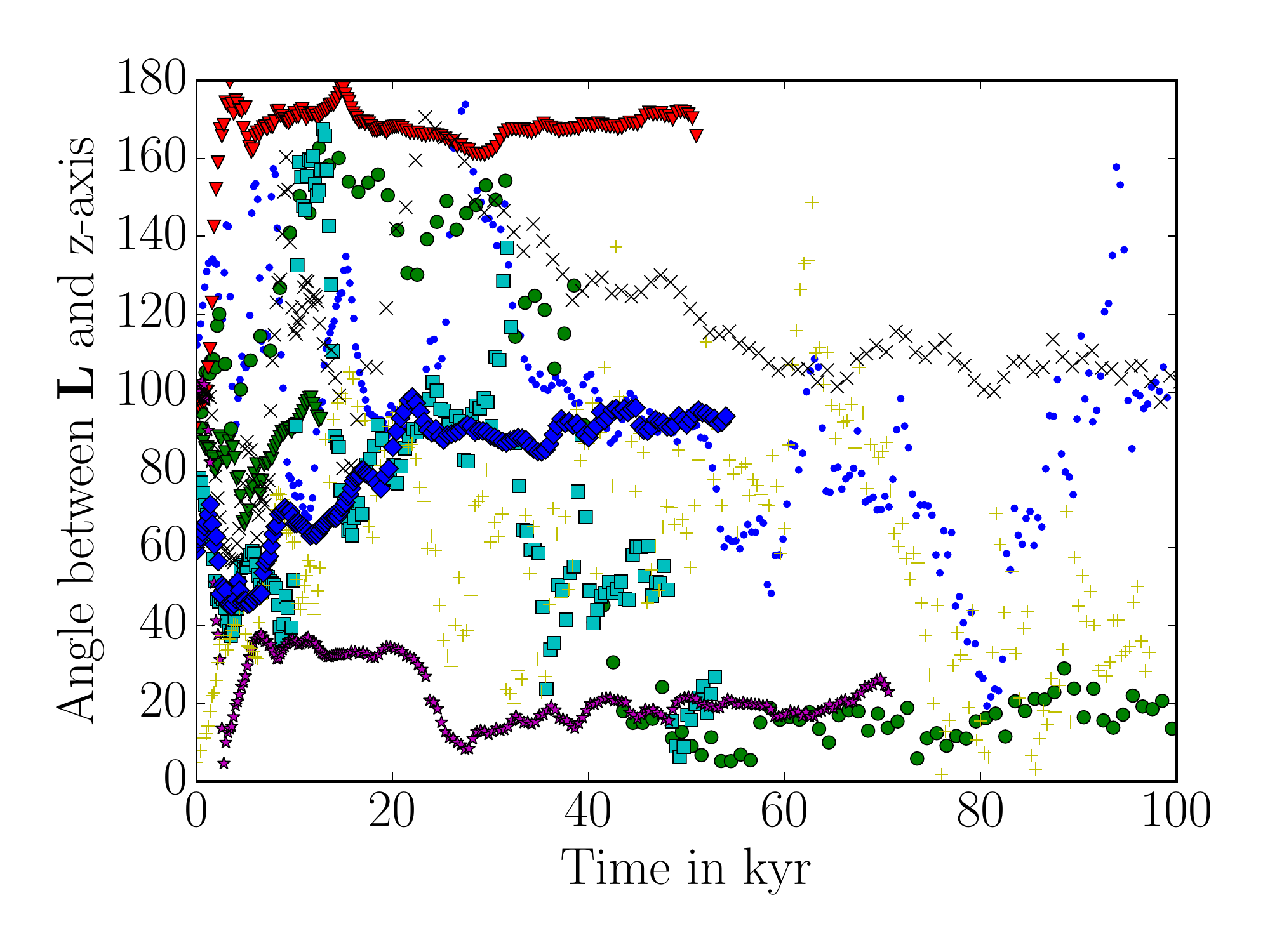} }
\protect\caption{\label{fig:L-x_100} Evolution of the angle between the total angular momentum vector of the gas within 100 AU from the sink
and the three coordinate-axes (left panel: x-axis, middle panel: y-axis, right panel: z-axis).
The symbols belong to the same sinks as in \Fig{acc_mass_run9}.}
\end{figure*}

Given the turbulent motions in the GMC, the mean angular momentum vector of the collapsing core
is expected to be randomly oriented with respect to the grid in our simulation.
\Fig{L-x_100} shows the orientation of the mean angular momentum vector calculated
within a sphere of 100 AU radius around the sinks during the evolution of the zoom runs.
At t=0, we find a variation of different orientations of the mean angular momentum vector
as expected from the underlying turbulence.
However, with evolving time we see a tendency of alignment of the mean angular momentum vector
with one of the coordinate axis for the runs that formed protoplanetary disks.
This alignment is caused by numerical effects in our grid code \ramses \ as pointed out by
\citet{2015MNRAS.450...53H}.
Certainly the presence of grid alignment makes a detailed study of the disk dynamics challenging.
The artificially aligned disk may induce additional torques on infalling gas
and in this way enhance (or suppress) the rotational support of a disk.
Previous local core collapse studies in grid codes circumvented this problem
by aligning the rotational axis of the gas motion with
one of the coordinate axes.

We point out that the problem of grid alignment is a fundamental one, likely
affecting also simulations that are carried out with other grid codes.
Fortunately, grid alignment can in principle be avoided.
The tendency of alignment is induced by calculating the velocity components at the center of cell surfaces
with 1D- or 2D-Riemann solvers.
This combination favors the momentum contribution along the coordinate axes.
A uniformly applied higher resolution may reduce this effect sufficiently,
although this would be computationally expensive and it would not address the core of the problem.
Applying a method where the stress terms are explicitly represented by a symmetric
tensor would likely reduce this problem significantly, as would
an implementation of a fully three-dimensional Riemann solver \citep{2015JCoPh.295....1B}.

Nevertheless, we are confident about the statistical differences induced by the variable stellar environments.
Numerical effects may influence the outcome of our runs to some degree, but the detected differences in the accretion
and protoplanetary disk formation process
are too large to be solely explained by numerical effects, whereas they are in line with the 'inherited' variations
from the stellar environments.

\end{document}